\documentclass[twocolumn, onecolappendix, times]{aastex63} 
\usepackage{amsmath}
\usepackage{soul}
\usepackage{natbib}
\usepackage{mathtools}
\graphicspath{{./}{Pictures/}}

\shorttitle{Dynamical Masses and Stellar Evolutionary Model Predictions of M-Stars}
\shortauthors{J. Pegues et al.}

\begin{document}

\title{Dynamical Masses and Stellar Evolutionary Model Predictions of M-Stars}



\author{Jamila Pegues}
\affiliation{
Center for Astrophysics $\mid$ Harvard \& Smithsonian,  Cambridge, MA 02138, USA}

\author{Ian Czekala}
\altaffiliation{NASA Hubble Fellowship Program Sagan Fellow}
\affiliation{Department of Astronomy and Astrophysics, 525 Davey Laboratory, The Pennsylvania State University, University Park, PA 16802, USA}
\affiliation{Center for Exoplanets and Habitable Worlds, 525 Davey Laboratory, The Pennsylvania State University, University Park, PA 16802, USA}
\affiliation{Center for Astrostatistics, 525 Davey Laboratory, The Pennsylvania State University, University Park, PA 16802, USA}
\affiliation{Institute for Computational \& Data Sciences, The Pennsylvania State University, University Park, PA 16802, USA}
\affiliation{Department of Astronomy, 501 Campbell Hall, University of California, Berkeley, CA 94720-3411, USA}

\author{Sean M. Andrews}
\affiliation{
Center for Astrophysics $\mid$ Harvard \& Smithsonian,  Cambridge, MA 02138, USA}

\author{Karin I. \"Oberg}
\affiliation{
Center for Astrophysics $\mid$ Harvard \& Smithsonian,  Cambridge, MA 02138, USA}

\author{Gregory J. Herczeg}
\affiliation{
Kavli Institute for Astronomy and Astrophysics, Peking University, Yiheyuan Lu 5, Haidian Qu, 100871 Beijing, People’s Republic of China}

\author{Jennifer B. Bergner}
\altaffiliation{NASA Hubble Fellowship Program Sagan Fellow}
\affiliation{
Department of Geophysical Sciences, University of Chicago, Chicago, IL 60637, USA}

\author{L. Ilsedore Cleeves}
\affiliation{
Astronomy Department, University of Virginia, Charlottesville, VA 22904, USA}

\author{Viviana V. Guzm\'an}
\affiliation{Instituto de Astrof{\'i}sica, Ponticia Universidad Cat{\'o}lica de Chile, Av.~Vicu{\~n}a Mackenna 4860, 7820436 Macul, Santiago, Chile}

\author{Jane Huang}
\altaffiliation{NASA Hubble Fellowship Program Sagan Fellow}
\affiliation{
Department of Astronomy, University of Michigan, 323 West Hall, 1085 S. University Avenue, Ann Arbor, MI 48109, USA}

\author{Feng Long}
\affiliation{
Center for Astrophysics $\mid$ Harvard \& Smithsonian,  Cambridge, MA 02138, USA}

\author{Richard Teague}
\affiliation{
Center for Astrophysics $\mid$ Harvard \& Smithsonian,  Cambridge, MA 02138, USA}

\author{David J. Wilner}
\affiliation{
Center for Astrophysics $\mid$ Harvard \& Smithsonian,  Cambridge, MA 02138, USA}

\begin{abstract}
In this era of \textit{Gaia} and ALMA, dynamical stellar mass measurements, derived from spatially and spectrally resolved observations of the Keplerian rotation of circumstellar disks, provide benchmarks that are independent of observations of stellar characteristics and their uncertainties.  These benchmarks can then be used to validate and improve stellar evolutionary models, the latter of which can lead to both imprecise and inaccurate mass predictions for pre-main-sequence, low-mass ($\leq$0.5M$_\Sun$) stars.
We present the dynamical stellar masses derived from disks around three M-stars (FP Tau, J0432+1827, and J1100-7619) using ALMA observations of $^{12}$CO (J=2--1) and $^{13}$CO (J=2--1) emission.  These are the first dynamical stellar mass measurements for J0432+1827 and J1100-7619 (0.192$\pm$0.005 M$_\Sun$ and 0.461$\pm$0.057 M$_\Sun$, respectively) and the most precise measurement for FP Tau (0.395$\pm$0.012 M$_\Sun$).
%
Fiducial stellar evolutionary model tracks, which do not include any treatment of magnetic activity, agree with the dynamical stellar mass measurement of J0432+1827 but underpredict the mass by $\sim$60\% for FP Tau and $\sim$80\% for J1100-7619.  Possible explanations for the underpredictions include inaccurate assumptions of stellar effective temperature, undetected binarity for J1100-7619, and that fiducial stellar evolutionary models are not complex enough to represent these stars.  In the former case, the stellar effective temperatures would need to be increased by amounts ranging from $\sim$40K to $\sim$340K to reconcile the fiducial stellar evolutionary model predictions with the dynamically-measured masses.  In the latter case, we show that the dynamical masses can be reproduced using results from stellar evolutionary models with starspots, which incorporate fractional starspot coverage to represent the manifestation of magnetic activity.
Folding in low-mass M-stars from the literature and assuming that the stellar effective temperatures are imprecise but accurate, we find tentative evidence of a relationship between fractional starspot coverage and observed effective temperature for these young, cool stars.

\end{abstract}


\section{Introduction}
\label{sec_introduction}

M-type stars are the most common hosts of planetary systems in the local Galaxy~\citep[e.g.,][]{cite_henryetal2006, cite_dressingetal2015, cite_muldersetal2015, cite_henryetal2016}.  Accurate characterization of the stars' masses at young ages (i.e., pre-main-sequence) provides critical insight into their thermal and radiative environment, and consequently the chemical conditions where planets form.

Stellar evolutionary models are often used to predict the masses of young stars from observational estimates of their luminosities and effective temperatures.  These masses are a key property for understanding not only individual stellar evolution, but also entire star-forming regions, as mass is crucial for building initial mass functions and other statistical analyses of large populations of stars~\citep[e.g.,][]{cite_hillenbrandetal1997}. 

Unfortunately, characterizing low-mass ($\leq$0.5M$_\Sun$), pre-main-sequence (pre-MS) stars is observationally and theoretically challenging.
Observationally, uncertainties on the observed effective temperatures and luminosities lead to imprecise model predictions.  Theoretically, these stars are convective, have not yet reached hydrostatic equilibrium, and can exhibit strong magnetic activity, all of which are difficult states to model.  Stellar evolutionary models that intrinsically simplify these complex physics can lead to inaccurate mass predictions.
Different models can also assign different luminosities and effective temperatures to the same stellar mass and age, making comparison across models difficult~\citep[e.g.,][]{cite_hillenbrandetal2004, cite_belletal2012, cite_stassunetal2014a, cite_feidenetal2016}.

These uncertainties highlight the importance of benchmarking stellar evolutionary model predictions for the young, low-mass stellar regime.  Benchmarks in this context are stars for which fundamental stellar properties, namely mass, have been independently, precisely, and accurately measured.
Pre-MS star systems with circumstellar disks can be ideal benchmarks, because the Keplerian rotation profiles of the disks, observed in molecular line emission, can be used to dynamically estimate the stellar masses~\citep[e.g.,][]{cite_koerneretal1993, cite_dutreyetal1998, cite_guilloteauetal1998, cite_simonetal2001, cite_rosenfeldetal2012b}.
Within the past half-decade, well-resolved molecular emission from the Atacama Large Millimeter/submillimeter Array (ALMA) has enabled tight constraints on these disk-based dynamical masses, while precise parallax measurements from Gaia~\citep[e.g.,][]{cite_gaia2016, cite_gaia2018b} have eliminated the linear distance degeneracy intrinsic to rotation-based dynamical estimates.  These technological advancements have greatly improved dynamical mass measurements from circumstellar disks, permitting fainter targets and increasing measurement precision~\citep[e.g.,][]{cite_lindbergetal2014, cite_czekalaetal2015, cite_czekalaetal2016, cite_whiteetal2016, cite_czekalaetal2017, cite_simonetal2017, cite_wuetal2017, cite_czekalaetal2019, cite_simonetal2019}.

Recently,~\cite{cite_simonetal2019} presented dynamical stellar masses of 0.10-2.11M$_\Sun$ measured from 38 disks around unary, binary, and ternary systems in the Taurus and Ophiuchus star-forming regions.  These dynamical masses were either (1) newly measured or (2) updated from previous publications using precise \textit{Gaia} distance measurements.  The authors found that stellar evolutionary models without any treatment of magnetic fields underpredicted masses in the dynamically-measured $\sim$0.4-1.4M$_\Sun$ range by $\sim$30\% relative to the dynamical mass.

To date, however, stellar evolutionary model performance for low-mass M-stars remains poorly characterized, simply because so few benchmarks exist for M-stars $\leq$0.5M$_\Sun$.  Disk-based dynamical mass measurements for low-mass M-stars are more difficult due to their relatively small and faint disks, often necessitating high-resolution and high-sensitivity observations to distinguish the Keplerian structure.  The aforementioned compilation of~\cite{cite_simonetal2019} presents dynamical masses for nine $\leq$0.5M$_\Sun$ unary stars: two have stellar masses of 0.1M$_\Sun$ and 0.3M$_\Sun$, another three are from 0.36-0.38M$_\Sun$, and the final four are from 0.41-0.47M$_\Sun$.  Five Keplerian disk candidates from the Orion Nebula cluster are presented in~\cite{cite_boydenetal2020} with dynamical stellar masses $\leq$0.5M$_\Sun$, but the authors noted that cloud contamination and low signal-to-noise observations are likely affecting their measurements, and they called for higher-sensitivity observations to confirm the measurements and reduce the mass uncertainties.  There are also existing low-mass benchmarks in multi-star systems, such as the low-mass stellar companion (DH Tau, 0.10M$_\Sun$) presented in the~\cite{cite_sheehanetal2019} study, but we focus on unary star systems in this work.
%
In light of the uncertainties and inaccuracies of model predictions found for existing pre-MS benchmarks~\citep[e.g.,][]{cite_hillenbrandetal2004, cite_lopezmoralesetal2005, cite_mathieuetal2007, cite_gennaroetal2012, cite_tognellietal2012, cite_stassunetal2014a, cite_feidenetal2016, cite_simonetal2019},
studies have called for (1) more precise constraints on luminosity and effective temperature, (2) more dynamical low-mass benchmarks, and (3) more concrete evaluation of stellar evolutionary model performance for young, cool stars~\cite[e.g.,][]{cite_mathieuetal2007, cite_simonetal2019}. 


In this study, we present in-depth analysis of dynamical masses, measured using the code \textsc{DiskJockey}~\citep[e.g.,][]{cite_czekalaetal2015, cite_czekalaetal2016}, for three unary low-mass M-stars ($\leq$0.5M$_\Sun$) with protoplanetary disks.  Two of these masses have not been dynamically estimated before, while the third is now measured with higher precision.  We also evaluate the performance of stellar evolutionary models with varying degrees of magnetic activity for these low-mass stars.  In Section~\ref{sec_data}, we briefly discuss the three star+disk systems, observations, data reduction, and data formatting.  In Section~\ref{sec_methodology}, we describe the disk model and our inference procedure, implemented in the \textsc{DiskJockey} codebase.  In Section~\ref{sec_results}, we present the constrained stellar masses based on the $^{12}$CO (J=2--1) and $^{13}$CO (J=2--1) emission, the median disk models, and the residuals, and we compare them to fiducial stellar evolutionary model predictions.  In Section~\ref{sec_discussion} we discuss our results in the context of the other existing low-mass benchmarks and of different stellar evolutionary models, and in Section~\ref{sec_summary} we summarize our findings.

\begin{deluxetable*}{lccccccccc}
\tablecaption{Characteristics of the Sample. \label{table_sample}}
\tablehead{
Disk                                  & R.A.$^{[0]}$ & Decl.$^{[0]}$  & Region       & Dist.$^{[0]}$   & Spectral   & $L_*$ & $M_\mathrm{fid}^{[1]}$  & $t_\mathrm{fid}^{[1]}$            & $T_{\mathrm{eff}}^{[2]}$ \\
          & {(}J2000{)}   & {(}J2000{)}   &   &  {(}pc{)}  & Type &  ($L_\Sun$)    & ($M_\Sun$)  &   {(}Myr{)}   &  (K) }
\startdata
FP Tau     & 04:14:47.31 & 26:46:26.06  & Taurus       & 128$^{+2}_{-2}$ & M4$^{[3]}$         & 0.269$^{+0.033}_{-0.029}$ & 0.240$^{+0.091}_{-0.066}$ & 1.0$^{+0.6}_{-0.7}$ & 3311$^{+156}_{-149}$ \\
J0432+1827 & 04:32:22.12 & 18:27:42.36  & Taurus       & 141$^{+3}_{-3}$ & M4.75$^{[4]}$      & 0.076$^{+0.024}_{-0.018}$ & 0.174$^{+0.083}_{-0.048}$ & 2.5$^{+3.8}_{-1.3}$ & 3162$^{+149}_{-142}$ \\
J1100-7619 & 11:00:40.14 & -76:19:28.00 & Cha. I & 190$^{+4}_{-4}$ & M4$^{[5]}$         & 0.145$^{+0.085}_{-0.053}$ & 0.251$^{+0.051}_{-0.042}$ & 2.0$^{+3.0}_{-1.0}$ & 3311$^{+77}_{-75}$  
\enddata
\tablecomments{Right ascension (R.A.) and declination (decl.) coordinates and distances (Dist.) are from \textit{Gaia}~\citep[e.g.,][]{cite_gaia2016, cite_gaia2018b}.  \textit{Cha. I} is an abbreviation of \textit{Chamaeleon I}.  Stellar effective temperatures ($T_\mathrm{eff}$) were estimated from the spectral types using the logarithmic conversion illustrated in Figure 9 of~\cite{cite_luhmanetal2003}.  \textit{See Section~\ref{sec_discussion_inputs} for discussion of spectral types and effective temperatures assumed in other works.}  The stellar luminosities ($L_*$) were calculated using the spectral types, effective temperatures, and \textit{Gaia} distances and the methodology of~\cite{cite_andrewsetal2018scale}.  The stellar masses ($M_\mathrm{fid}$) and stellar ages ($t_\mathrm{fid}$) are predictions from tracks given by the MESA Isochrones and Stellar Tracks code~\citep[MIST;][]{cite_mist1, cite_mist2}, which used the listed effective temperatures and luminosities as inputs.  These fiducial model estimates do \textit{not} account for magnetic fields.  $*$: J0432+1827 and J1100-7619 are abbreviations of J04322210+1827426 and J11004022-7619280, respectively.  They are also known as MHO 6 and T10, respectively, in the literature.  [0]~\cite{cite_gaia2016,cite_gaia2018b}; [1]~\cite{cite_mist1, cite_mist2}; [2]~\cite{cite_luhmanetal2003}; [3]~\cite{cite_kenyonetal1995} using optical photometry; [4]~\cite{cite_bricenoetal2002} using optical spectroscopy; [5]~\cite{cite_manaraetal2017} using optical spectroscopy.}
\end{deluxetable*}

\begin{deluxetable}{l|ccc}
\tablecaption{Priors for Model Parameters. \label{table_modinit}}
\tablehead{
Parameters (Units)                               & \multicolumn{3}{c}{Priors} \\
   \     & FP Tau & J0432+1827   & J1100-7619    
}
\startdata
Distance (pc)                              & 128.4                & 141.9           & 191.5                    \\
$M_*$ (M$_\Sun$)                                   & 0.01 - 5.0      & 0.01 - 5.0      & 0.01 - 5.0      \\
$\theta_\mathrm{inc}$ ($^\circ$)                           & 10 - 80         & 10 - 80         & 0.01 - 80       \\
$\theta_\mathrm{P.A.}$ ($^\circ$)                    & 180 - 540       & 180 - 540       & 180 - 540       \\
$v_\mathrm{sys}$ (km s$^\mathrm{-1}$)                 & 7.55 - 9.05   & 4.85 - 6.35   & 4.0 - 5.5 \\
$\mu_\mathrm{RA}$ (")                      & (-1) - 1        & (-1) - 1        & (-1) - 1             \\
$\mu_\mathrm{Dec.}$ (")                    & (-1) - 1        & (-1) - 1        & (-1) - 1            \\
\hline
$q$                                      & 0.01 - 5.0      & 0.01 - 5.0      & 0.01 - 5.0      \\
$T_\mathrm{10}$ (K)                       & 5 - 80          & 5 - 80          & 5 - 80          \\
$\gamma$              &            1.0 & 1.0 & 1.0                      \\
$r_\mathrm{c}$ (AU)                 & 0 - 250       & 5 - 250       & 5 - 250   \\
$\log_{10} (M_\mathrm{gas}   (M_\Sun) )$ & (-8.0) - (-1.0) & (-8.0) - (-1.0) & (-8.0) - (-1.0) \\
$\xi$ (km s$^\mathrm{-1}$)             & 0.01 - 0.6      & 0.01 - 0.6      & 0.01 - 0.6         
\enddata
\tablecomments{Uniform priors for the thirteen model parameters of the \textsc{DiskJockey} code.  The distances and $\gamma$ values were fixed.  \textit{Gaia} distance uncertainties were incorporated into the final model constraints for the stellar masses \textit{after} fitting was complete (see Section~\ref{sec_methodology_parameters}).}
\end{deluxetable}

\begin{deluxetable*}{l|cc|cc|c}
\tablecaption{Stellar Mass and Primary Parameter Constraints. \label{table_params}}
\tablehead{
\                   & \multicolumn{2}{c|}{FP Tau}      & \multicolumn{2}{c|}{J0432+1827}      & \multicolumn{1}{c}{J1100-7619}    
}
\startdata
\ & $^{12}$CO   & $^{13}$CO & $^{12}$CO   & $^{13}$CO & $^{12}$CO \\   
$M_\mathrm{dyn}$ (M$_\Sun$)                    & \multicolumn{2}{c|}{0.395 $\pm$ 0.012}        & \multicolumn{2}{c|}{0.192 $\pm$ 0.005} & \multicolumn{1}{c}{0.461 $\pm$ 0.057}    \\
$M_*$ (M$_\Sun$)                    & 0.399$^{+0.0072}_{-0.007}$   & 0.379$^{+0.015}_{-0.014}$        & 0.185$^{+0.0041}_{-0.0041}$  & 0.203$^{+0.0052}_{-0.0051}$ & 0.461$^{+0.057}_{-0.046}$    \\
$\theta_\mathrm{inc}$   ($^\circ$)  & 65$^{+0.48}_{-0.46}$         & 63.5$^{+2}_{-2.2}$               & 56.9$^{+0.29}_{-0.28}$       & 59.8$^{+0.74}_{-0.76}$      & 15.5$^{+0.86}_{-0.91}$  \\
$\theta_\mathrm{P.A.}^*$   ($^\circ$) & 330$^{+0.32}_{-0.31}$        & 334$^{+1.4}_{-1.4}$              & 24$^{+0.18}_{-0.18}$        & 24$^{+0.53}_{-0.52}$       & 71$^{+0.16}_{-0.15}$        \\
$v_\mathrm{sys}$ (km s$^\mathrm{-1}$)     & 8.34$^{+0.0053}_{-0.0055}$   & 8.39$^{+0.029}_{-0.03}$          & 5.62$^{+0.0019}_{-0.0019}$   & 5.64$^{+0.0061}_{-0.0061}$  & 4.74$^{+0.00069}_{-0.00067}$ \\
$\mu_\mathrm{RA}$ (")         & 0.0816$^{+0.0011}_{-0.0011}$ & 0.0721$^{+0.0034}_{-0.0036}$     & 0.24$^{+0.0013}_{-0.0014}$   & 0.244$^{+0.0026}_{-0.0026}$ & -0.077$^{+0.0017}_{-0.0017}$  \\
$\mu_\mathrm{Dec.}$ (")           & -0.436$^{+0.0013}_{-0.0013}$ & -0.426$^{+0.004}_{-0.0038}$      & -0.264$^{+0.0015}_{-0.0015}$ & -0.267$^{+0.0031}_{-0.003}$ & 0.153$^{+0.0015}_{-0.0015}$ 
\enddata
\tablecomments{The final dynamical masses ($M_\mathrm{dyn}$) and the final constraints for the primary \textsc{DiskJockey} parameters.  All parameters were measured from $^{12}$CO and $^{13}$CO emission independently and are described in Section~\ref{sec_methodology_model}.  The listed values are the medians of the final sampling distributions, while the lower and upper uncertainties are the 16$^\mathrm{th}$ and 84$^\mathrm{th}$ percentiles, respectively (see Section~\ref{sec_methodology_parameters}).  No results are shown for $^{13}$CO toward J1100-7619 because the corresponding model was unconstrained.  The $M_\mathrm{dyn}$ values are described in Section~\ref{sec_results_mass}.  $*$: These $\theta_\mathrm{PA}$ values have been wrapped as necessary to be in the 0-360$^\circ$ range.  The wrapped median values were originally 384$^\circ$ for $^{12}$CO and $^{13}$CO toward J0432+1827 and 431$^\circ$ for $^{12}$CO toward J1100-7619.  No $\theta_\mathrm{PA}$ values were wrapped for FP Tau.}
\end{deluxetable*}




\section{Data}
\label{sec_data}

\subsection{Sample}

Our sample (Table~\ref{table_sample}) contains three of the five star+disk systems (FP Tau, J0432+1827, J1100-7619, J1545-3417, and Sz 69) presented in Pegues et al. (in rev.).  These systems were known to be bright in $^{12}$CO (J=3--2) and $^{13}$CO (J=3--2) from existing ALMA disk surveys~\citep[e.g.,][van der Plas et al. in prep.]{cite_ansdelletal2016, cite_longetal2017}.  We have excluded the two systems (J1545-3417 and Sz 69) that are severely cloud contaminated in both $^{12}$CO (J=2--1) and $^{13}$CO (J=2--1) emission.
The system with the biggest disk, J1100-7619, is located in the Chamaeleon I star-forming region and hosts an M4 star~\citep{cite_manaraetal2017}.  The other two systems, FP Tau and J0432+1827, are both located in the Taurus star-forming region and host M4 and M4.75 stars, respectively~\citep{cite_kenyonetal1995, cite_bricenoetal2002}.  
The assumed spectral types, stellar luminosities, and stellar effective temperatures were extracted from the literature and are listed in Table~\ref{table_sample}.
Using these assumed effective temperatures and luminosities as model inputs, tracks from the fiducial stellar evolutionary model~\citep[the MIST code;][]{cite_mist1, cite_mist2} predict stellar masses for the three systems of 0.240$^{+0.091}_{-0.066}$M$_\Sun$ (FP Tau), 0.174$^{+0.083}_{-0.048}$M$_\Sun$ (J0432+1827), and 0.251$^{+0.051}_{-0.042}$M$_\Sun$ (J1100-7619).  ``Fiducial'' here refers to models that do not include any treatment of magnetic activity.  As stellar evolutionary model predictions of ages are especially uncertain for these stars, due to nearly vertical model tracks in the low-mass pre-MS regime, we focus on the model predictions of mass within this paper.

\subsection{Observations, Reduction, and Formatting}

All three star+disk systems were observed during the Atacama Large Millimeter/submillimeter Array (ALMA) Project 2017.1.01107.S from December 2017 to September 2018.  On-source integration times per observation execution were $\sim$19-40min.  The native spectral resolution was 70.56kHz ($\sim$0.09km/s) and 141.11kHz ($\sim$0.18km/s) for $^{12}$CO (J=2--1) and $^{13}$CO (J=2--1), respectively.  The angular resolution was 0.13-0.27" and the maximum angular scale was 2.49-3.94".  For more details of the observations and a description of the data reduction, refer to Pegues et al. (in rev.).

After data reduction, the $^{12}$CO and $^{13}$CO emission was imaged with beam sizes of 0.24-0.41".  The channel maps are displayed in Appendix~\ref{sec_appendix_chans}.  The channel spacings were averaged and the channel ranges were restricted to reduce the total data volume of the measurement set (ms) files, while still preserving sensitivity to the Keplerian rotation of the disk.  The channel spacings and ranges are illustrated in Figure~\ref{fig_series}.  We used channel spacings of 0.40km s$^\mathrm{-1}$ for $^{12}$CO and $^{13}$CO toward FP Tau, 0.20km s$^\mathrm{-1}$ for $^{12}$CO and $^{13}$CO toward J0432+1827, and finally 0.08km s$^\mathrm{-1}$ for $^{12}$CO and 0.10km s$^\mathrm{-1}$ for $^{13}$CO toward J1100-7619.  J1100-7619 required the highest spectral resolution due to its lowest inclination angle.  For FP Tau, we considered velocity channels within 1.00-7.80km s$^\mathrm{-1}$ and 8.40-15.00km s$^\mathrm{-1}$.  These ranges exclude the cloud contamination seen near the systemic velocity in the channel maps for this disk.  For J0432+1827, we considered channels within 0.60-10.40km s$^\mathrm{-1}$.  Finally we considered channels within 3.15-6.27km s$^\mathrm{-1}$ and 2.75-6.55km s$^\mathrm{-1}$ for $^{12}$CO and $^{13}$CO, respectively, toward J1100-7619.


\section{Modeling}
\label{sec_methodology}

\subsection{DiskJockey}


We used the code \textsc{DiskJockey}~\citep{cite_czekalaetal2015, cite_diskjockey} to infer the stellar masses of the three star+disk systems.  \textsc{DiskJockey} works by forward-modeling the Keplerian rotation of a given disk.  The underlying parametric disk model is described in Section~\ref{sec_methodology_model}.  We represent the Keplerian rotation for each disk using the observed $^{12}$CO (J=2--1) and $^{13}$CO (J=2--1) molecular line emission.
\textsc{DiskJockey} performs all model fits and comparisons directly on the interferometric visibilities of the observations.  It uses \textsc{RADMC-3D}~\citep{cite_radmc3D} to perform radiative transfer and simulate the line emission.
Imaging of the median model and residuals is performed as the very last step of inspection.

\subsection{The Disk Model}
\label{sec_methodology_model}

\textsc{DiskJockey} adopts the disk model described in detail by~\cite{cite_rosenfeldetal2012b}.  Here we briefly describe the assumptions and the relevant parameters of the model.  Possible systematic errors due to the assumptions are discussed in Section~\ref{sec_results_residuals}.

The model assumes that the disk is geometrically thin and axisymmetric.  The gas is assumed to be in hydrostatic equilibrium with a vertically isothermal temperature structure.  The disk model is Keplerian, with stellar mass $M_*$ and systemic velocity $v_\mathrm{sys}$ in the Local Standard of Rest (LSR) frame.  $\xi$ is the non-thermal contribution to the line broadening, assumed to be constant across the disk.  The disk is a distance $d$ away.  Its orientation is described completely by the position and inclination angles $\theta_\mathrm{PA}$ and $\theta_\mathrm{inc}$, respectively.  $\theta_\mathrm{PA}$ is measured East of North, and $\theta_\mathrm{inc}$ is defined such that $\theta_\mathrm{inc}$=0$^\circ$ is face-on and $\theta_\mathrm{inc}$=90$^\circ$ is edge-on.  The disk center is offset from the image center along the R.A. and Dec. axes by $\mu_\mathrm{RA}$ and $\mu_\mathrm{Dec}$, respectively.

The radial gas temperature $T\{r\}$ of the disk model is described by a power law, with exponent $q$ and a fixed temperature at 10 AU of $T_{10}$.  The volume density is assumed to be Gaussian, with a scale height dependent on the radial temperature profile.  The surface density profile $\Sigma\{r\}$ is the similarity solution for a viscous accretion disk~\citep{cite_lyndenbelletal1974}.  The profile takes the form of an exponentially tapered power law, with exponent $\gamma$, characteristic radius $r_\mathrm{c}$, and characteristic surface density $\Sigma_\mathrm{c}$.  $\Sigma_\mathrm{c}$ is calculated from the mass of the molecular gas disk, $M_\mathrm{gas}$.

To calculate the molecular gas disk mass ($M_\mathrm{gas}$), \textsc{DiskJockey} assumes that the abundance ratio of $^{12}$CO relative to hydrogen nuclei is a constant equal to $1.5\times10^{-4}$.  The abundance ratios of $^{13}$CO and C$^{18}$O relative to $^{12}$CO are assumed to be constants equal to 1/69 and 1/557, respectively.

Altogether, this parametric disk model contains 13 parameters: seven primary geometric parameters \{$M_*$, $d$, $\theta_\mathrm{PA}$, $\theta_\mathrm{inc}$, $\mu_\mathrm{RA}$, $\mu_\mathrm{Dec}$, v$_\mathrm{sys}$\} and six structural profile parameters \{$\xi$, $T_{10}$, $q$, $r_\mathrm{c}$, $\gamma$, $M_\mathrm{gas}$\}.  The $M_\mathrm{gas}$ parameter is treated in log-space within \textsc{DiskJockey}, while all other parameters are treated in linear space.  Given that the $^{12}$CO and $^{13}$CO emission is likely optically thick for the three star+disk systems, the six structural parameters only describe the optically thick surfaces of the observed emission.  They are \textit{not} representative of the disks' true midplane temperature and surface density profiles.  
The dynamical mass measurement is not significantly affected by this approach to parametrizing the CO surface (see discussion in Section~\ref{sec_results_residuals}).

%
%

\subsection{Modeling Procedure}
\label{sec_methodology_parameters}

\textsc{DiskJockey} infers the parameters of the model using \textsc{emcee}~\citep{cite_emcee}, a Markov Chain Monte Carlo (MCMC) approach.  All parameters and Uniform priors are listed in Table~\ref{table_modinit}.  We fixed $d$ to the precise distances provided by \textit{Gaia}.  We also fixed $\gamma$ to 1.0, similarly to previous works~\citep[e.g.,][]{cite_rosenfeldetal2012b, cite_czekalaetal2015, cite_czekalaetal2019}, which reduces the surface density profile into a tapered power law.  We then used \textsc{DiskJockey} to infer the remaining 11 parameters.  We assigned a total of 275 MCMC chains for each disk and each CO line.  Initial parameter values for each chain were selected randomly from broad Uniform distributions, which were informed by/adjusted from inspection of the $^{12}$CO emission and estimates for solar-type disks~\citep[e.g.,][]{cite_andrewsetal2011}.

We ran the chains for each disk for at least 1000 steps, with the number of steps increased per model until apparent convergence.  We qualitatively estimated sufficient convergence using the percent slope changes (i.e., the slope relative to the final parameter value, described below) of the median, 16$^\mathrm{th}$ percentile, and 84$^\mathrm{th}$ percentile across all chains for each parameter.  For all disks and all parameters (including structural parameters), the percent slope changes over the last 500 steps of the chains were less than 0.02\%.  For $M_*$ and the other primary parameters, the percent slope changes over the last 500 steps were $<$0.005\% for all lines and disks.  These final 500 steps formed the sampling distribution for each line and disk.  The full chains, along with corner plots of the sampling distributions, are available as Figure Sets for $^{12}$CO and $^{13}$CO in Appendix~\ref{sec_appendix_MCMC_12CO} and Appendix~\ref{sec_appendix_MCMC_13CO}, respectively.
%

We used the median parameter values of the sampling distributions as the final parameter estimates and to construct the median model (interpreted here as the ``best" model) for each disk.  We then used the 16$^\mathrm{th}$ and 84$^\mathrm{th}$ percentiles of the sampling distributions as the lower and upper uncertainties, respectively.  Given that stellar mass estimates scale linearly with distance, we added the relative \textit{Gaia} distance uncertainties from Table~\ref{table_sample} in quadrature to the relative uncertainties of each stellar mass estimate (i.e., $M_*\sqrt{(\sigma_\mathrm{d}/d)^2 + (\sigma_\mathrm{M}/M_*)^2}$, where $\sigma_\mathrm{d}$ and $\sigma_\mathrm{M}$ are the distance and stellar mass errors, respectively) to account for distance uncertainties in the stellar mass.  The distance contribution to the combined error is comparable to the mass contribution for FP Tau and J0432+1827, while the combined error for J1100-7619 is dominated by the mass contribution.


\begin{figure*}
\centering
\resizebox{0.99\hsize}{!}{
    \includegraphics[trim=70pt 50pt 95pt 45pt, clip]{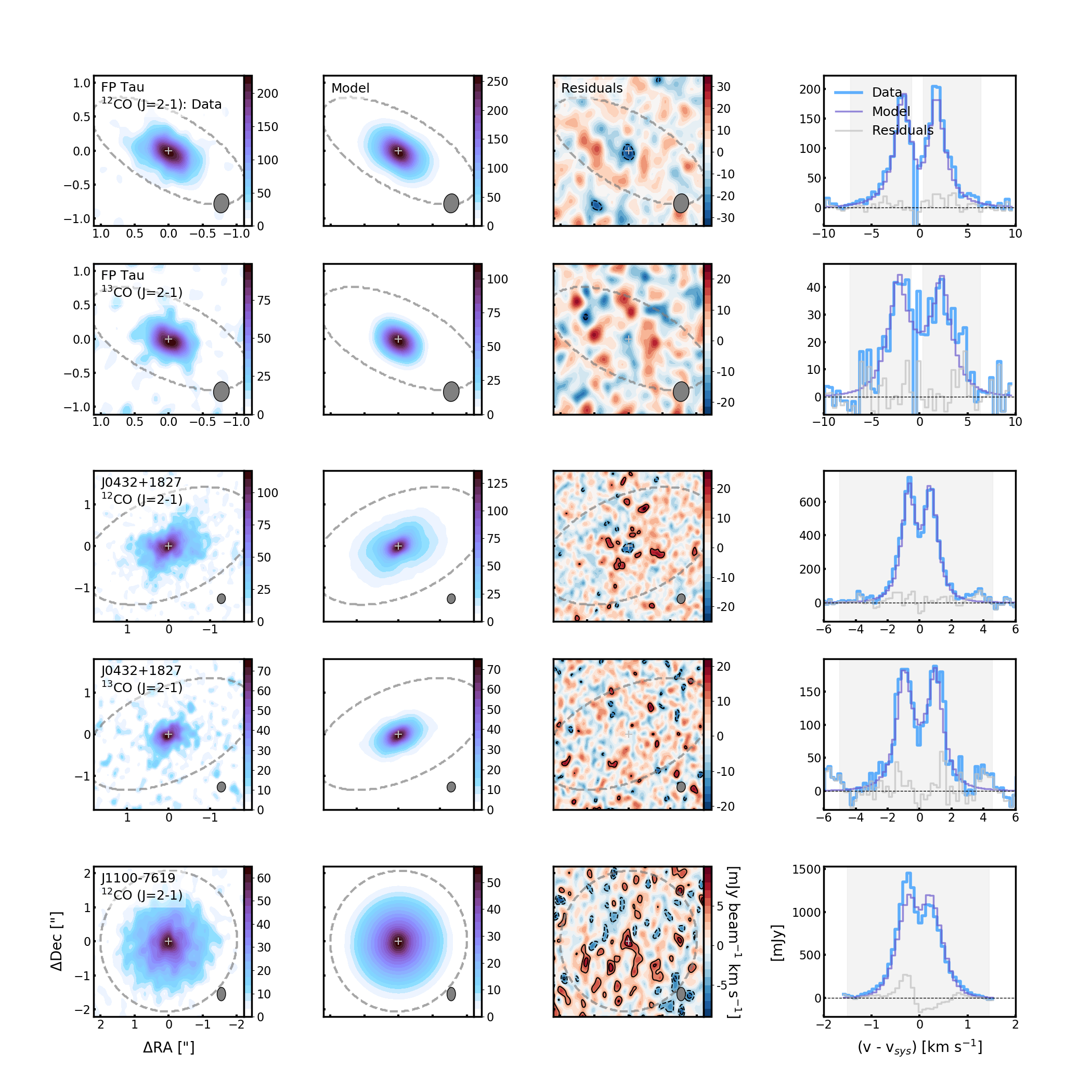}}
\caption{Comparisons of the data and constrained \textsc{DiskJockey} models for the $^{12}$CO and $^{13}$CO emission.  The rows from top to bottom show the results for FP Tau $^{12}$CO and $^{13}$CO, J0432+1827 $^{12}$CO and $^{13}$CO, and J1100-7619 $^{12}$CO, respectively.  Columns 1, 2, and 3 show velocity-integrated emission maps of the data, model, and residuals, respectively, while column 4 shows the corresponding spectra.  Beam sizes are drawn in the lower right corners of each map.  The spectra were extracted from within the dashed gray contours overlaid in columns 1 through 3.  These gray contours extend to the Keplerian mask boundaries used for $^{12}$CO toward these disks in Pegues et al. (in rev.).  The velocity channels included in each fit are highlighted in light gray in column 4.  Contours for the residuals in column 3 correspond to $\pm$[3$\sigma$, 5$\sigma$, 10$\sigma$, 20$\sigma$...], where $\sigma$ is estimated roughly as $\sigma = \Delta V \sqrt{N_\mathrm{chan} \sigma_\mathrm{chan}^2}$, where $\Delta V$ is the channel spacing, $N_\mathrm{chan}$ is the number of channels included in the fit, and $\sigma_\mathrm{chan}$ is the channel rms (as estimated in Pegues et al. in rev.).  No model is shown for $^{13}$CO toward J1100-7619 because that model was unconstrained.
\label{fig_series}}
\end{figure*}

\begin{figure*}
\centering
\resizebox{0.925\hsize}{!}{
    \includegraphics[trim=50pt 20pt 30pt 45pt, clip]{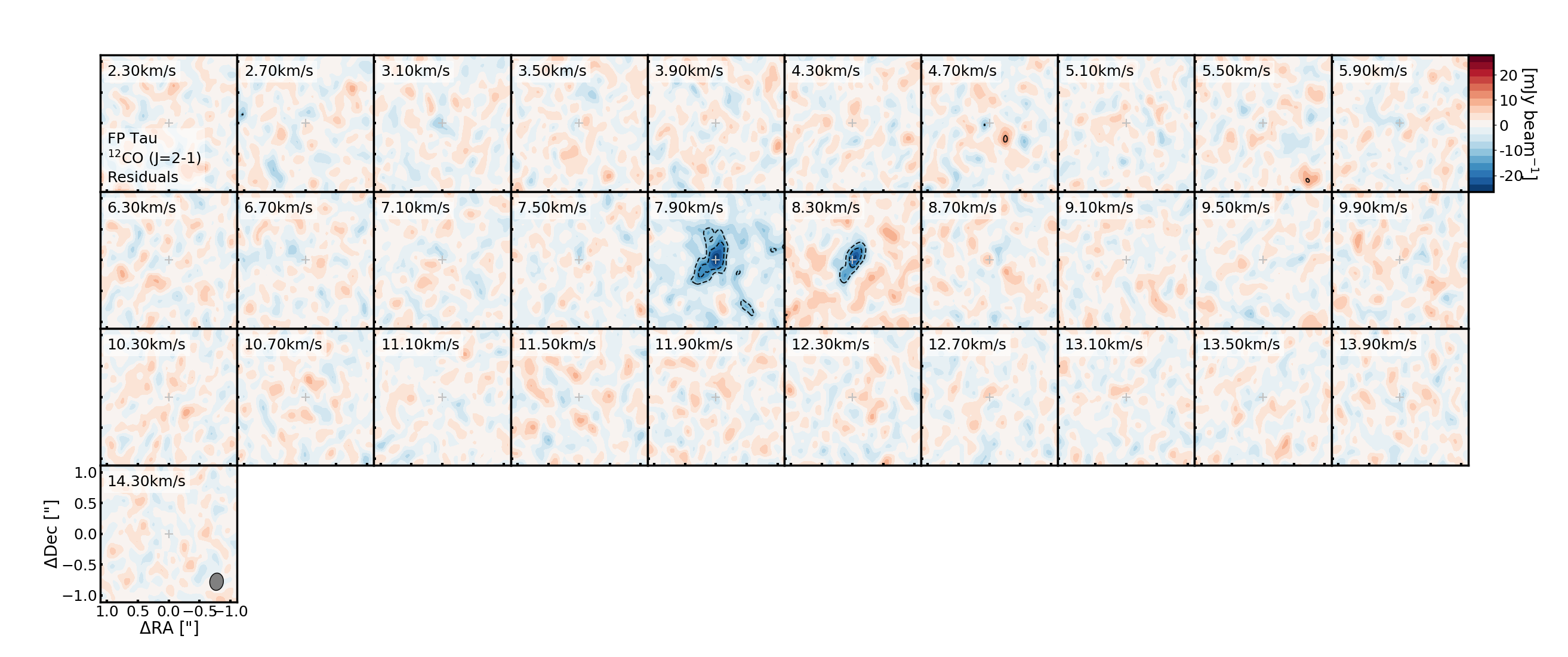}}
\resizebox{0.925\hsize}{!}{
    \includegraphics[trim=50pt 20pt 30pt 45pt, clip]{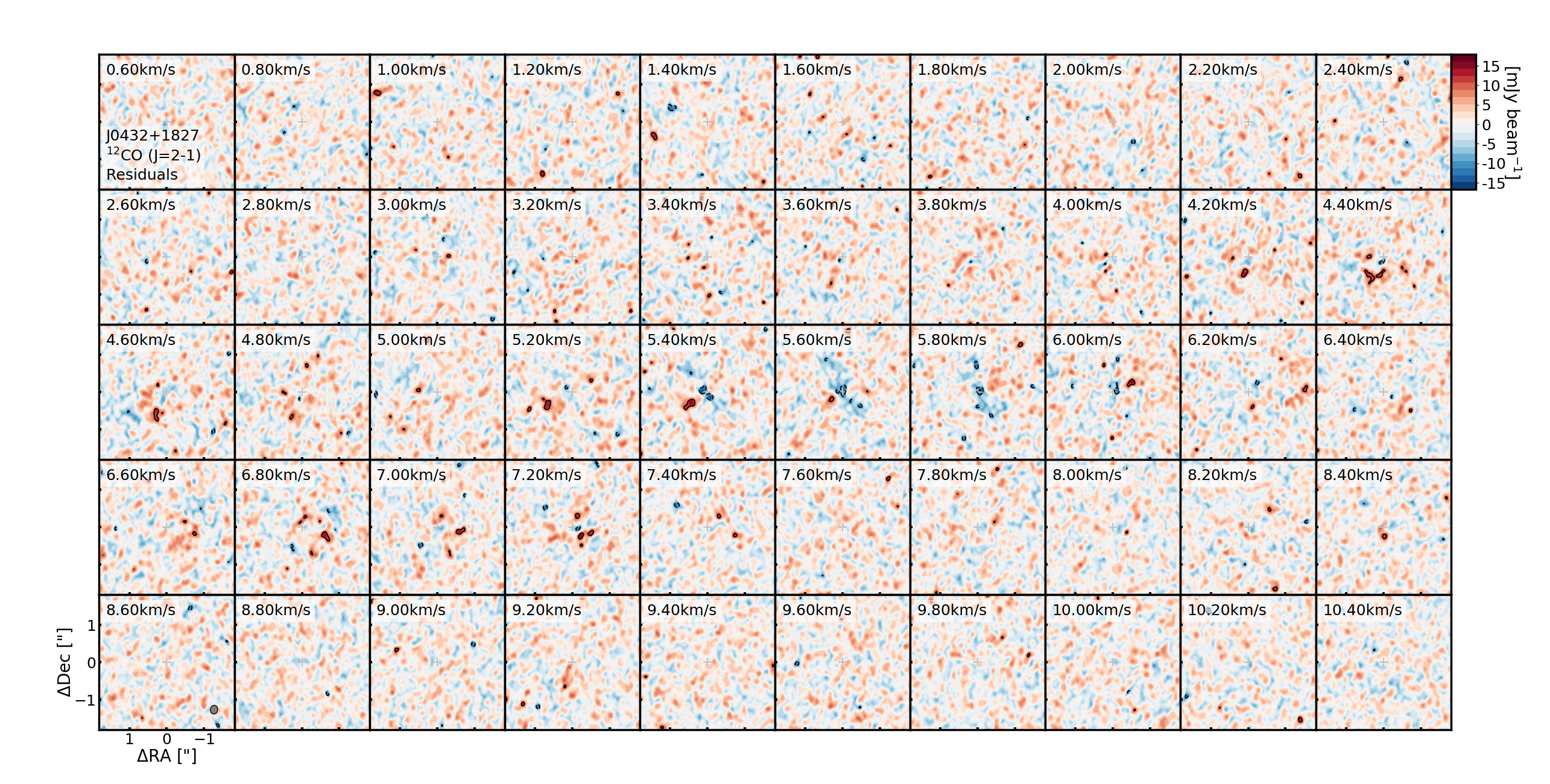}}
\resizebox{0.925\hsize}{!}{
    \includegraphics[trim=50pt 20pt 30pt 45pt, clip]{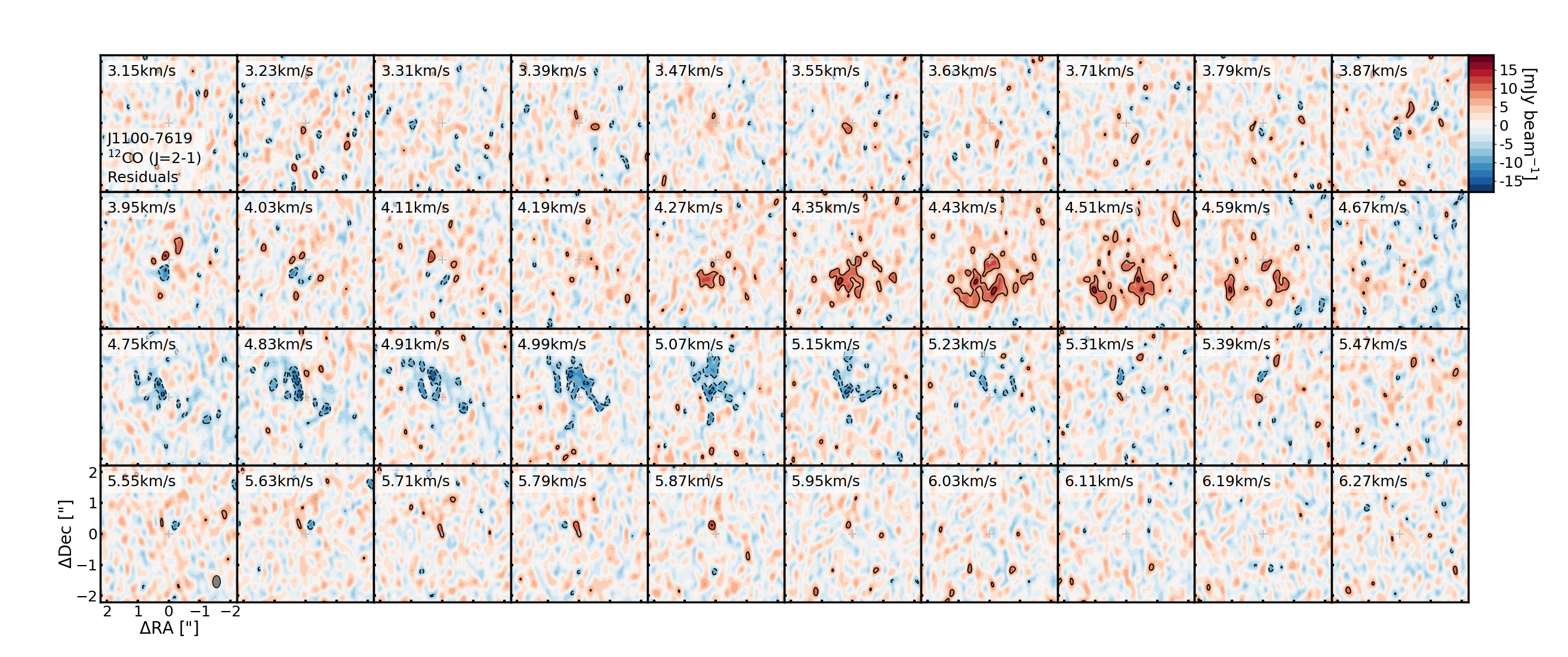}}
\caption{Channel maps of the data vs. model residuals for $^{12}$CO toward FP Tau (top), J0432+1827 (middle), and J1100-7619 (bottom).  Contours correspond to $\pm$[3$\sigma$, 5$\sigma$, 10$\sigma$, 20$\sigma$...], where $\sigma$ is the estimated channel rms (from Pegues et al. in rev.).  Beam sizes are drawn in the lower left corners of each panel.  The significant residuals toward FP Tau are due to cloud contamination near the systemic velocity, while significant residuals toward J1100-7619 are due to asymmetries in the $^{12}$CO emission.
\label{fig_chanex}}
\end{figure*}


\begin{figure}
\centering
\resizebox{0.9\hsize}{!}{
    \includegraphics[]{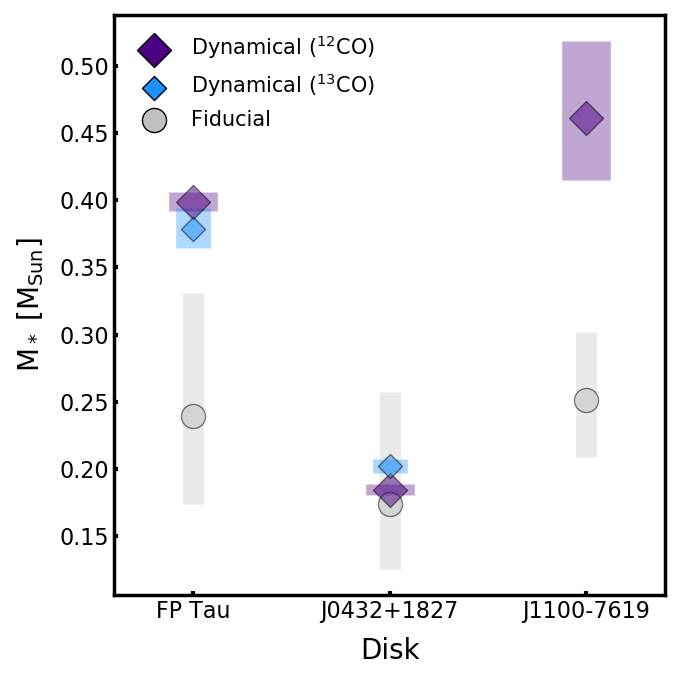}}
\caption{Comparison of the $^{12}$CO and $^{13}$CO stellar mass measurements (marked with dark purple and light blue diamonds, respectively) to the stellar mass predictions of the fiducial stellar evolutionary model (Table~\ref{table_sample}; marked with gray circles).  The star+disk systems are written along the x-axis, while the stellar mass values are given along the y-axis.  Stellar mass uncertainties are shown as vertical shaded bars, where the colors match those of the corresponding measurement.  The horizontal widths of the bars have been varied so that the bars for each measurement are easier to distinguish.  They are \textit{not} representations of x-axis error.
\label{fig_mass}}
\end{figure}

\begin{figure*}
\centering
\resizebox{0.99\hsize}{!}{
    \includegraphics[trim=70pt 25pt 70pt 75pt, clip]{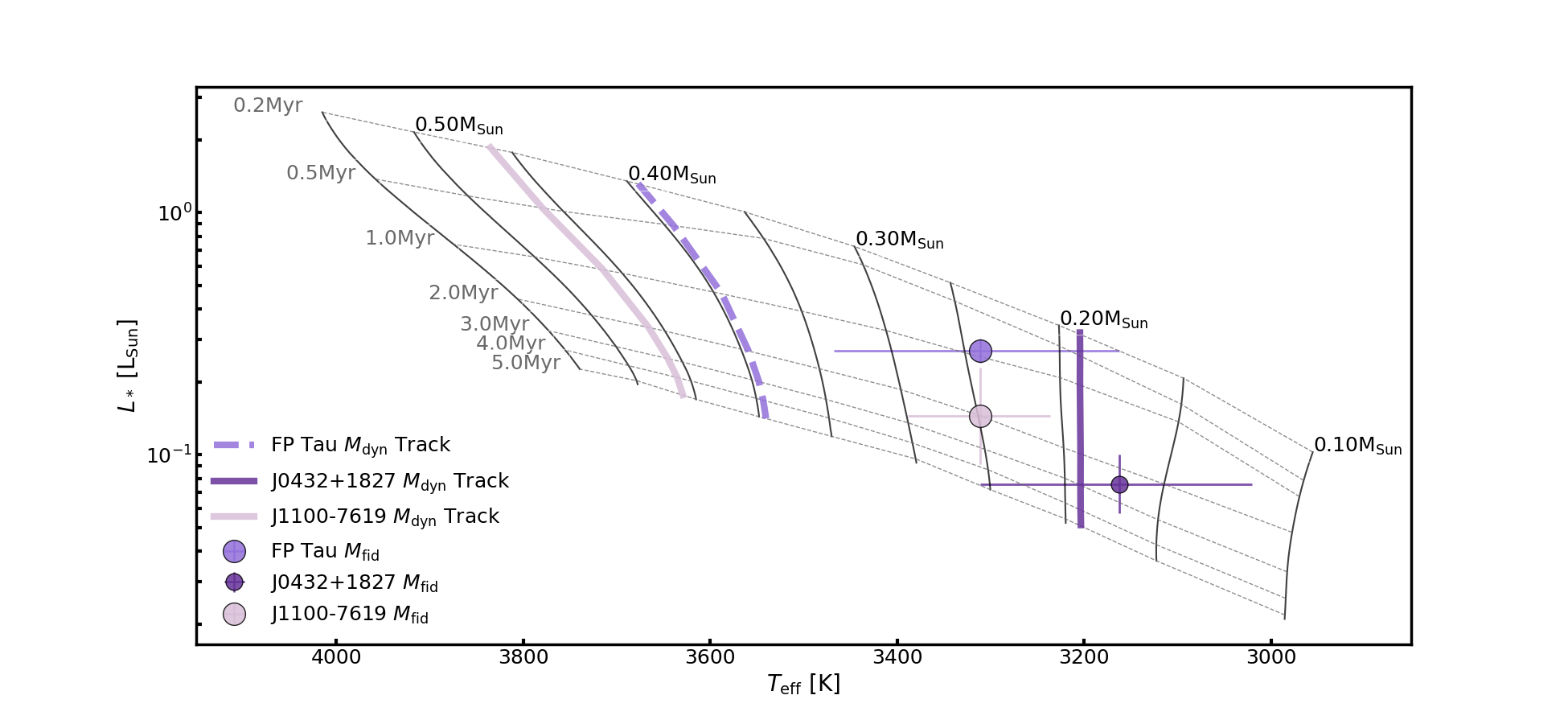}}
\caption{Stellar tracks across stellar effective temperature ($T_\mathrm{eff}$) and stellar luminosity ($L_*$).  The tracks drawn in thin, solid gray lines follow the 0.2-5.0Myr evolution of stars with masses of 0.10-0.55M$_\Sun$ at 0.05M$_\Sun$ intervals.  The thin, dashed gray lines are linearly interpolated age tracks from 0.2-5.0Myr.  These tracks were extracted from the online MESA Isochrones and Stellar Tracks database~\citep[MIST;][]{cite_mist1, cite_mist2}.  The purple points mark the placement of our low-mass M-stars among these tracks.  The thick purple lines trace the interpolated mass tracks corresponding to the dynamically-estimated stellar masses.
\label{fig_HR}}
\end{figure*}

\begin{figure}
\centering
\resizebox{0.75\hsize}{!}{
    \includegraphics[trim=70pt 5pt 90pt 10pt, clip]{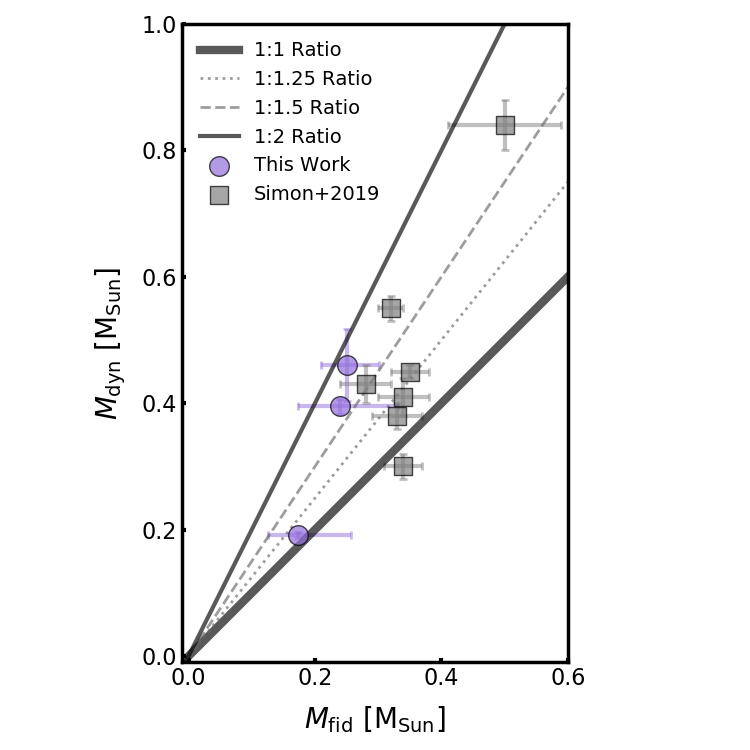}}
\caption{Dynamical mass measurements ($M_\mathrm{dyn}$) plotted as a function of mass predictions from fiducial stellar evolutionary models ($M_\mathrm{fid}$) for $M_\mathrm{fid}\leq$0.5M$_\Sun$.  Measurements from this work and from the compilation of~\cite{cite_simonetal2019} (described in Section~\ref{sec_discussion_fid}) are marked with purple circles and gray squares, respectively.  The errorbars are the 1$\sigma$ uncertainties.  To help guide the eye, lines are drawn for the 1:1 (thick, solid black), 1:1.25 (dotted gray), 1:1.5 (dashed gray), and 1:2 (thin, solid black) ratios.
\label{fig_massdiff}}
\end{figure}



\section{Results}
\label{sec_results}

\subsection{Dynamical Masses}
\label{sec_results_mass}

Figure~\ref{fig_series} presents velocity-integrated emission maps and spectra comparing the observations, median disk models, and the model residuals for $^{12}$CO and $^{13}$CO toward the three star+disk systems.  Appendix~\ref{sec_appendix_chans} presents the corresponding channel maps.  Channel maps of the residuals for $^{12}$CO toward the three star+disk systems are additionally shown in Figure~\ref{fig_chanex}.  No model is shown for $^{13}$CO toward J1100-7619, because the constraints on the emission are not informative (see the chain plot in Appendix~\ref{sec_appendix_MCMC_13CO} and discussion later in this section).  Otherwise, we see that our simple disk models are generally in good agreement with the data.  The residuals are typically below 3$\sigma$, with some exceptions due to more complex disk and environmental structures not covered by the models.  These exceptions are discussed in Section~\ref{sec_results_residuals}.

Figure~\ref{fig_mass} displays the $^{12}$CO and $^{13}$CO stellar mass measurements toward FP Tau and J0432+1827 and the $^{12}$CO stellar mass measurement toward J1100-7619.  Table~\ref{table_params} lists the primary parameter constraints measured for each disk using the $^{12}$CO and $^{13}$CO emission.  Results for the structural parameters are presented separately in Appendix~\ref{sec_appendix_params} to avoid any interpretation of them as measured structures of the disks rather than the CO.

Upper and lower uncertainties across all $^{12}$CO and $^{13}$CO stellar mass measurements are 1.8-4.0\% for FP Tau and J0432+1827.
However, the measurements derived independently from the $^{12}$CO and $^{13}$CO observations have relative differences of 5.1\% and 9.6\% for FP Tau and J0432+1827, respectively.
As these are independent measurements, the relative differences between the measurements provide a more realistic understanding of the systematic uncertainties on the stellar masses~\citep[e.g.,][]{cite_premnathetal2020}.  The statistical uncertainties are much higher for the $^{12}$CO stellar mass measurement for J1100-7619, spanning 10-12\%.  These uncertainties are due to the disk's low inclination angle and are discussed in Section~\ref{sec_results_residuals}.

For FP Tau and J0432+1827, we take the weighted averages of the $^{12}$CO and $^{13}$CO stellar mass estimates as the final dynamical stellar mass measurement, $M_\mathrm{dyn}$, for each disk.  Each average is weighted by $(1/\sigma^2)$, where $\sigma$ is either the difference between the median and 16$^\mathrm{th}$ percentile estimate or between the median and 84$^\mathrm{th}$ percentile estimate, whichever difference is larger (the differences are nearly symmetrical for both disks).  For J1100-7619, we use the $^{12}$CO stellar mass estimate as the final $M_\mathrm{dyn}$ measurement, and we use the maximum of the nearly symmetrical upper and lower uncertainties as the final uncertainty for clarity.  The resulting $M_\mathrm{dyn}$ estimates for FP Tau, J0432+1827, and J1100-7619 are 0.395$\pm$0.012 M$_\Sun$, 0.192$\pm$0.005 M$_\Sun$, and 0.461$\pm$0.057 M$_\Sun$, respectively, and are listed in Table~\ref{table_params}.  The dynamical mass measurement for FP Tau is roughly consistent with the previous dynamical mass measurement given in the literature~\citep[0.36$\pm$0.02 M$_\Sun$,][]{cite_simonetal2019}.

\subsection{Sources of Model Residuals and Systematic Uncertainties}
\label{sec_results_residuals}

\subsubsection{FP Tau and J0432+1827}

For $^{12}$CO toward FP Tau, the significant residuals are due to cloud contamination of the data in the central velocity channels.  These channels were excluded from the model fits, although they are still included in the velocity-integrated emission maps of Figure~\ref{fig_series} for completeness.  For $^{12}$CO toward J0432+1827, there are significant residuals that appear systematic in Figure~\ref{fig_series} but actually appear scattered in the corresponding channel map in Figure~\ref{fig_chanex}.

\subsubsection{J1100-7619}

For $^{12}$CO toward J1100-7619, there are significant residuals due to asymmetries in the peaks of the actual spectrum.  It is not clear what is causing these asymmetries toward J1100-7619, as we do not see any cloud contamination within the residual channel maps, and the same asymmetries do not appear to be present in the $^{13}$CO emission (Appendix~\ref{sec_appendix_chans}).  We note that these asymmetries are not the primary cause of the relatively large statistical uncertainties in the $^{12}$CO mass measurement.
These uncertainties are instead due to the disk's low inclination angle.  The line-of-sight velocities are proportional to $(\sqrt{M_*} \sin \theta_\mathrm{inc})$, where $M_*$ and $\theta_\mathrm{inc}$ are the stellar mass and inclination angle, respectively~\citep[e.g.,][]{cite_dutreyetal1994}.  For nearly face-on disks, even small changes in inclination thus lead to significant changes in the stellar mass estimate (Appendix~\ref{sec_appendix_Msin}).  The $^{13}$CO stellar mass measurement for J1100-7619 was unconstrained for the same reason.

\subsubsection{The Disk Model}
\label{sec_results_residuals_model}

We note that the underlying disk model (Section~\ref{sec_methodology_model}) makes several assumptions on disk structure (e.g., that the disks are geometrically thin instead of flared), which lead to intrinsic systematic uncertainties in our results.  That being said, these disk-based dynamical mass measurements are derived from the velocity field of each disk rather than its temperature or density structures.
We thus do not expect our simple assumptions on disk structure to significantly affect our dynamical mass measurements.

Indeed, Appendices~\ref{sec_appendix_MCMC_12CO} and~\ref{sec_appendix_MCMC_13CO} show that our stellar mass measurements are not significantly correlated with any of the structural parameters (listed in Appendix~\ref{sec_appendix_params}).  The measurements are degenerate only with the inclination angle $\theta_\mathrm{inc}$ (Appendix~\ref{sec_appendix_Msin}).  Previous works have also found that more complex disk models with vertical temperature gradients and treatments of CO freeze-out yield similar mass estimates to simple power-law disk models~\citep[e.g.,][]{cite_czekalaetal2017}, and that the mass precision is driven by the sensitivity and signal-to-noise of the observations rather than by the assumed disk model~\citep[e.g.,][]{cite_rosenfeldetal2012b, cite_simonetal2019, cite_boydenetal2020}.
Even when there are asymmetries in the emission that violate the symmetric assumptions of the disk model, such as due to moderate cloud contamination, modest disk eccentricities, or disk warping, the models can still recover an underlying Keplerian pattern and yield consistent dynamical mass measurements~\citep[e.g.,][]{cite_czekalaetal2015, cite_simonetal2019}.

%

Inaccurate estimates of the inclination angle on the other hand, such as from structural complexities or emission asymmetries not captured by the model, would have some effect on estimates of the dynamical mass.  This effect is illustrated for the three star+disk systems in Appendix~\ref{sec_appendix_Msin}.  For FP Tau and J0432+1827, the consistency between the independent $^{12}$CO and $^{13}$CO-based angle and mass measurements is evidence against such inaccuracy.  J1100-7619, for which the effect of the correlation is most significant due to the nearly face-on inclination angle, would be worth follow-up in CO isotopologue emission at higher signal-to-noise.

%

\subsection{Comparisons to Fiducial Stellar Evolutionary Model Predictions}
\label{sec_results_sem}

Figure~\ref{fig_mass} also displays the stellar mass predictions from the fiducial stellar evolutionary model~\citep[tracks from the MIST code;][]{cite_mist1, cite_mist2} given in Table~\ref{table_sample}, which does not incorporate any treatment of magnetic activity.  Fiducial model estimates ($M_\mathrm{fid}$) for FP Tau, J0432+1827, and J1100-7619 are 0.240$^{+0.091}_{-0.066}$M$_\Sun$, 0.174$^{+0.083}_{-0.048}$M$_\Sun$, and 0.251$^{+0.051}_{-0.042}$M$_\Sun$, respectively.
When we compare the fiducial model mass predictions to the dynamical mass estimates for each star, we find that the fiducial model prediction agrees with the dynamical mass measurement for one star+disk system (J0432+1827).  However, the fiducial model underestimates the dynamical masses by significant factors for the two remaining systems (underpredictions of $\sim$60\% for FP Tau and $\sim$80\% for J1100-7619).  For these two systems, the dynamical masses exceed the 1$\sigma$ fiducial model uncertainties (by $\sim$1.7$\sigma$ and $\sim$4.1$\sigma$, respectively).

The mass underprediction for J1100-7619 may be explained by undetected binarity.  \cite{cite_nguyenetal2012} conducted a multiplicity study of the Chamaeleon I and Taurus star-forming regions using high-resolution spectroscopy and radial velocity techniques.  They did not identify J1100-7619 as a system hosting a close or wide companion.  That being said, the low inclination of the J1100-7619 star+disk system makes this analysis more challenging, and J1100-7619 could still possibly host an undetected companion within the intermediate regime between ``close'' and ``wide''.  We assume that J1100-7619 hosts a unary star for the remainder of the paper, but we note that J1100-7619 is worth follow-up in the future at higher spatial and spectral resolution (see also discussion in Section~\ref{sec_results_residuals}).

Figure~\ref{fig_HR} compares the fiducial model mass predictions to interpolated dynamical mass tracks, based on stellar mass tracks extracted from the online MIST database~\citep{cite_mist1, cite_mist2}\footnote{Available as packaged model grids containing Equivalent Evolutionary Phase (EEP) tracks from http://waps.cfa.harvard.edu/MIST/.}.  Reconciling the fiducial mass predictions with the dynamical measurements would require (1) a shift in the assumed stellar effective temperatures and/or (2) a shift in the model tracks.  We consider these two cases, and their implications, in Section~\ref{sec_discussion}.

\section{Discussion}
\label{sec_discussion}

\begin{figure*}
\centering
\resizebox{0.325\hsize}{!}{
    \includegraphics[trim=10pt 10pt 10pt 10pt, clip]{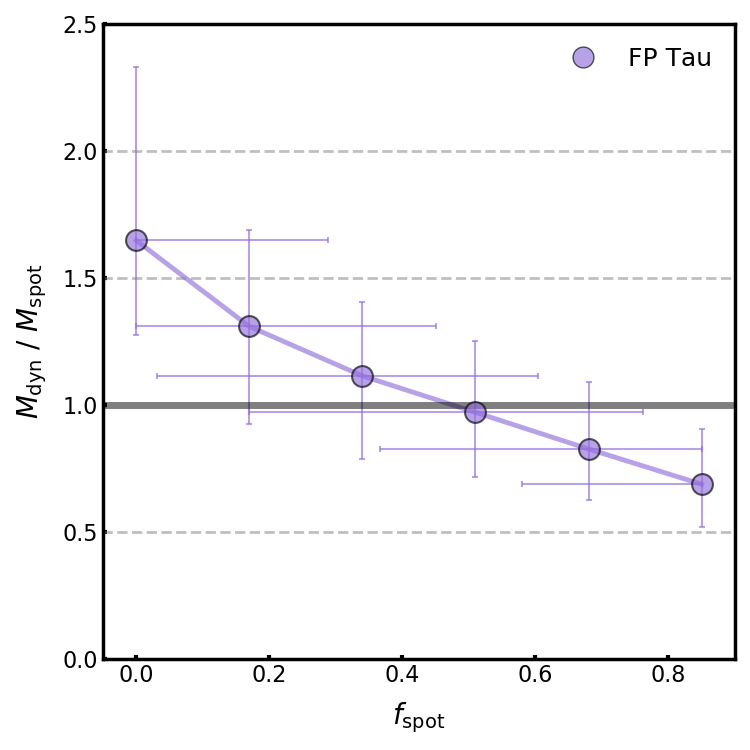}}
\resizebox{0.325\hsize}{!}{
    \includegraphics[trim=10pt 10pt 10pt 10pt, clip]{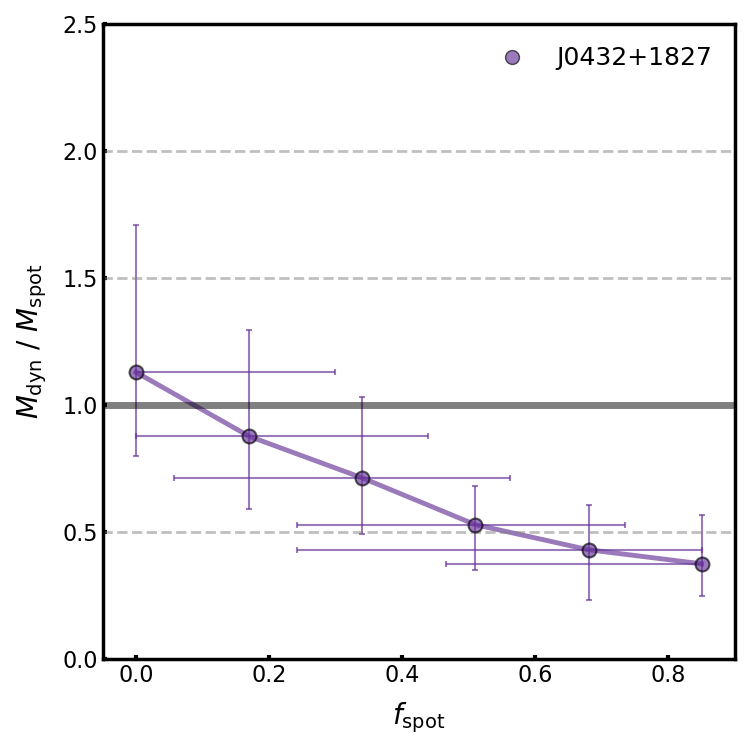}}
\resizebox{0.325\hsize}{!}{
    \includegraphics[trim=10pt 10pt 10pt 10pt, clip]{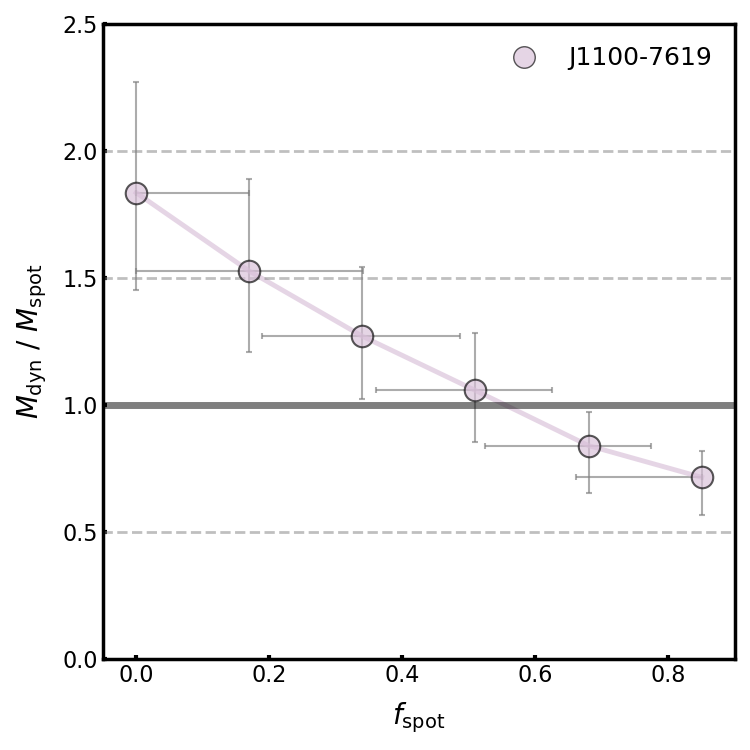}}
\caption{Performance of the starspot stellar evolutionary models for FP Tau (left), J0432+1827 (middle), and J1100-7619 (right).  The plotted points are the discrete results from the starspot model tracks, while the paths connecting the points are interpolated.  The x-axis gives the fractional starspot coverage $f_\mathrm{spot}$, while the y-axis gives the ratio between the dynamical masses ($M_\mathrm{dyn}$) and the starspot model predictions ($M_\mathrm{spot}$).  Perfect prediction ($M_\mathrm{dyn}/M_\mathrm{pred}$=1) is drawn horizontally in solid black, while factors of 0.5, 1.5, and 2.0 are drawn in dashed gray.  Errors along the x-axis span the $f_\mathrm{spot}$ values interpolated for the error in M$_\mathrm{spot}$, i.e. the interpolated $f_\mathrm{spot}$ values that best describe the lower and upper bounds on $M_\mathrm{spot}$.  Errors along the y-axis are calculated using standard error propagation, i.e. $\sigma^\pm = (M_\mathrm{dyn} / M_\mathrm{spot}) \times \sqrt{(\sigma_\mathrm{spot}^\pm / M_\mathrm{spot})^2 + (\sigma_\mathrm{dyn}^\pm / M_\mathrm{dyn})^2}$, where $\sigma_\mathrm{spot}^\pm$ and $\sigma_\mathrm{dyn}^\pm$ are the $\pm$ error in the $M_\mathrm{spot}$ and $M_\mathrm{dyn}$ predictions/measurements, respectively.
\label{fig_spotperf}}
\end{figure*}

\begin{figure*}
\centering
\resizebox{0.99\hsize}{!}{
    \includegraphics[trim=70pt 25pt 70pt 75pt, clip]{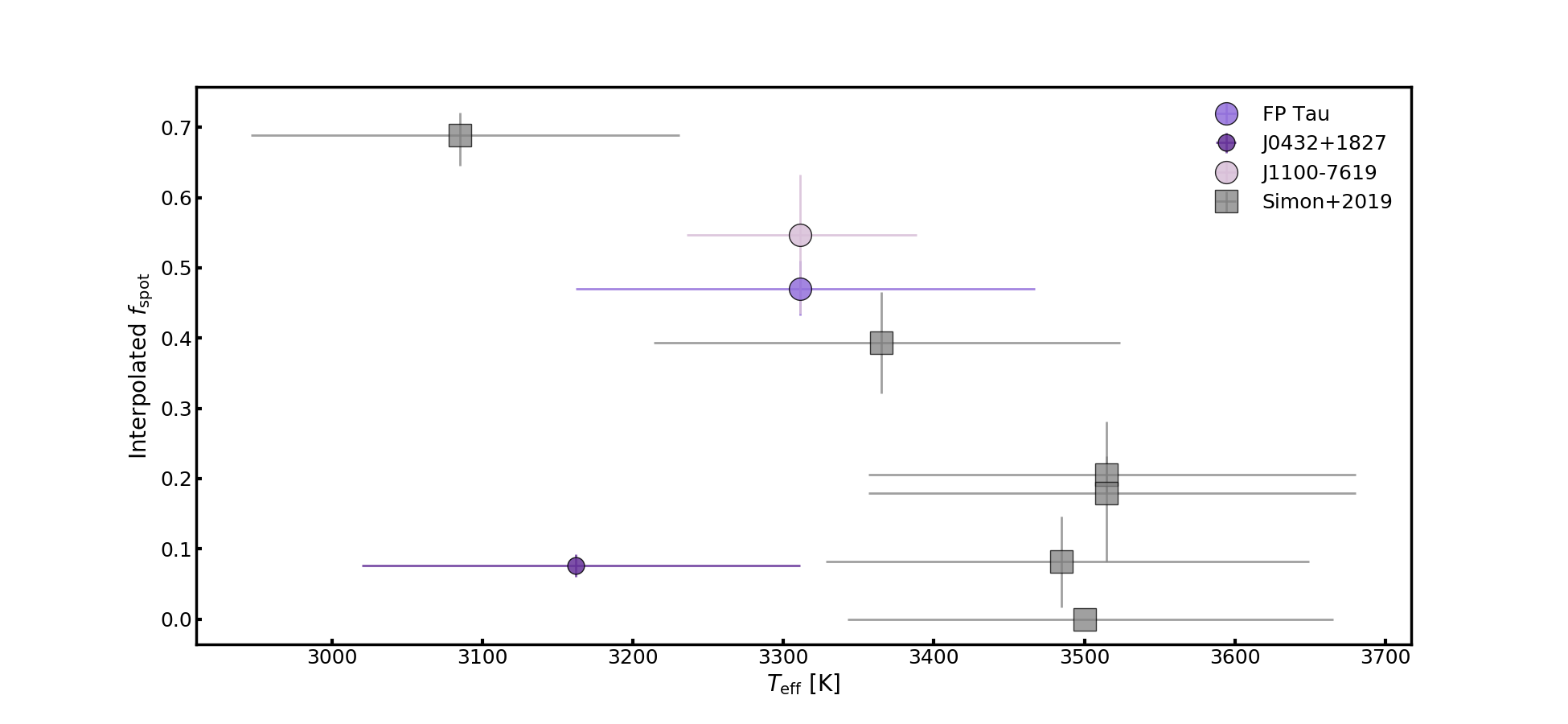}}
\caption{The interpolated $f_\mathrm{spot}$ values (i.e., the values that minimize $|M_\mathrm{dyn}-M_\mathrm{spot}|$), plotted as a function of stellar effective temperature ($T_\mathrm{eff}$).  The purple circles and gray squares are stars from this work and from~\cite{cite_simonetal2019}, respectively (the latter sample is described in Section~\ref{sec_discussion_sems}).  Errors along the y-axis span the $f_\mathrm{spot}$ values interpolated for the error in M$_\mathrm{dyn}$, i.e. the interpolated $f_\mathrm{spot}$ values that best describe the lower and upper bounds on $M_\mathrm{dyn}$.  \textit{Note that the errors along the y-axis assume perfect recovery from the starspot model tracks.}
\label{fig_bestfspot}}
\end{figure*}

%

\subsection{Performance of Fiducial Stellar Evolutionary Models for Low-Mass Unary Stars}
\label{sec_discussion_fid}

Our results show that the fiducial stellar evolutionary model predicts the stellar mass for J0432+1827 within 10\%, but significantly underpredicts the stellar masses for the remaining 2/3 low-mass M-stars.  Specifically, we find that FP Tau and J1100-7619, which were predicted from the fiducial model to have stellar masses $M_\mathrm{fid}$ of 0.24-0.25M$_\Sun$ (Table~\ref{table_sample}), have dynamically-estimated stellar masses that belong to an entirely different mass bracket, with $M_\mathrm{dyn}\gtrsim0.4$M$_\Sun$.

To further investigate systematic differences between fiducial model predictions and the dynamical mass measurements, we plot M$_\mathrm{dyn}$ as a function of M$_\mathrm{fid}$ in Figure~\ref{fig_massdiff}.  We also plot unary stars from~\cite{cite_simonetal2019} with fiducial model mass predictions $\leq$0.5M$_\Sun$.
\cite{cite_simonetal2019} used the~\cite{cite_baraffeetal2015} stellar evolutionary models, which do not incorporate magnetic fields, as their fiducial stellar evolutionary models.  This provides another point of comparison to the fiducial stellar evolutionary models used in this work~\cite[tracks from the MIST code;][]{cite_mist1, cite_mist2}.
Note that we use the 7/13 unary stars (AA Tau, CX Tau, CY Tau, DE Tau, DM Tau, GO Tau, and HO Tau) from~\cite{cite_simonetal2019} with fiducial model mass predictions labeled as MB$_\mathrm{hrd}$ in that paper.  The MB$_\mathrm{hrd}$ predictions drew their spectral types, stellar effective temperatures, and stellar luminosities from a single source~\citep{cite_herczegetal2014}.  This allows us to maintain consistency in the sets of stellar characteristics input to the fiducial models.  We have also excluded the low-mass M-star (YLW 58) in the Ophiuchus region, 
because~\cite{cite_simonetal2019} gave concerns about the reliability of the characterization and model predictions for the Ophiuchus cloud.  Finally, FP Tau was also presented in~\cite{cite_simonetal2019} (M$_\mathrm{dyn}$=0.36$\pm$0.02 M$_\Sun$).  Since the two $M_\mathrm{dyn}$ measurements for FP Tau are roughly consistent, we use the $M_\mathrm{dyn}$ and $M_\mathrm{fid}$ measurements from this work.


As shown in Figure~\ref{fig_massdiff}, 7/10 of the combined sample of $M_\mathrm{fid}$ predictions underpredict the $M_\mathrm{dyn}$ measurements by $\gtrsim$20\%.  Half (5/10) of the $M_\mathrm{dyn}$ measurements are underpredicted by 50-100\%.  The underpredictions are not distinguishable between the two sources of fiducial models~\citep{cite_baraffeetal2015, cite_mist1, cite_mist2}, implying that these underpredictions are not specific to any particular model.

\subsection{Possible Remedies for the Underpredictions}

These consistent underpredictions from fiducial models may be due to a combination of two factors.  On the one hand, inaccurate observations of the stellar characteristics that are fed into these models (i.e., the effective temperature and luminosity) lead to inaccurate predictions of the stellar mass.  On the other hand, stellar evolutionary models that are missing key underlying physics, such as treatment of magnetic activity, can also lead to inaccurate predictions of the stellar mass.  We are currently unable to distinguish individual contributions from these two sources of inaccuracy to the fiducial model predictions.  Instead, we treat each case scenario separately in subsequent discussion.

\subsubsection{Adjustments to the Stellar Characteristics}
\label{sec_discussion_inputs}

Inaccuracies in spectral type, and consequently effective temperature, are especially critical for low-mass pre-MS M-stars.  Across this low-mass regime, changes in effective temperature have a greater effect on the mass predictions of stellar evolutionary models than changes in the luminosity (e.g., Figure~\ref{fig_HR}).
Here we consider the implications of our results on the spectral types and effective temperatures of our three stars.  \textit{Note that in doing so, we are assuming that the fiducial stellar evolutionary models incorporate physics that are sufficient to predict the stellar masses.}  As noted previously, we \textit{cannot} distinguish the effects of inaccurate stellar characteristics from the effects of inaccurate stellar evolutionary models on our predictions.

We assumed spectral types and stellar effective temperatures of M4 and 3311$^{+156}_{-149}$K for FP Tau, M4.75 and 3162$^{+149}_{-142}$K for J0432+1827, and M4 and $\sim$3311$^{+77}_{-75}$K for J1100-7619~\citep[Table~\ref{table_sample};][]{cite_kenyonetal1995, cite_bricenoetal2002, cite_manaraetal2017, cite_luhmanetal2003}.
A few other independent spectral type and effective temperature measurements exist in the literature.
\cite{cite_herczegetal2014} estimated spectral types and effective temperatures of M2.6 and $\sim$3410-3560K for FP Tau and M5 and $\sim$2980K for J0432+1827, respectively, while \cite{cite_luhmanetal2007} estimated M3.75 and $\sim$3306K for J1100-7619.
The effective temperature estimates for J1100-7619 across the different literature sources are consistent with each other.  The M2.6 estimate for FP Tau, however, is warmer than our assumptions by $\sim$100-250K, while the M5 estimate for J0432+1827 is cooler by $\sim$180K.  Interpreting these values with respect to Figure~\ref{fig_HR}, we see that the warmer M2.6 estimate for FP Tau would shift the star so that the fiducial model prediction would be in agreement with the dynamical mass track within uncertainties.  At the same time, the cooler M5 estimate for J0432+1827 would have the opposite effect, shifting the star so that the fiducial model prediction would no longer agree with the dynamical mass track.  \textit{These differences highlight the importance of accurate constraints on the effective temperatures for these low-mass stars.}  Without them, we cannot use these models to accurately predict the masses.

\textit{If we assume the fiducial stellar evolutionary models are sufficient to accurately describe these stars}, we can estimate new spectral types and stellar effective temperatures for our stars using Figure~\ref{fig_HR} and the conversion table (Table 5) in~\cite{cite_herczegetal2014}.
In order for the three stars to match their dynamical masses, we find that the stars require increases in stellar effective temperature of $\sim$250-260K for FP Tau, $\sim$40K for J0432+1827, and $\sim$320-340K for J1100-7619.
FP Tau would thus be better represented by a spectral type within M1-M2 (3720-3560K), which is somewhat warmer than the M2.6 estimate from~\cite{cite_herczegetal2014}.
%
J0432+1827 would be better described by a spectral type within M3-M4 (3410-3190K).  The effective temperature range overlaps with the uncertainties of our assumed values in Table~\ref{table_sample}, although the spectral type is earlier than the values from the literature.
Finally, J1100-7619 would be better described by a spectral type within M0-M2 (3900-3560K), with larger uncertainties due to a more significant dependence on stellar luminosity along tracks at larger masses.

We cannot overemphasize that these new estimates assume the fiducial stellar evolutionary models are sufficient to represent the physics of these young, cool stars.
It is also possible that the stellar evolutionary models are missing key physics that will shift the model tracks to cooler temperatures.  We consider this alternative case scenario in the next section.

\subsubsection{Treatment of Magnetic Activity} 
\label{sec_discussion_sems}

\cite{cite_simonetal2019} used the stellar evolutionary model of~\cite{cite_feidenetal2016model}, which includes internal magnetic fields, to investigate if the incorporation of magnetic activity could bridge the gap between the dynamical mass measurements and the stellar evolutionary model predictions.  Using the same stellar luminosities and effective temperatures as inputs, they compared the performance of their fiducial model~\citep{cite_baraffeetal2015} to this magnetic model~\citep[the mass predictions are denoted as MB$_\mathrm{hrd}$ and MC$_\mathrm{hrd}$, respectively, in][]{cite_simonetal2019}.  Across the $\sim$0.4-1.4 M$_\Sun$ range,~\cite{cite_simonetal2019} found that the magnetic model typically provided more accurate mass predictions.
They noted, however, that this improved performance was not perfect.  Focusing on the seven stars from the literature with fiducial model mass predictions $\leq$0.5M$_\Sun$ (discussed in Section~\ref{sec_discussion_fid}),~\cite{cite_simonetal2019} show that the magnetic model predicts the dynamical masses within uncertainties for 3/7 of these stars.  Another 3/7 stellar masses are either underpredicted or overpredicted by the magnetic model by $\geq$20\% with respect to the dynamical masses.  For 2/7 stars, the use of the magnetic model actually decreases prediction accuracy relative to the fiducial model~\citep[see Table 5 of][]{cite_simonetal2019}.  In discussing possible sources of error from using the magnetic model across their entire sample,~\cite{cite_simonetal2019} noted that this magnetic model might have to be adjusted on a star-by-star basis, as different stars might have weaker or stronger internal magnetic fields.

We now use tracks from the Stellar Parameters of Tracks with Starspots (SPOTS) code of~\cite{cite_spots} as a way to flexibly treat magnetic activity for our stars.  Starspots are born from concentrated regions of magnetic activity, and so higher fractional starspot coverage essentially indicates more magnetic activity~\citep[e.g., discussion by][]{cite_somersetal2015}.  Observationally, starspots are expected to be common for low-mass pre-MS stars~\citep[e.g.,][]{cite_herbstetal1994, cite_codyetal2014}.  Theoretically, the inclusion of starspots in the models lead to predicted stars that are cooler, older, and larger in radius than stellar evolutionary models without starspots~\citep{cite_spots}.  The SPOTS code provides tracks with different fractional starspot coverage (denoted as $f_\mathrm{spot}$) of the stellar surface.  $f_\mathrm{spot}$ values range discretely from 0.00 to 0.85 at intervals of 0.17.

\textit{Assuming that the stellar effective temperatures in Table~\ref{table_sample} are accurate,} we compare the mass predictions $M_\mathrm{spot}$ of these starspot stellar evolutionary models for different values of $f_\mathrm{spot}$ to the dynamical masses ($M_\mathrm{dyn}$) for our three stars in Figure~\ref{fig_spotperf}.  The results for the starspot models at $f_\mathrm{spot}=0$ are consistent with the results from the fiducial models (Appendix~\ref{sec_appendix_fspot}).
Figure~\ref{fig_spotperf} illustrates two key findings.  First, the starspot model mass predictions \textit{can} agree with the dynamically-measured masses of these stars.  Second, the three stars require different $f_\mathrm{spot}$ values to match the dynamical masses.  $M_\mathrm{spot}$ predictions for different values of $f_\mathrm{spot}$ are listed in Appendix~\ref{sec_appendix_fspot}.


We also predict masses for a subset of six Taurus stars with M$_\mathrm{dyn}$ $\leq$0.5M$_\Sun$ from the~\cite{cite_simonetal2019} compilation, using tracks from the starspot model and our fiducial (MIST) model.
For these stars from the literature, we adapted the stellar characteristics given in~\cite{cite_herczegetal2014}.  Details, assumed characteristics, and the $M_\mathrm{spot}$ predictions for the literature are listed in Appendix~\ref{sec_appendix_fspot}.
We linearly interpolated the $f_\mathrm{spot}$ values that minimized the difference between $M_\mathrm{dyn}$ and $M_\mathrm{spot}$ for the combined sample of nine stars.  These interpolated $f_\mathrm{spot}$ values are plotted as a function of stellar effective temperature in Figure~\ref{fig_bestfspot}, assuming that the effective temperatures are accurate.
%



Across the combined sample of stars from this work and from~\cite{cite_simonetal2019}, stars are represented by fractional starspot coverages from $\sim$0-70\%.
It is important to consider if these percentages are physically reasonable.  \cite{cite_somersetal2015} noted that most measurements of starspot coverage have been for giants, subgiants, and dwarf stars, and that few such measurements exist for pre-MS stars in general, let alone pre-MS M-stars.  Based on measurements for main-sequence stars from the literature, \cite{cite_somersetal2015} assumed a tentative upper limit on $f_\mathrm{spot}$ of $\sim$50\%.
More recently, however,~\cite{cite_gullysantiagoetal2017} conducted a detailed observational study of the heavily-spotted LkCa 4, a pre-MS star, and found a fractional spot coverage of $\sim$80$\pm$5\%.  The authors further noted that there is spectroscopic and photometric evidence that implies significant starspot coverage for other young, low-mass stars.  This evidence is discussed in excellent detail in~\cite{cite_gullysantiagoetal2017}.  Here we summarize by saying that significant starspot coverage for pre-MS, low-mass stars may explain discrepancies in the literature, such as the differences between observed effective temperatures derived from different optical/near-infrared wavelength bands.

Assuming the interpolated $f_\mathrm{spot}$ values for the combined sample are reasonable, we return to Figure~\ref{fig_bestfspot}.  Since the combined sample size is small, and the uncertainties in effective temperature are large, we do not perform any quantitative analysis with this data.
That being said, we qualitatively note that the interpolated $f_\mathrm{spot}$ values appear to decrease as observed effective temperature increases, with the exception of the outlying system J0432+1827.  To statistically evaluate this trend, we would need (1) a much larger sample of dynamically-measured masses for low-mass unary M-stars and (2) more precise and confidently accurate constraints on observed stellar characteristics (particularly the effective temperatures).  If such a trend can be securely identified and quantified for low-mass M-stars, we will be able to more accurately predict masses using starspot stellar evolutionary models in the low-mass regime.

%

\section{Summary}
\label{sec_summary}

Using precise \textit{Gaia} distances and spatially resolved ALMA observations of CO isotopologue emission observed toward Keplerian-rotating protoplanetary disks, we have dynamically measured the stellar masses of three low-mass ($\leq$0.5M$_\Sun$) unary M-stars.  Two of these masses have not been measured before and the third is now measured with higher precision.  We have evaluated the performance of fiducial and starspot stellar evolutionary model predictions for our three stars and on a subset of low-mass M-stars extracted from the literature.  We summarize our main findings below:

\begin{enumerate}
    \item By forward-modeling the $^{12}$CO (J=2--1) and $^{13}$CO (J=2--1) emission, we dynamically measure stellar masses of 0.395$\pm$0.012 M$_\Sun$ for FP Tau, 0.192$\pm$0.005 M$_\Sun$ for J0432+1827, and 0.461$\pm$0.057 M$_\Sun$ for J1100-7619.
    \item We find that tracks from the fiducial stellar evolutionary model, which does not account for magnetic activity, accurately predicts the dynamical mass of J0432+1827, but underpredicts the dynamical masses by $\sim$60\% for FP Tau and $\sim$80\% for J1100-7619.  Possible explanations for these underpredictions include inaccurate assumptions on stellar effective temperatures, the lack of magnetic treatment in the fiducial models, and undetected binarity for J1100-7619.
    \item Our three stars would require warm shifts in stellar effective temperature of $\sim$250-260K for FP Tau, $\sim$40K for J0432+1827, and $\sim$320-340K for J1100-7619, respectively, to reconcile the fiducial stellar evolutionary model predictions with the dynamical masses.  These changes would correspond to warmer spectral types of M1-M2, M3-M4, and M0-M2, respectively.
    \item We evaluate stellar evolutionary models with starspots as a flexible treatment of magnetic activity for our stars.  We find that interpolated mass results for these starspot models are capable of reproducing the dynamical masses of our stars.
    Folding in a subset of the Taurus unary low-mass M-stars presented in~\cite{cite_simonetal2019}, we find a suggestive dependence of fractional starspot coverage on observed effective temperature across the combined sample.
    %
\end{enumerate}

These results and our discussion suggest that starspot stellar evolutionary models can and should be used to model low-mass pre-MS stars.  That being said, more work needs to be done before we can shift from qualitative to quantitative evaluations of starspot model performance.  Specifically, we must: (1) confirm that assumed stellar characteristics for low-mass pre-MS stars, particularly the effective temperatures, are accurate; (2) reduce the imprecision of the observed stellar characteristics; and (3) build a larger sample of dynamical masses for low-mass M-stars, measured from high-sensitivity observations (e.g., ALMA observations) of CO isotopologue emission toward circumstellar disks.
When these tasks are fulfilled, it would be extremely worthwhile to perform a larger investigative study of starspot models for young, cool stars.

%


\vspace{-15pt}

\acknowledgments

Jamila Pegues gratefully acknowledges support from the National Science Foundation (NSF) Graduate Research Fellowship through Grant Numbers DGE1144152 and DGE1745303.  Karin I. \"Oberg gratefully acknowledges support from the Simons Foundation through a Simons Collaboration on the Origins of Life (SCOL) PI Grant (Number 321183).  Gregory J. Herczeg gratefully acknowledges support from general grant 11773002 awarded by the National Science Foundation of China.  L. Ilsedore Cleeves gratefully acknowledges support from the David and Lucille Packard Foundation, the Virginia Space Grant Consortium, and Johnson \& Johnson’s WiSTEM2D Award.
Support for this work was also provided by NASA through the NASA Hubble Fellowship grants \#HST-HF2-51405.001-A, \#HST-HF2-51429.001-A, and \#HST-HF2-51460.001-A awarded by the Space Telescope Science Institute, which is operated by the Association of Universities for Research in Astronomy, Inc., for NASA, under contract NAS5-26555.

This paper makes use of the following ALMA data: ADS/JAO.ALMA\#2017.1.01107.S.  ALMA is a partnership of ESO (representing its member states), NSF (USA) and NINS (Japan), together with NRC (Canada), MOST and ASIAA (Taiwan), and KASI (Republic of Korea), in cooperation with the Republic of Chile. The Joint ALMA Observatory is operated by ESO, AUI/NRAO and NAOJ.  The National Radio Astronomy Observatory is a facility of the National Science Foundation operated under cooperative agreement by Associated Universities, Inc.
 
This work has made use of data from the European Space Agency (ESA) mission
{\it Gaia} (\url{https://www.cosmos.esa.int/gaia}), processed by the {\it Gaia}
Data Processing and Analysis Consortium (DPAC,
\url{https://www.cosmos.esa.int/web/gaia/dpac/consortium}). Funding for the DPAC
has been provided by national institutions, in particular the institutions
participating in the {\it Gaia} Multilateral Agreement.

This paper made use of tracks from the MESA Isochrones and Stellar Tracks code~\citep[MIST;][]{cite_mist1, cite_mist2, cite_mistmesa1, cite_mistmesa2, cite_mistmesa3, cite_mistmesa4} and from the Stellar Parameters of Tracks with Starspots code~\citep[SPOTS;][]{cite_spots}.

All data reduction scripts and computer code used for this research were written in Python (version 2.7).  All plots were generated using Python's Matplotlib package~\citep{cite_matplotlib}.  This research also made use of Astropy (\url{http://www.astropy.org}), a community-developed core Python package for Astronomy~\citep{cite_astropy2013, cite_astropy2018}, as well as the NumPy~\citep{cite_numpy} and SciPy~\citep{cite_scipy} Python packages.  This research additionally made use of the measurement set (ms) and hierarchical data format (version 5; hdf5) conversion scripts that are publicly available on the Astrochem Github page (\url{https://github.com/AstroChem/UVHDF5}).


\clearpage

\appendix
\vspace{-20pt}

\section{Channel Maps}
\label{sec_appendix_chans}

\begin{figure*}
\centering
\resizebox{0.99\hsize}{!}{
    \includegraphics{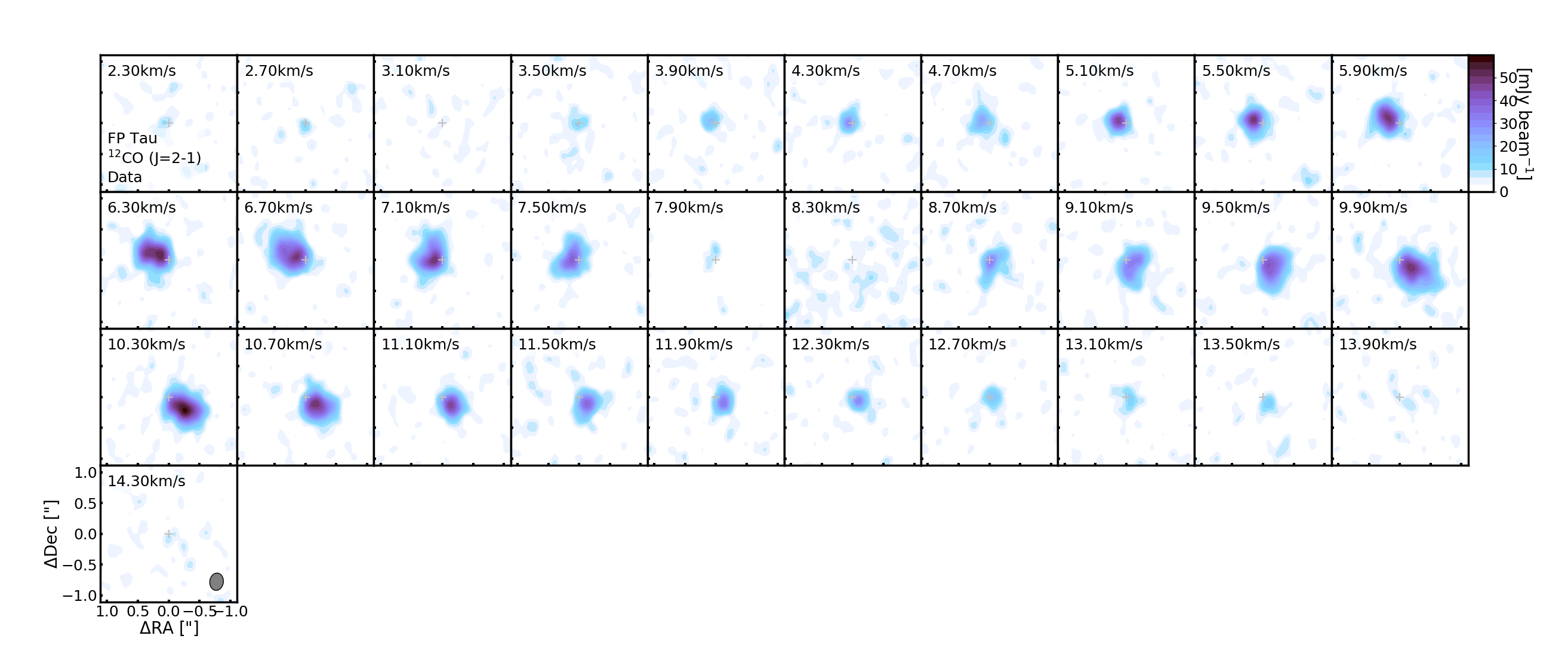}}
\caption{Channel maps of the $^{12}$CO data toward FP Tau.  Beam sizes are drawn in the lower left corners of the entire panels.
\label{fig_A1_1}}
\end{figure*}

\begin{figure*}
\centering
\resizebox{0.99\hsize}{!}{
    \includegraphics{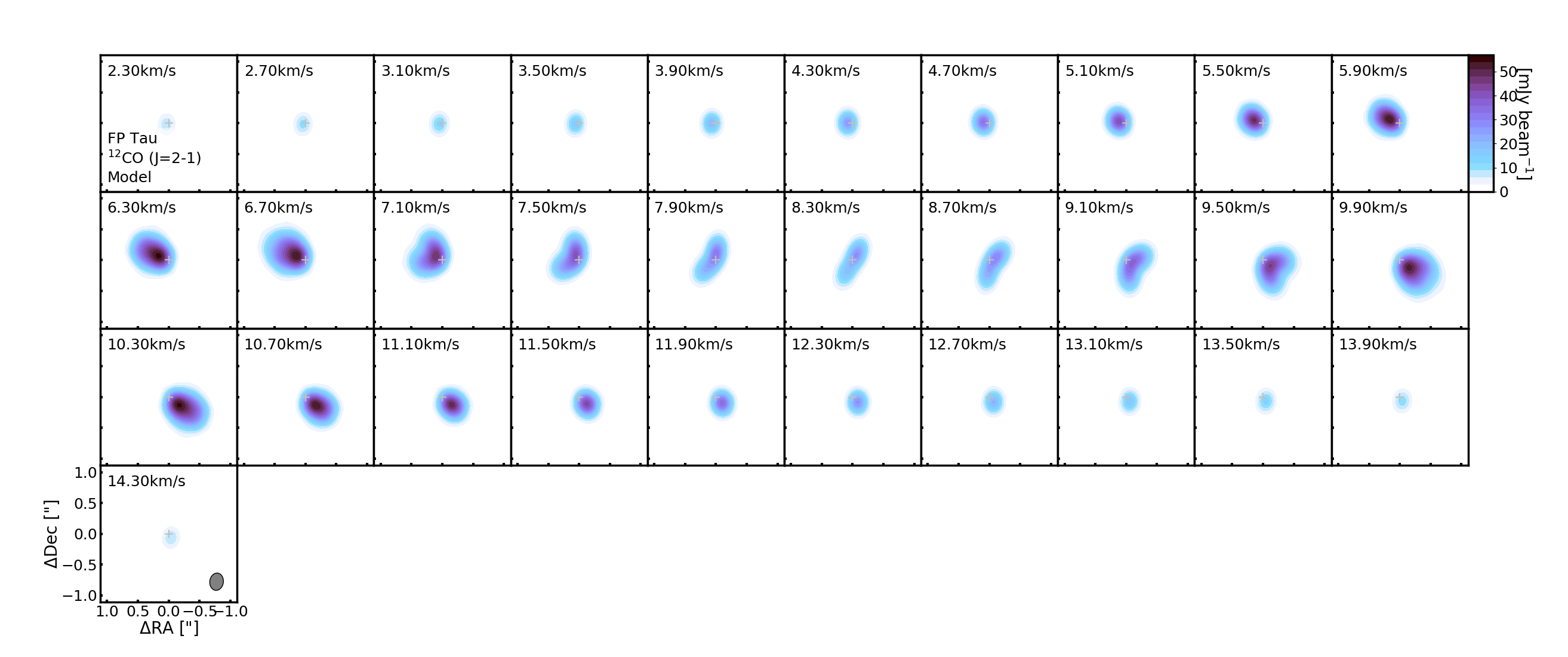}}
\caption{Channel maps of the $^{12}$CO model toward FP Tau.  Beam sizes are drawn in the lower left corners of the entire panels.
\label{fig_A1_2}}
\end{figure*}

\begin{figure*}
\centering
\resizebox{0.99\hsize}{!}{
    \includegraphics{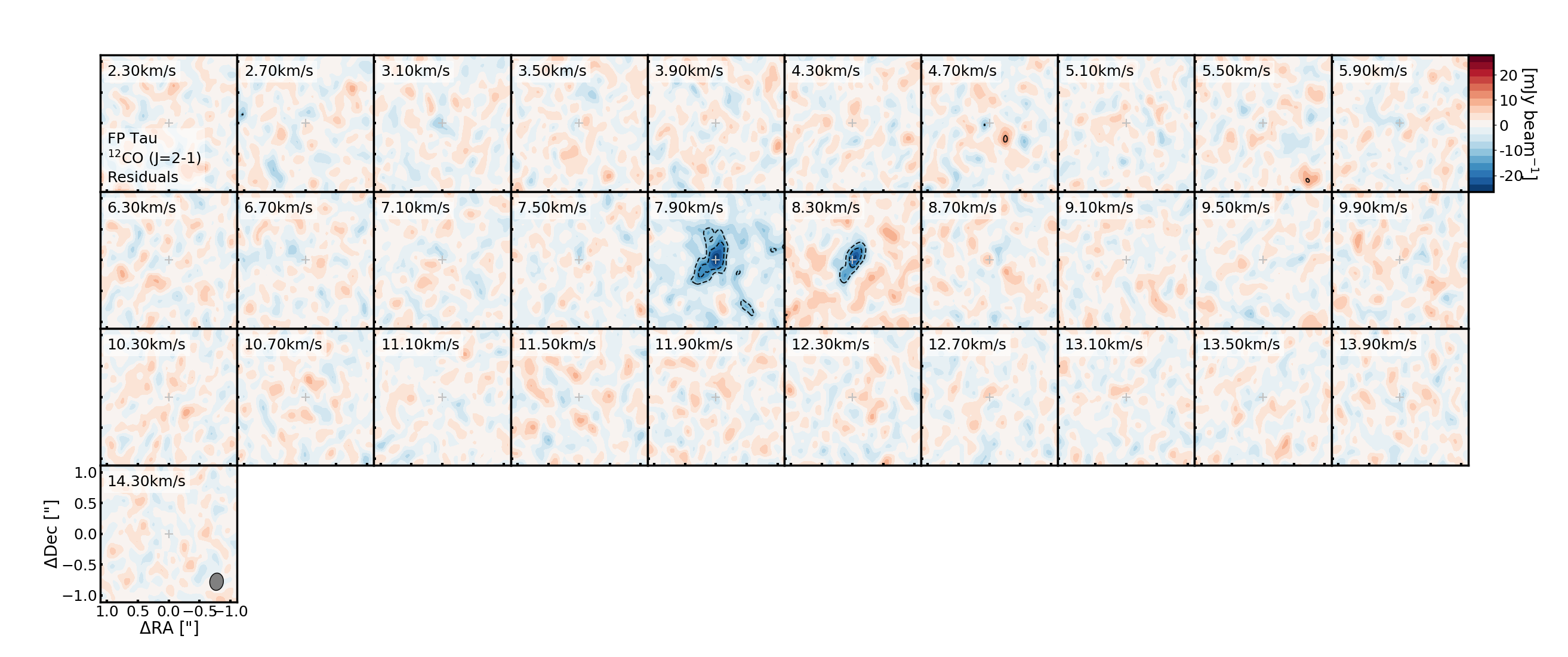}}
\caption{Channel maps of the $^{12}$CO residuals toward FP Tau.  The contours for the residual panel are [3$\sigma$, 5$\sigma$, 10$\sigma$, 20$\sigma$...], where $\sigma$ is the channel rms.  Beam sizes are drawn in the lower left corners of the entire panels.
\label{fig_A1_3}}
\end{figure*}

\begin{figure*}
\centering
\resizebox{0.99\hsize}{!}{
    \includegraphics{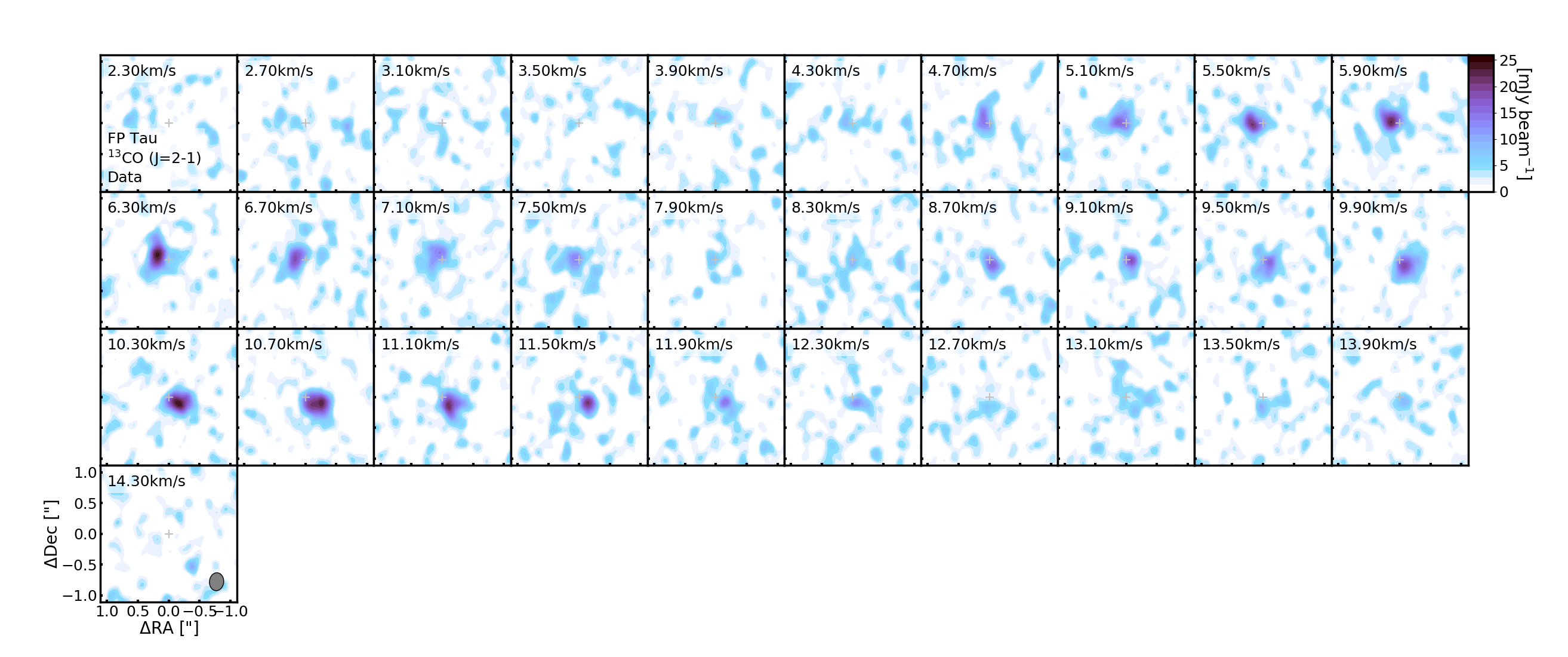}}
\caption{Channel maps of the $^{13}$CO data toward FP Tau.  Beam sizes are drawn in the lower left corners of the entire panels.
\label{fig_A1_4}}
\end{figure*}

\begin{figure*}
\centering
\resizebox{0.99\hsize}{!}{
    \includegraphics{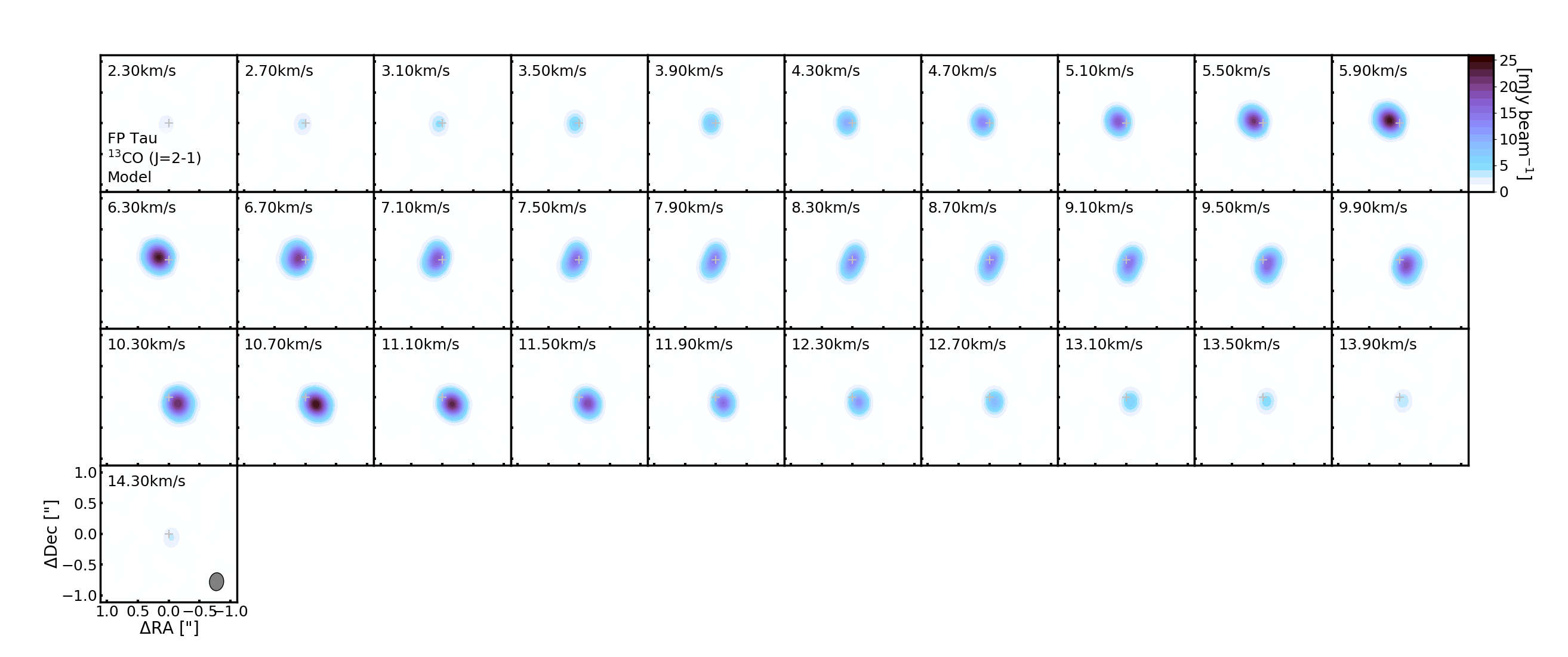}}
\caption{Channel maps of the $^{13}$CO model toward FP Tau.  Beam sizes are drawn in the lower left corners of the entire panels.
\label{fig_A1_5}}
\end{figure*}

\begin{figure*}
\centering
\resizebox{0.99\hsize}{!}{
    \includegraphics{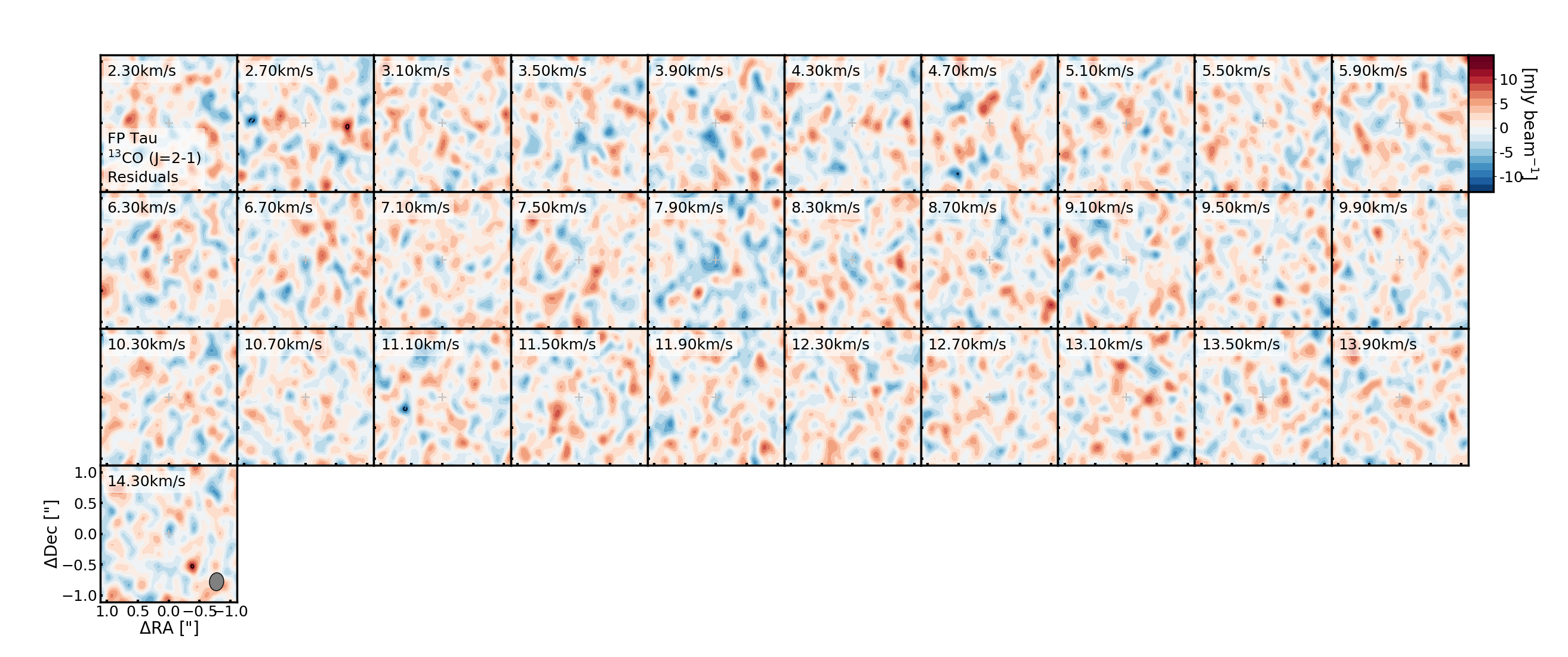}}
\caption{Channel maps of the $^{13}$CO residuals toward FP Tau.  The contours for the residual panel are [3$\sigma$, 5$\sigma$, 10$\sigma$, 20$\sigma$...], where $\sigma$ is the channel rms.  Beam sizes are drawn in the lower left corners of the entire panels.
\label{fig_A1_6}}
\end{figure*}

\begin{figure*}
\centering
\resizebox{0.99\hsize}{!}{
    \includegraphics{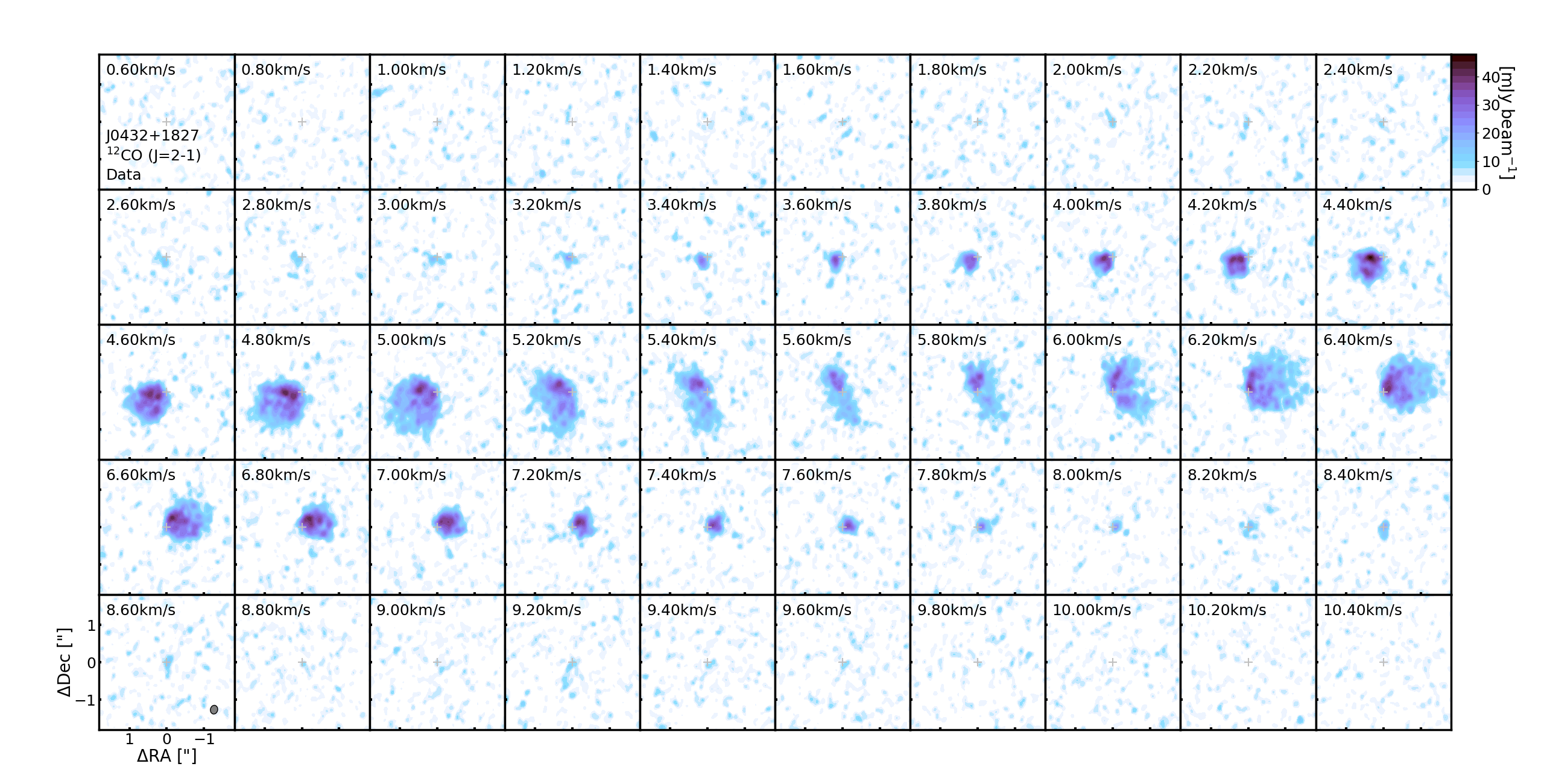}}
\caption{Channel maps of the $^{12}$CO 2--1 data toward J0432+1827.  Beam sizes are drawn in the lower left corners of the entire panels.
\label{fig_A2_1}}
\end{figure*}

\begin{figure*}
\centering
\resizebox{0.99\hsize}{!}{
    \includegraphics{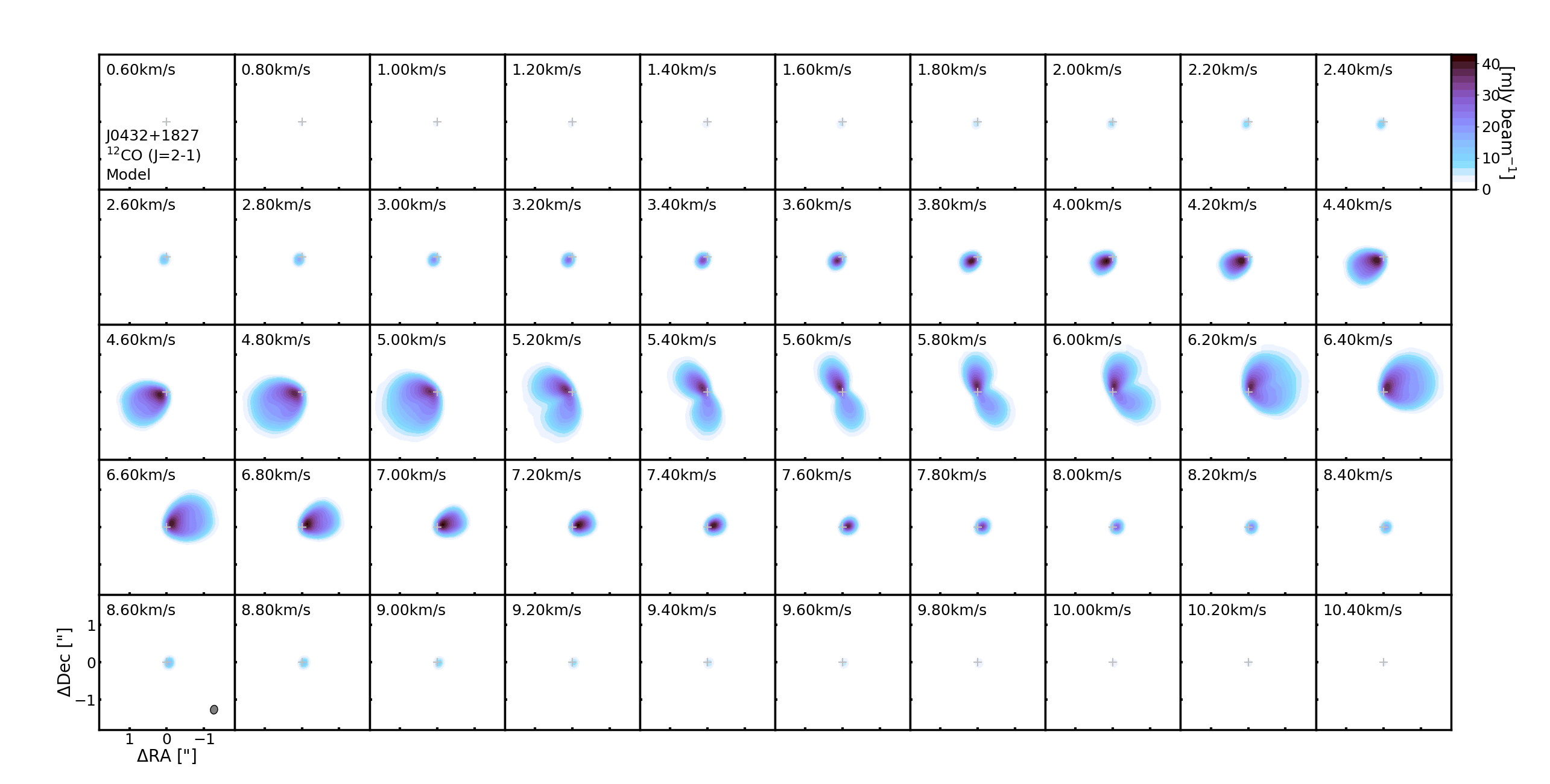}}
\caption{Channel maps of the $^{12}$CO 2--1 model toward J0432+1827.  Beam sizes are drawn in the lower left corners of the entire panels.
\label{fig_A2_2}}
\end{figure*}

\begin{figure*}
\centering
\resizebox{0.99\hsize}{!}{
    \includegraphics{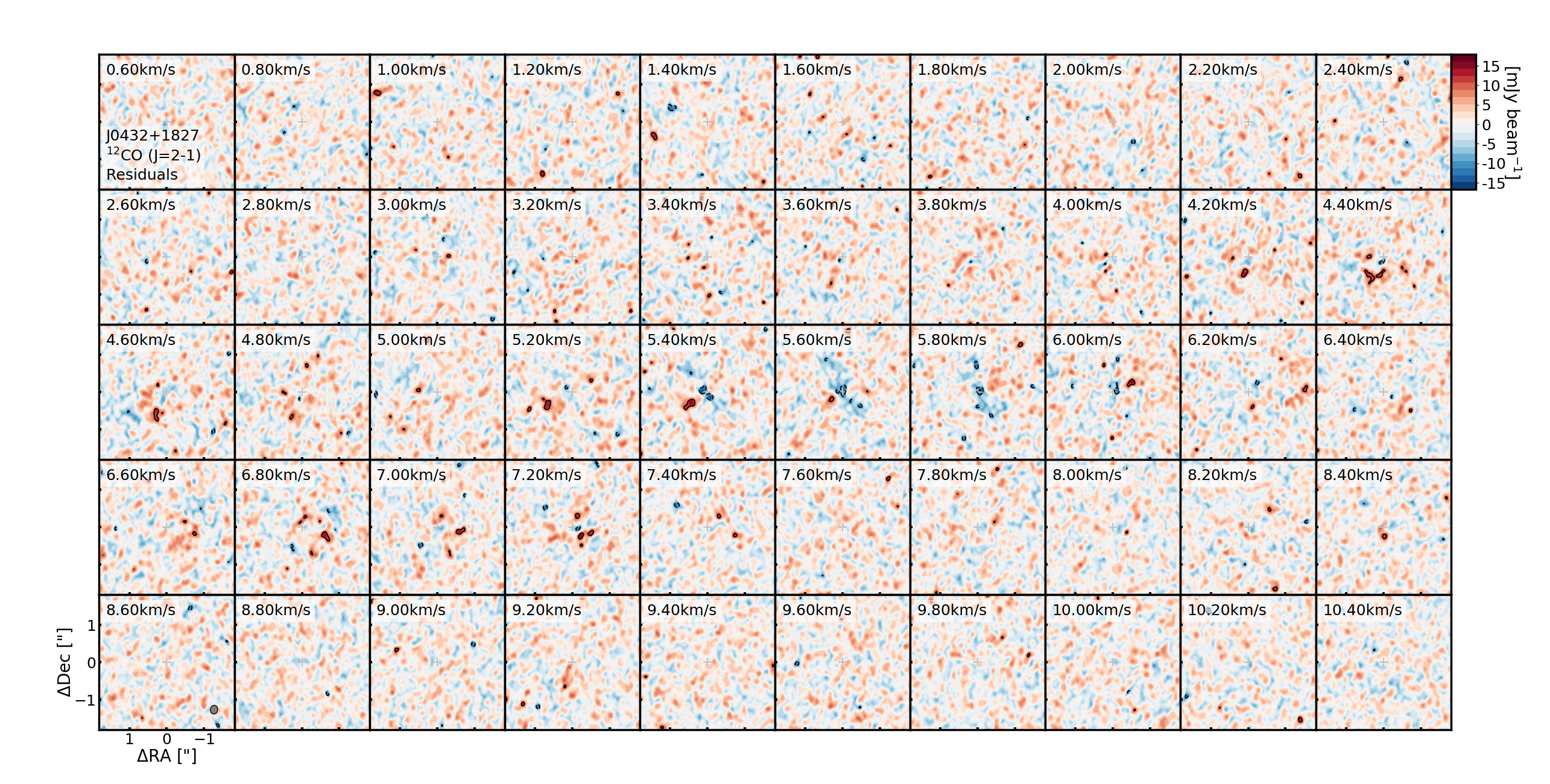}}
\caption{Channel maps of the $^{12}$CO 2--1 residuals toward J0432+1827.  The contours for the residual panel are [3$\sigma$, 5$\sigma$, 10$\sigma$, 20$\sigma$...], where $\sigma$ is the channel rms.  Beam sizes are drawn in the lower left corners of the entire panels.
\label{fig_A2_3}}
\end{figure*}

\begin{figure*}
\centering
\resizebox{0.99\hsize}{!}{
    \includegraphics{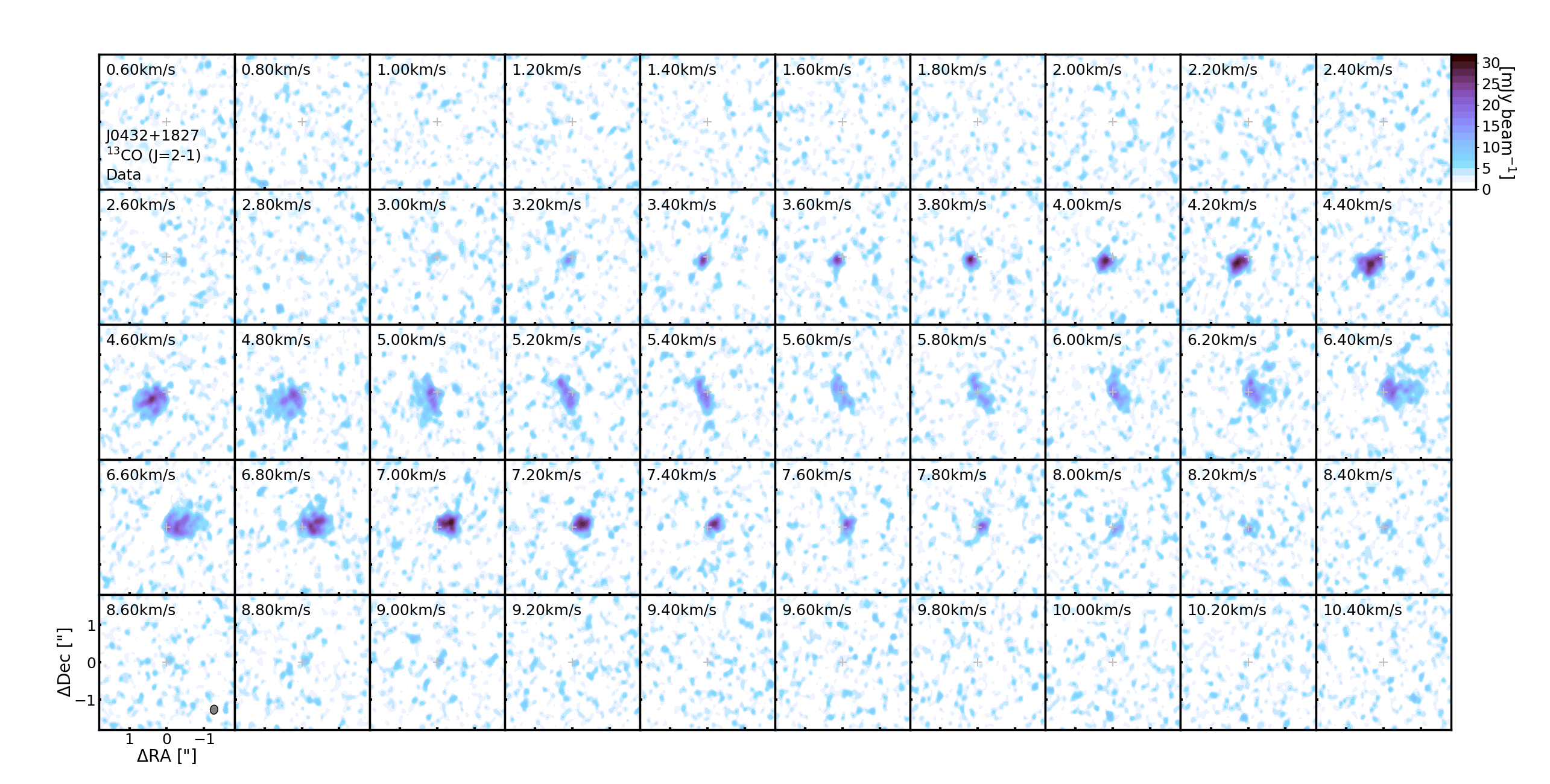}}
\caption{Channel maps of the $^{13}$CO 2--1 data toward J0432+1827.  Beam sizes are drawn in the lower left corners of the entire panels.
\label{fig_A2_4}}
\end{figure*}

\begin{figure*}
\centering
\resizebox{0.99\hsize}{!}{
    \includegraphics{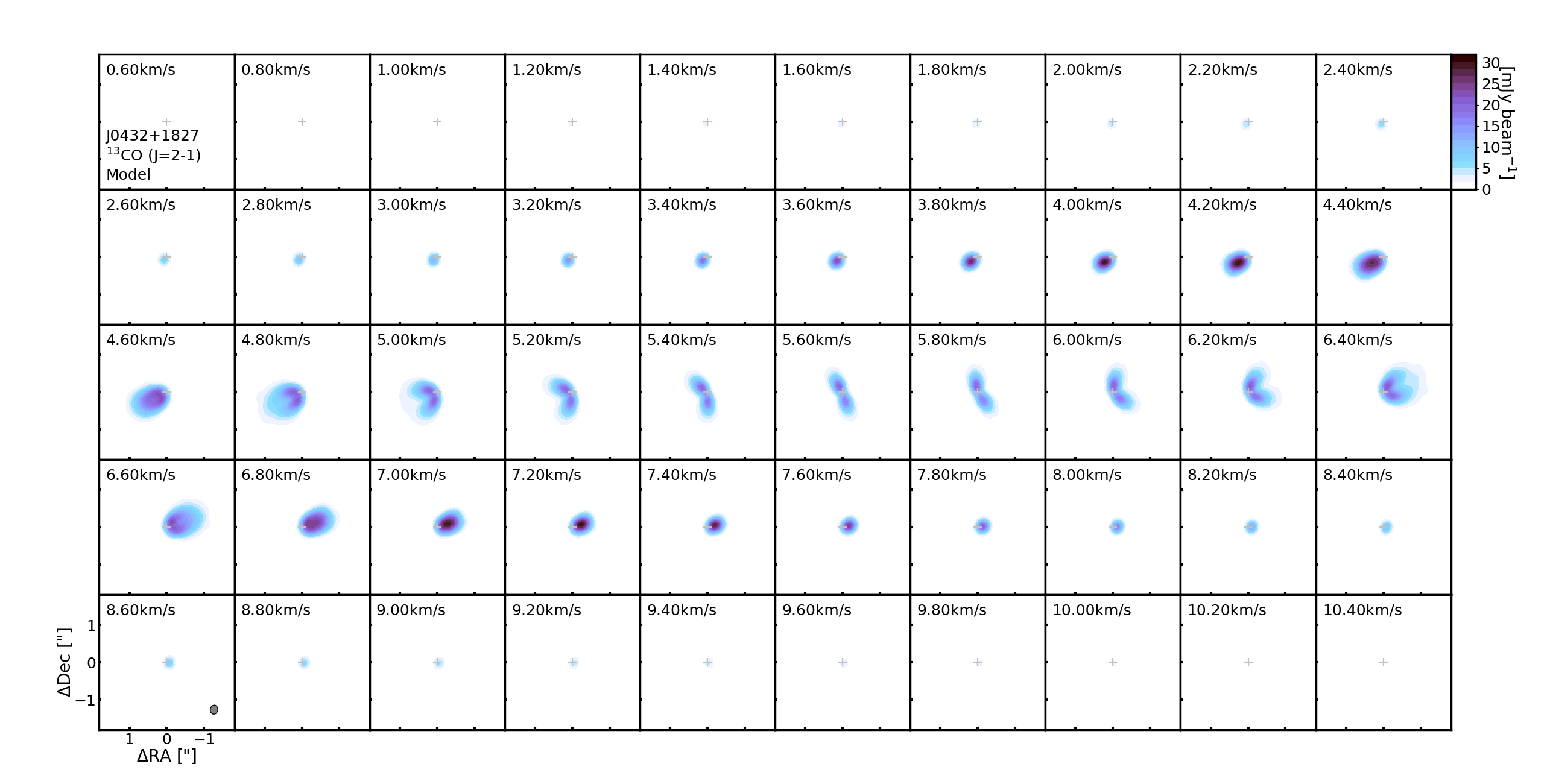}}
\caption{Channel maps of the $^{13}$CO 2--1 model toward J0432+1827.  Beam sizes are drawn in the lower left corners of the entire panels.
\label{fig_A2_5}}
\end{figure*}

\begin{figure*}
\centering
\resizebox{0.99\hsize}{!}{
    \includegraphics{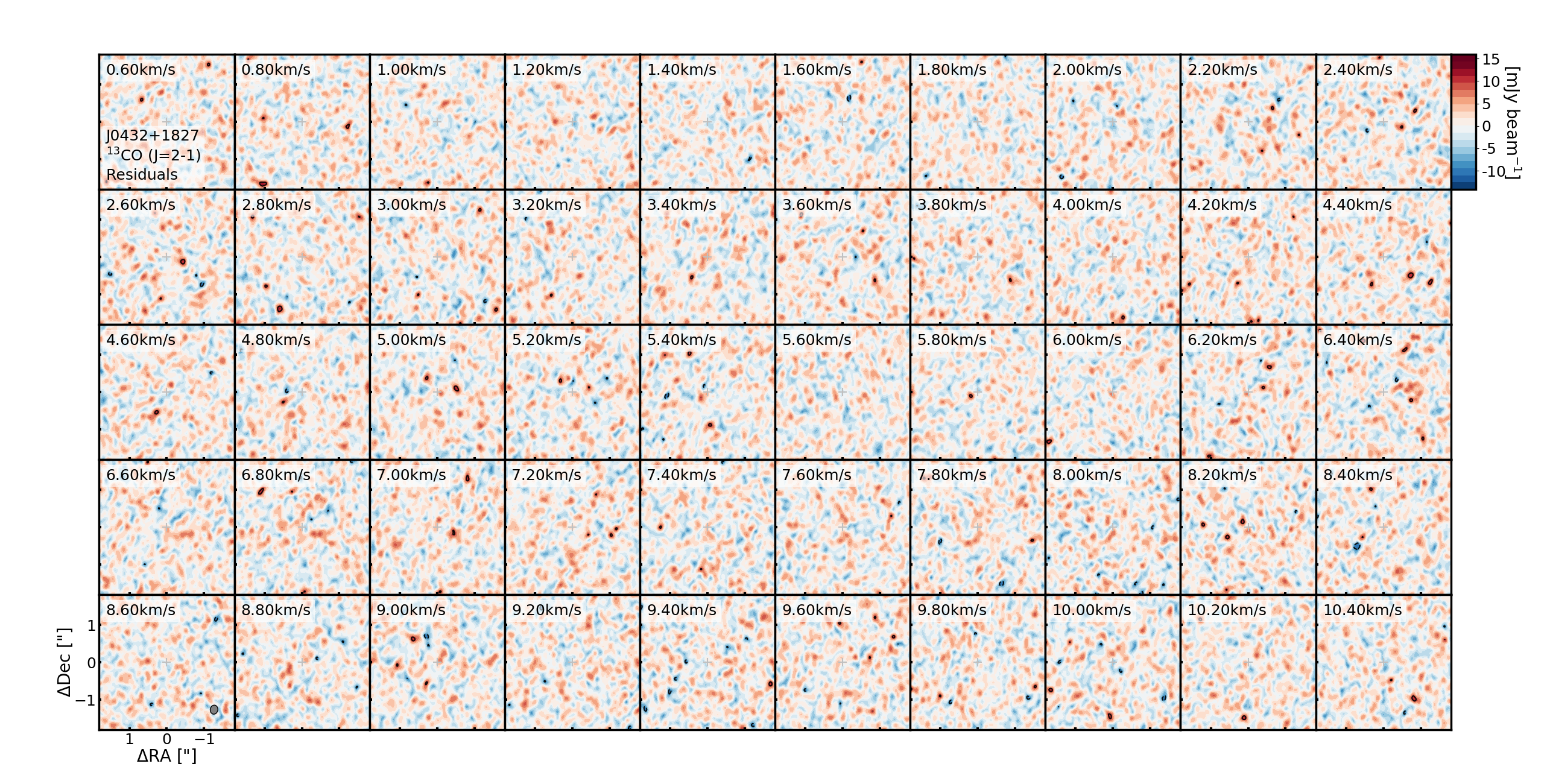}}
\caption{Channel maps of the $^{13}$CO 2--1 residuals toward J0432+1827.  The contours for the residual panel are [3$\sigma$, 5$\sigma$, 10$\sigma$, 20$\sigma$...], where $\sigma$ is the channel rms.  Beam sizes are drawn in the lower left corners of the entire panels.
\label{fig_A2_6}}
\end{figure*}

\begin{figure*}
\centering
\resizebox{0.99\hsize}{!}{
    \includegraphics{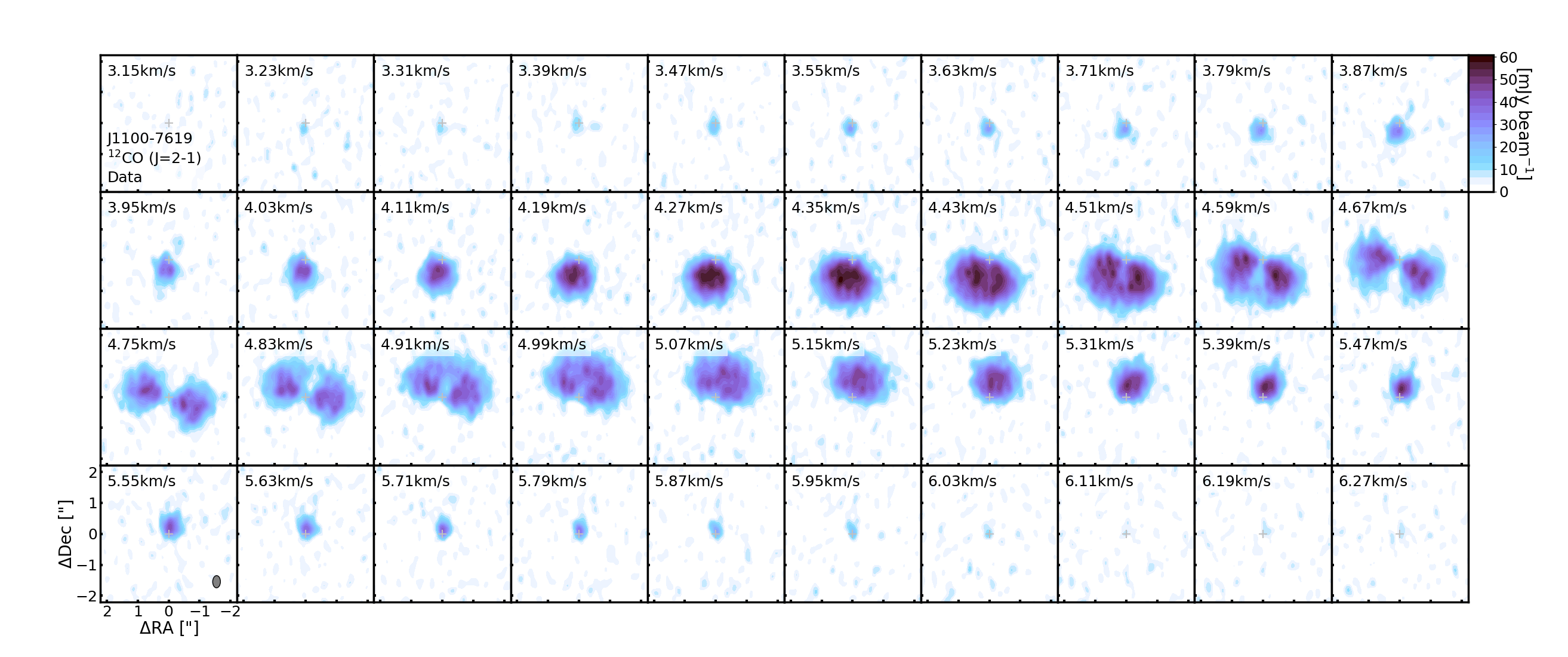}}
\caption{Channel maps of the $^{12}$CO 2--1 data toward J1100-7619.  Beam sizes are drawn in the lower left corners of the entire panels.
\label{fig_A3_1}}
\end{figure*}

\begin{figure*}
\centering
\resizebox{0.99\hsize}{!}{
    \includegraphics{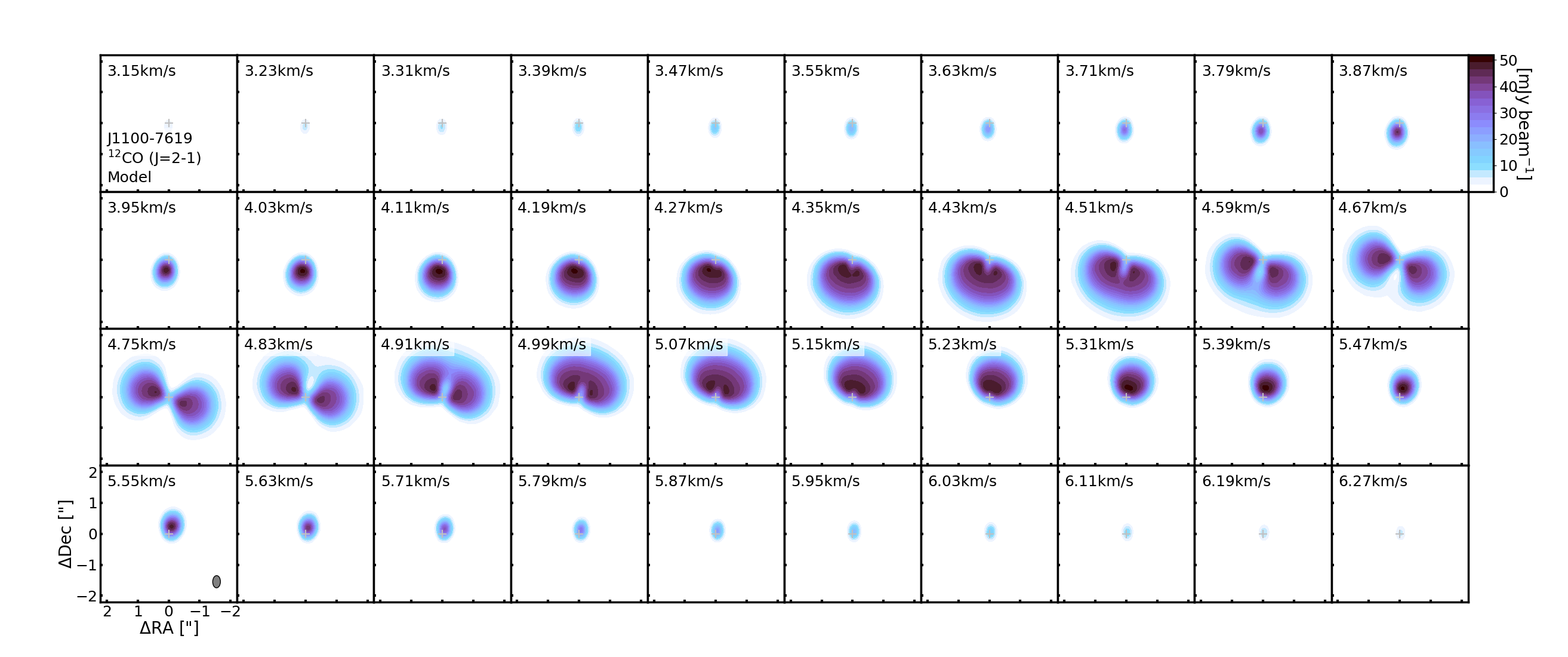}}
\caption{Channel maps of the $^{12}$CO 2--1 model toward J1100-7619.  Beam sizes are drawn in the lower left corners of the entire panels.
\label{fig_A3_2}}
\end{figure*}

\begin{figure*}
\centering
\resizebox{0.99\hsize}{!}{
    \includegraphics{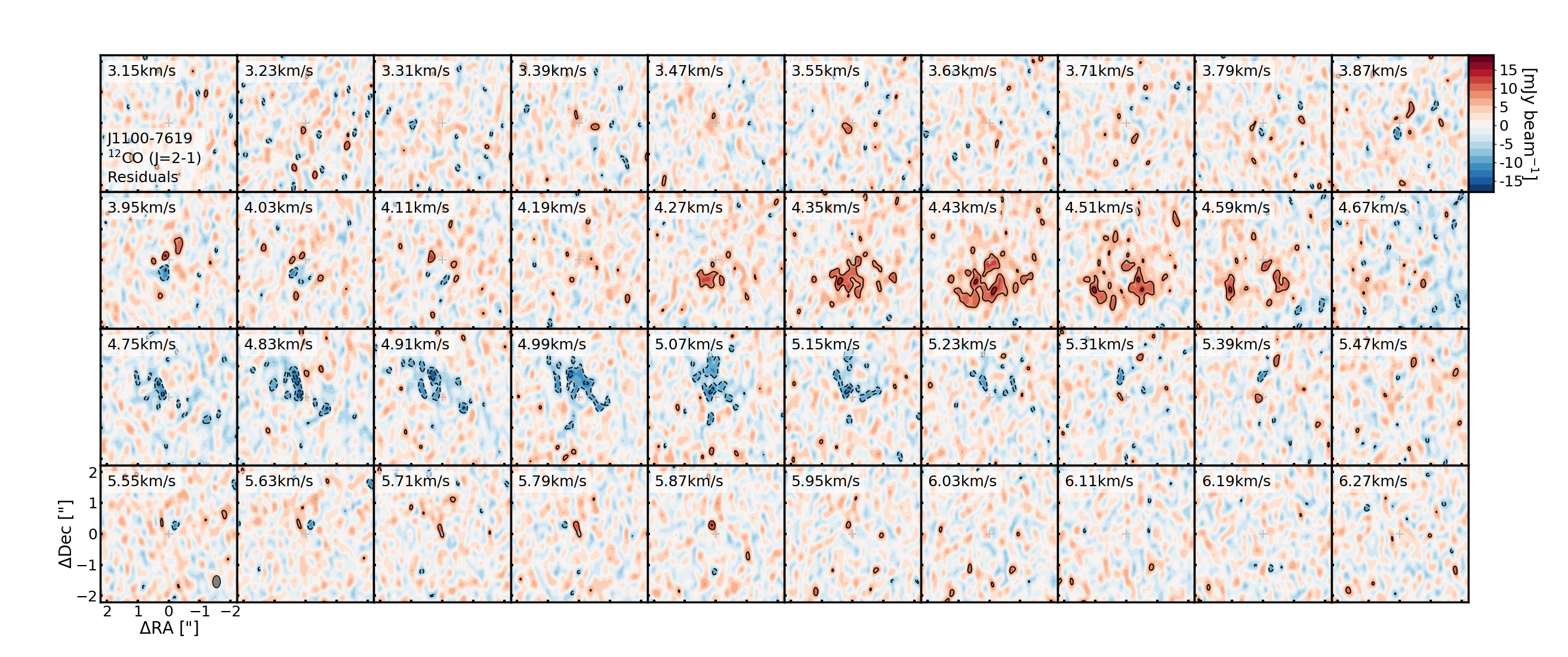}}
\caption{Channel maps of the $^{12}$CO 2--1 residuals toward J1100-7619.  The contours for the residual panel are [3$\sigma$, 5$\sigma$, 10$\sigma$, 20$\sigma$...], where $\sigma$ is the channel rms.  Beam sizes are drawn in the lower left corners of the entire panels.
\label{fig_A3_3}}
\end{figure*}

\begin{figure*}
\centering
\resizebox{0.99\hsize}{!}{
    \includegraphics{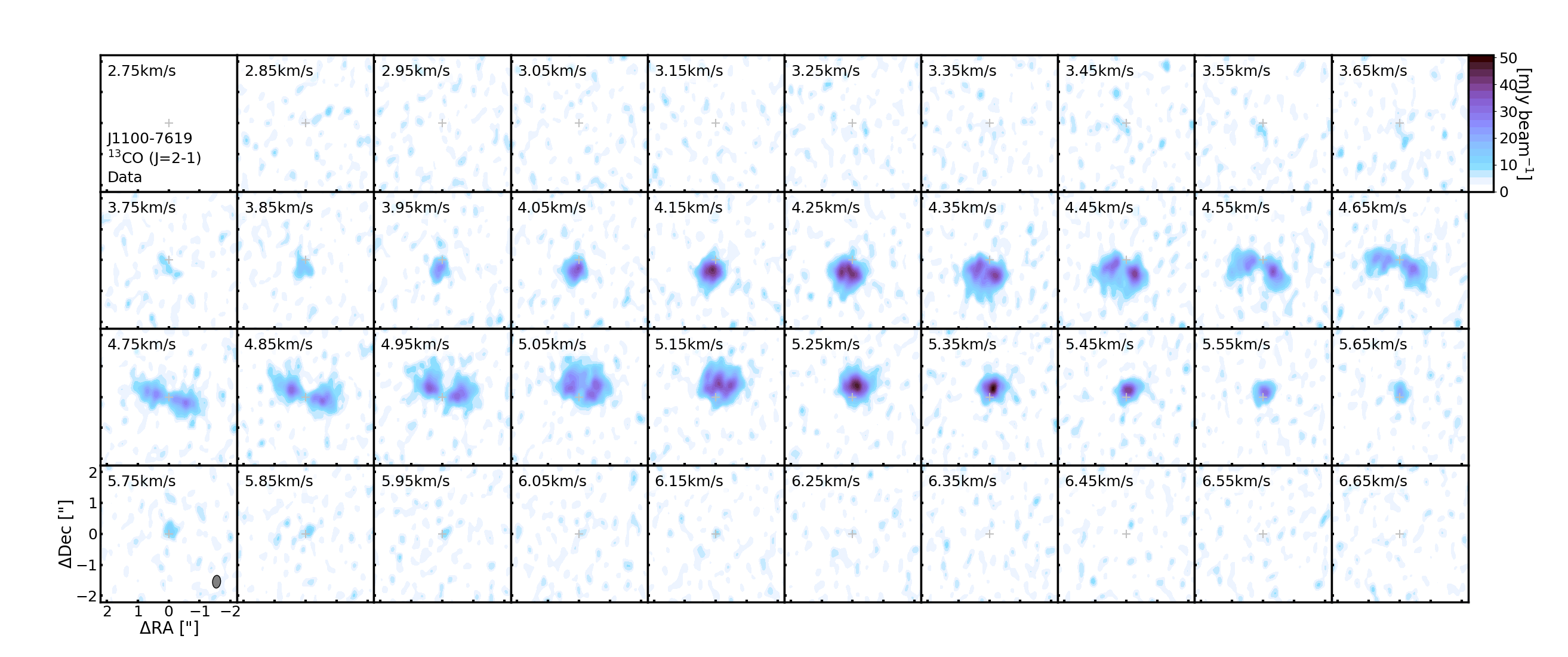}}
\caption{Channel maps of the $^{13}$CO 2--1 data toward J1100-7619.  Beam sizes are drawn in the lower left corners of the entire panels.
\label{fig_A3_4}}
\end{figure*}

Figures~\ref{fig_A1_1} through~\ref{fig_A1_6} show channel maps of the data, model, and residuals for $^{12}$CO 2--1 and $^{13}$CO 2--1 toward FP Tau.  Figures~\ref{fig_A2_1} through~\ref{fig_A2_6} show the same set of channel maps, but for $^{12}$CO 2--1 and $^{13}$CO 2--1 toward J0432+1827.  Figures~\ref{fig_A3_1} through~\ref{fig_A3_3} show the same set of channel maps for $^{12}$CO 2--1 toward J1100-7619, while Figure~\ref{fig_A3_4} shows channel maps of the $^{13}$CO 2--1 data toward J1100-7619.

%

%

%


\section{$^{12}$CO 2--1 MCMC Chains and Corner Plots}
\label{sec_appendix_MCMC_12CO}

\begin{figure*}
\centering
\resizebox{0.99\hsize}{!}{
    \includegraphics{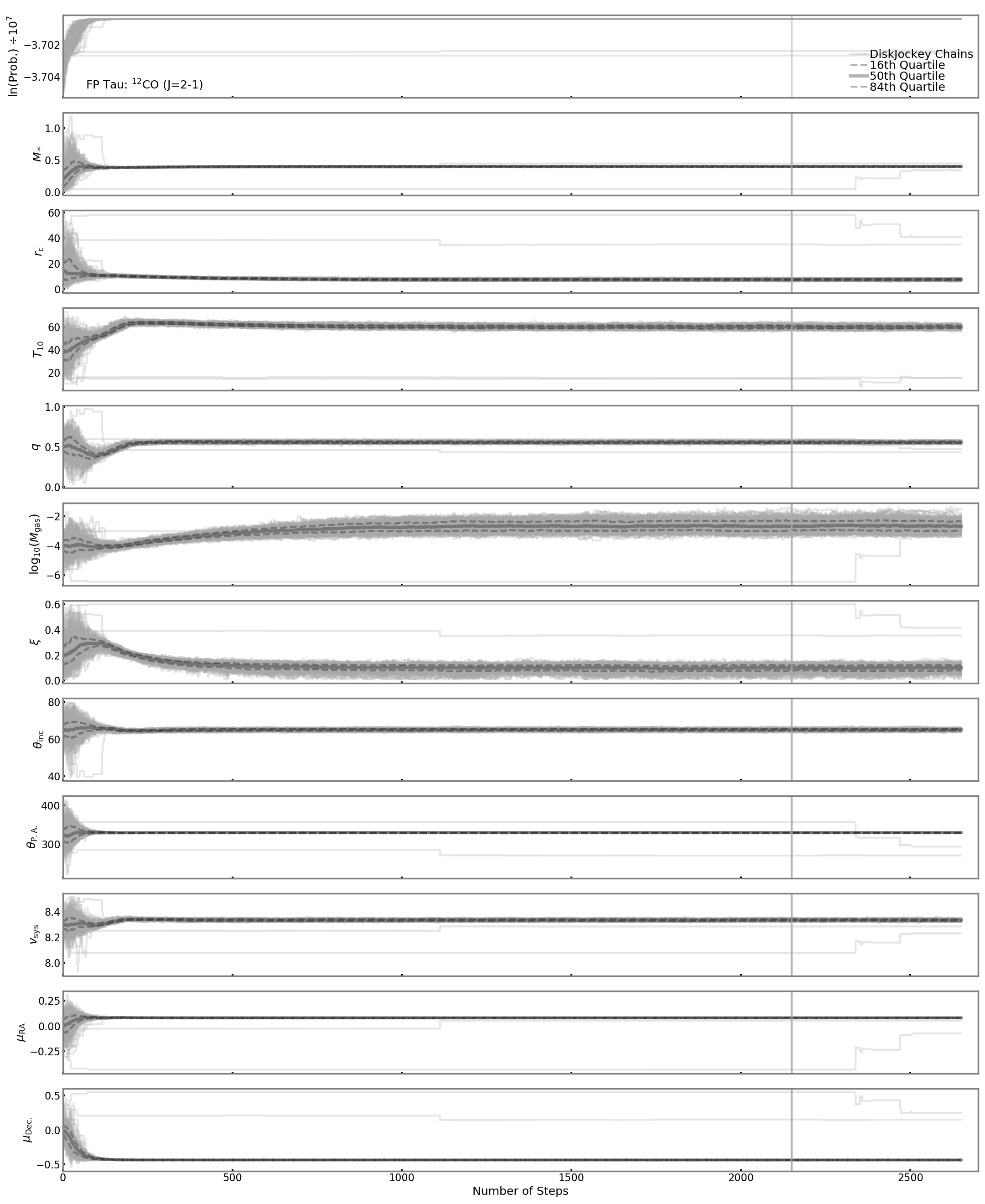}}
\caption{MCMC chain plot for $^{12}$CO toward FP Tau.  The outlying chains with probabilities significantly lower than all other chains were excluded from the final sampling distributions.
\label{fig_B1_1}}
\end{figure*}

\begin{figure*}
\centering
\resizebox{0.99\hsize}{!}{
    \includegraphics{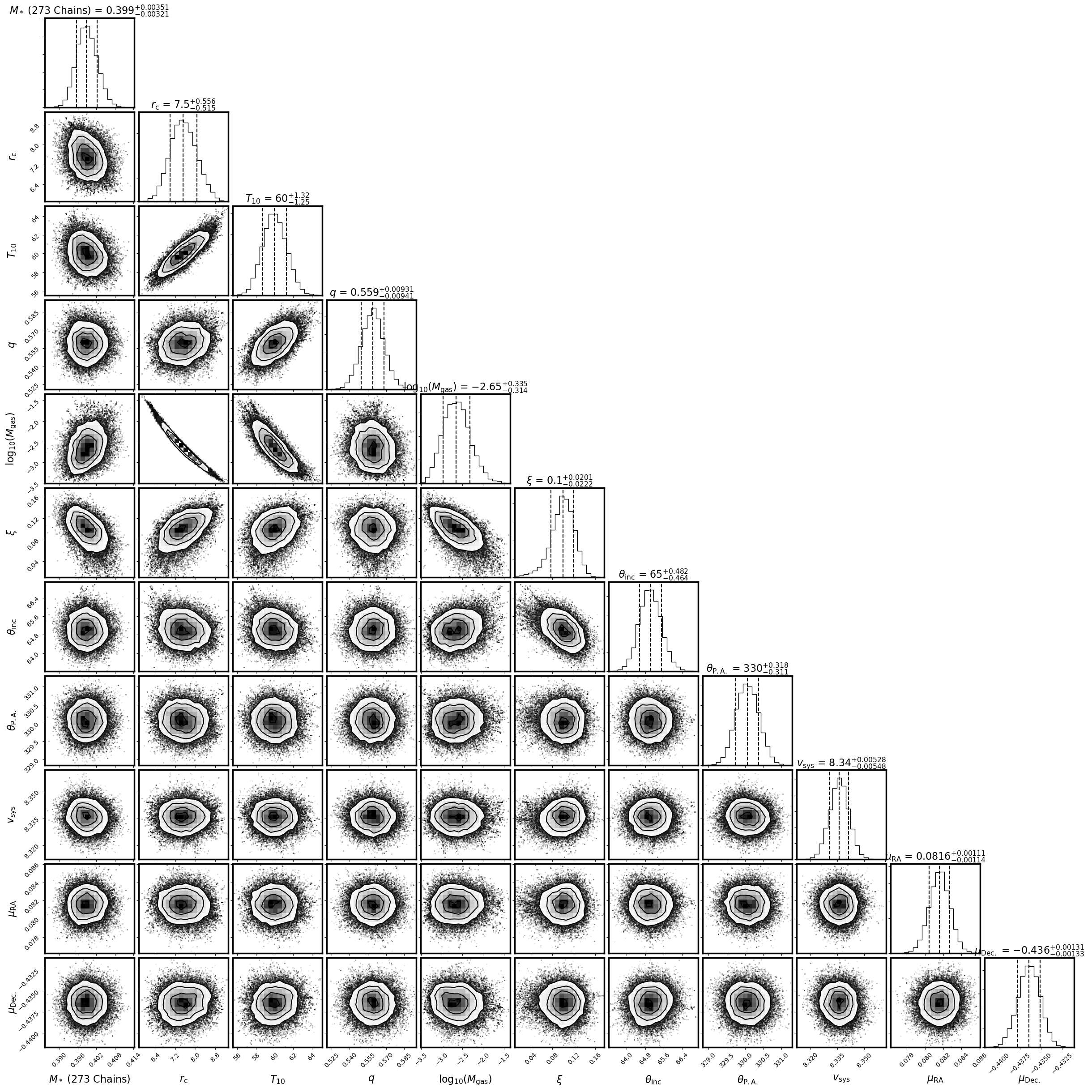}}
\caption{Corner plot for $^{12}$CO toward FP Tau.  The number of included MCMC chains (maximum of 275) is listed at the bottom of the figure.
\label{fig_B1_2}}
\end{figure*}

\begin{figure*}
\centering
\resizebox{0.99\hsize}{!}{
    \includegraphics{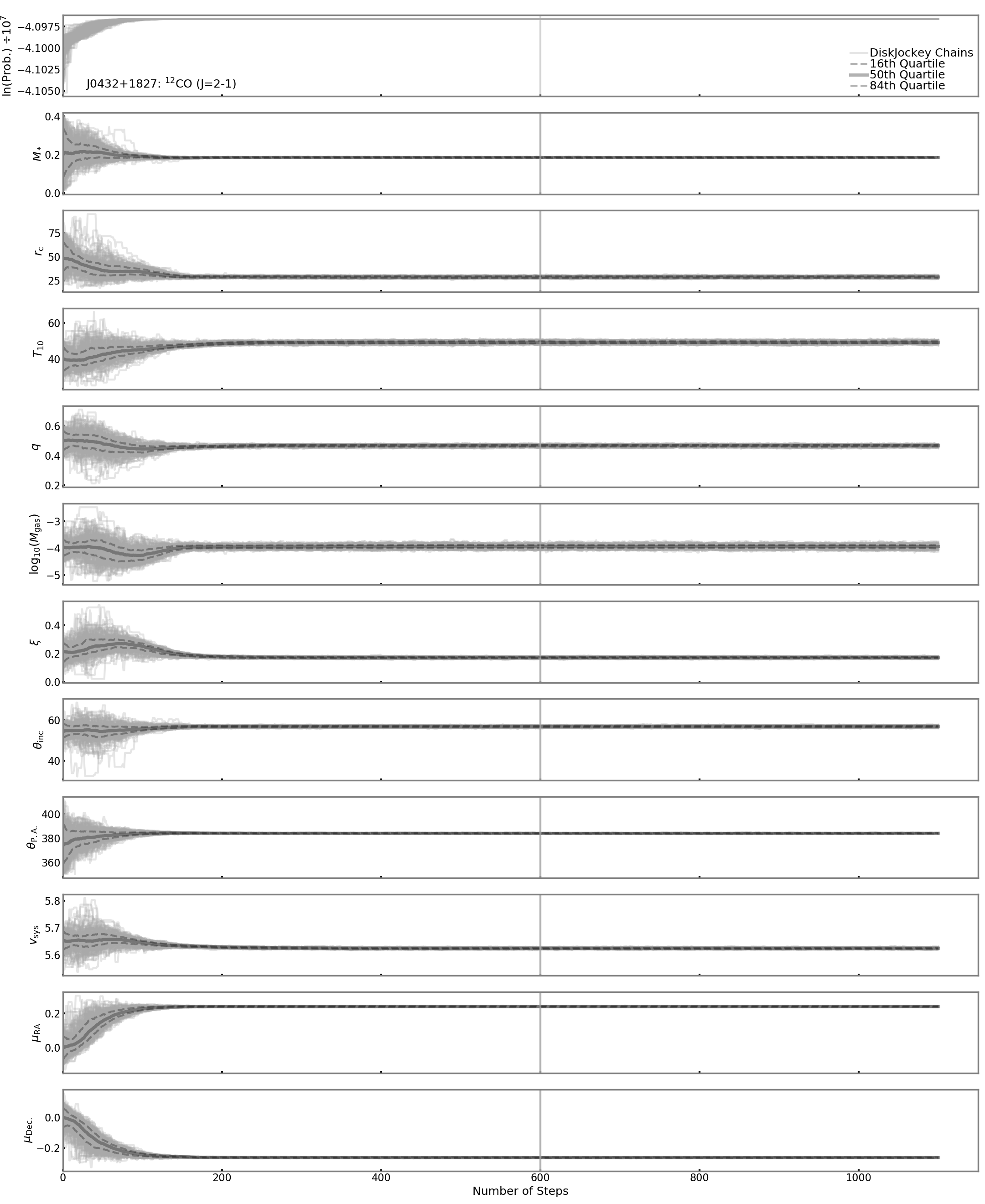}}
\caption{MCMC chain plot for $^{12}$CO toward J0432+1827.
\label{fig_B1_3}}
\end{figure*}

\begin{figure*}
\centering
\resizebox{0.99\hsize}{!}{
    \includegraphics{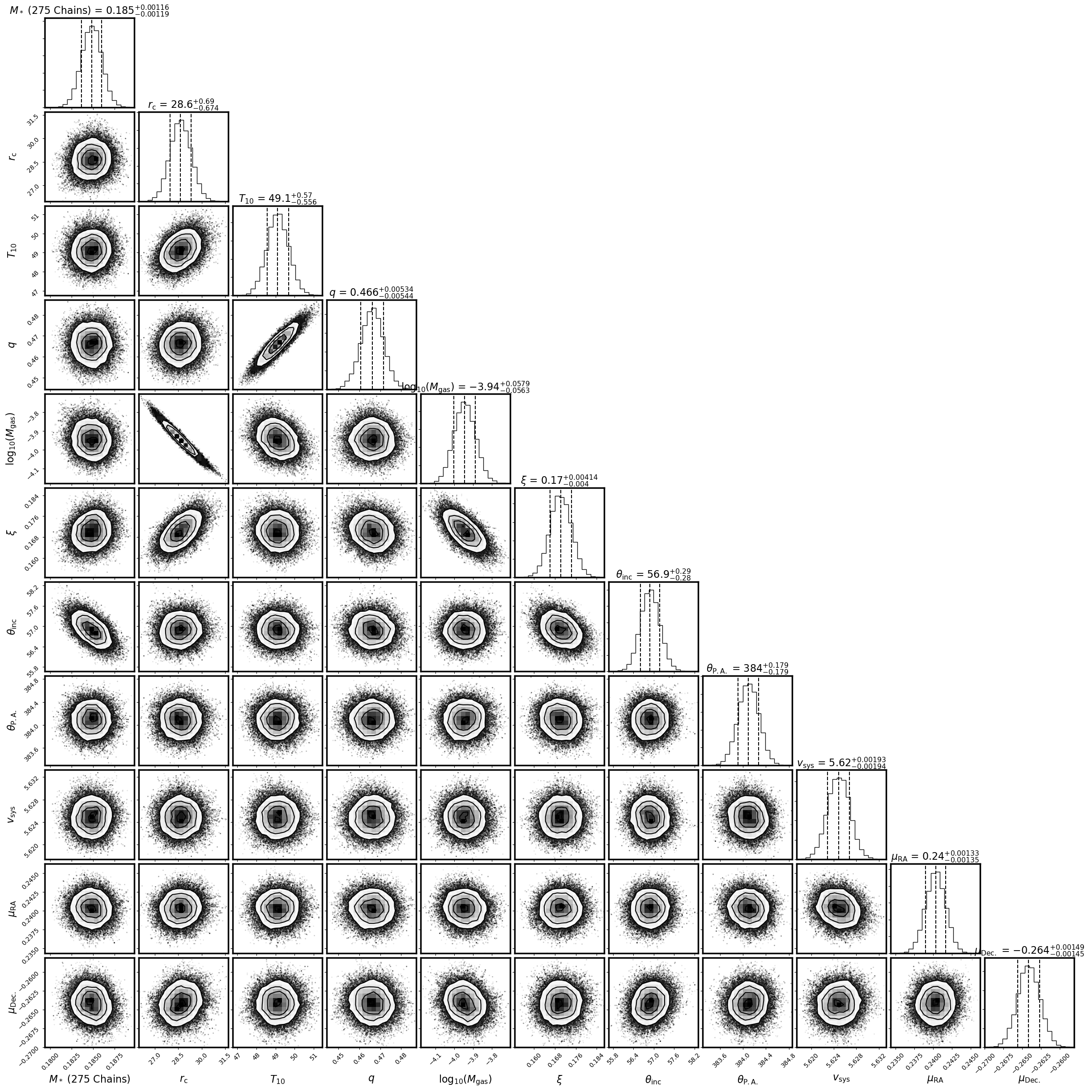}}
\caption{Corner plot for $^{12}$CO toward J0432+1827.  The number of included MCMC chains (maximum of 275) is listed at the bottom of the figure.
\label{fig_B1_4}}
\end{figure*}

\begin{figure*}
\centering
\resizebox{0.99\hsize}{!}{
    \includegraphics{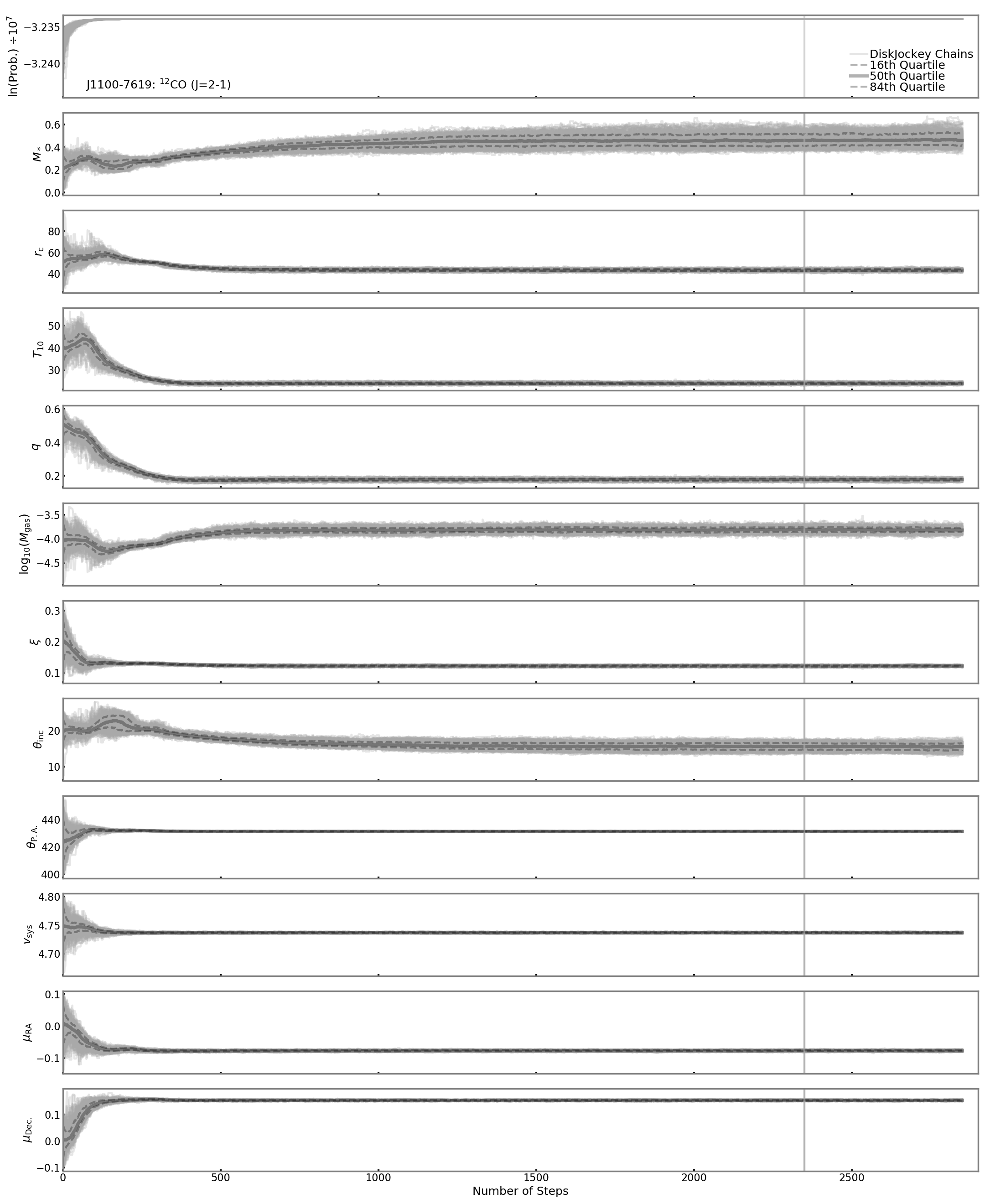}}
\caption{MCMC chain plot for $^{12}$CO toward J1100-7619.
\label{fig_B1_5}}
\end{figure*}

\begin{figure*}
\centering
\resizebox{0.99\hsize}{!}{
    \includegraphics{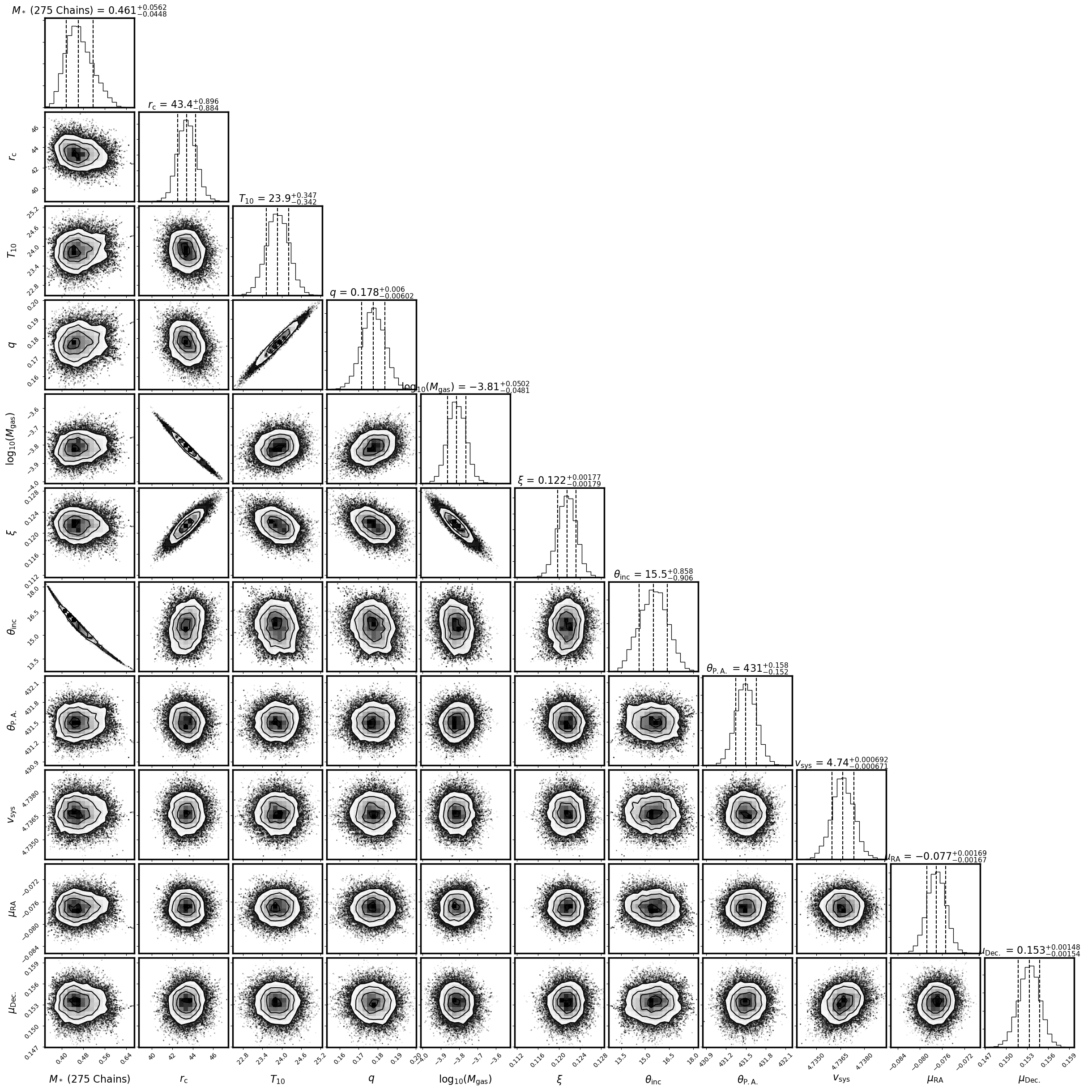}}
\caption{Corner plot for $^{12}$CO toward J1100-7619.  The number of included MCMC chains (maximum of 275) is listed at the bottom of the figure.
\label{fig_B1_6}}
\end{figure*}

Figures~\ref{fig_B1_1} through~\ref{fig_B1_6} present the MCMC chain plots and corner plots for the $^{12}$CO 2--1 emission.

%


\section{$^{13}$CO 2--1 MCMC Chains and Corner Plots}
\label{sec_appendix_MCMC_13CO}

\begin{figure*}
\centering
\resizebox{0.99\hsize}{!}{
    \includegraphics{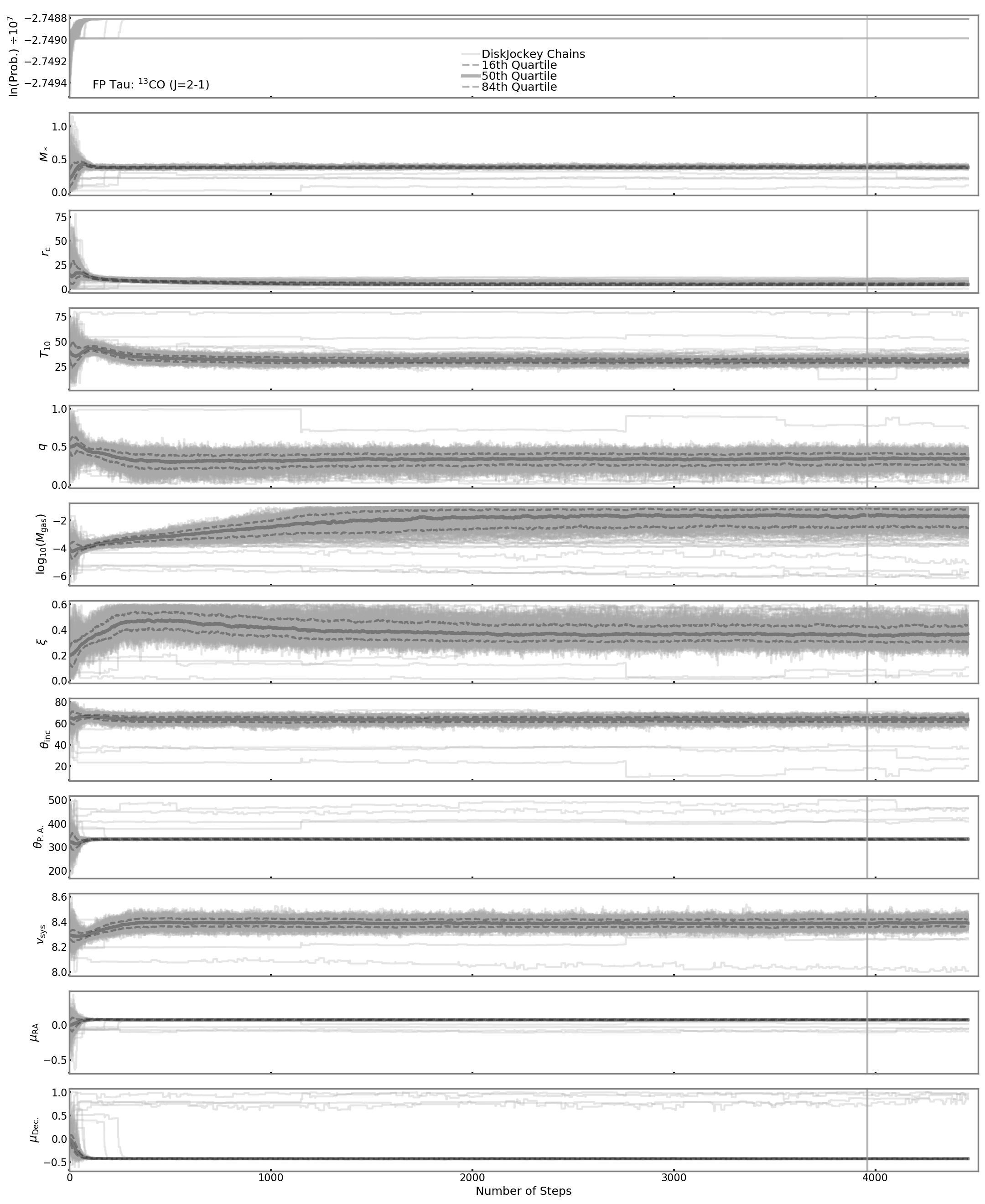}}
\caption{MCMC chain plot for $^{13}$CO toward FP Tau.  The outlying chains with probabilities significantly lower than all other chains were excluded from the final sampling distributions.
\label{fig_C1_1}}
\end{figure*}

\begin{figure*}
\centering
\resizebox{0.99\hsize}{!}{
    \includegraphics{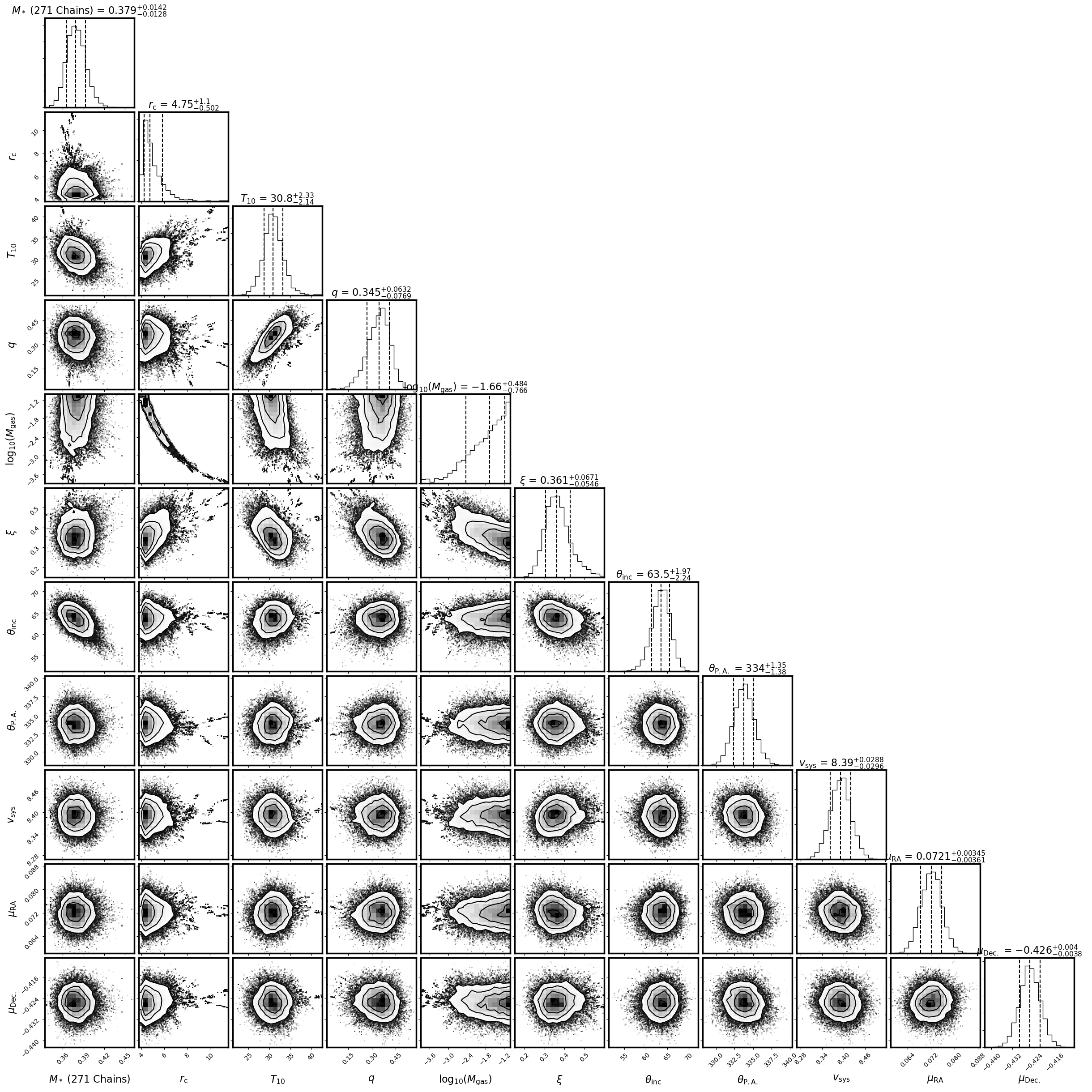}}
\caption{Corner plot for $^{13}$CO toward FP Tau.  The number of included MCMC chains (maximum of 275) is listed at the bottom of the figure.
\label{fig_C1_2}}
\end{figure*}

\begin{figure*}
\centering
\resizebox{0.99\hsize}{!}{
    \includegraphics{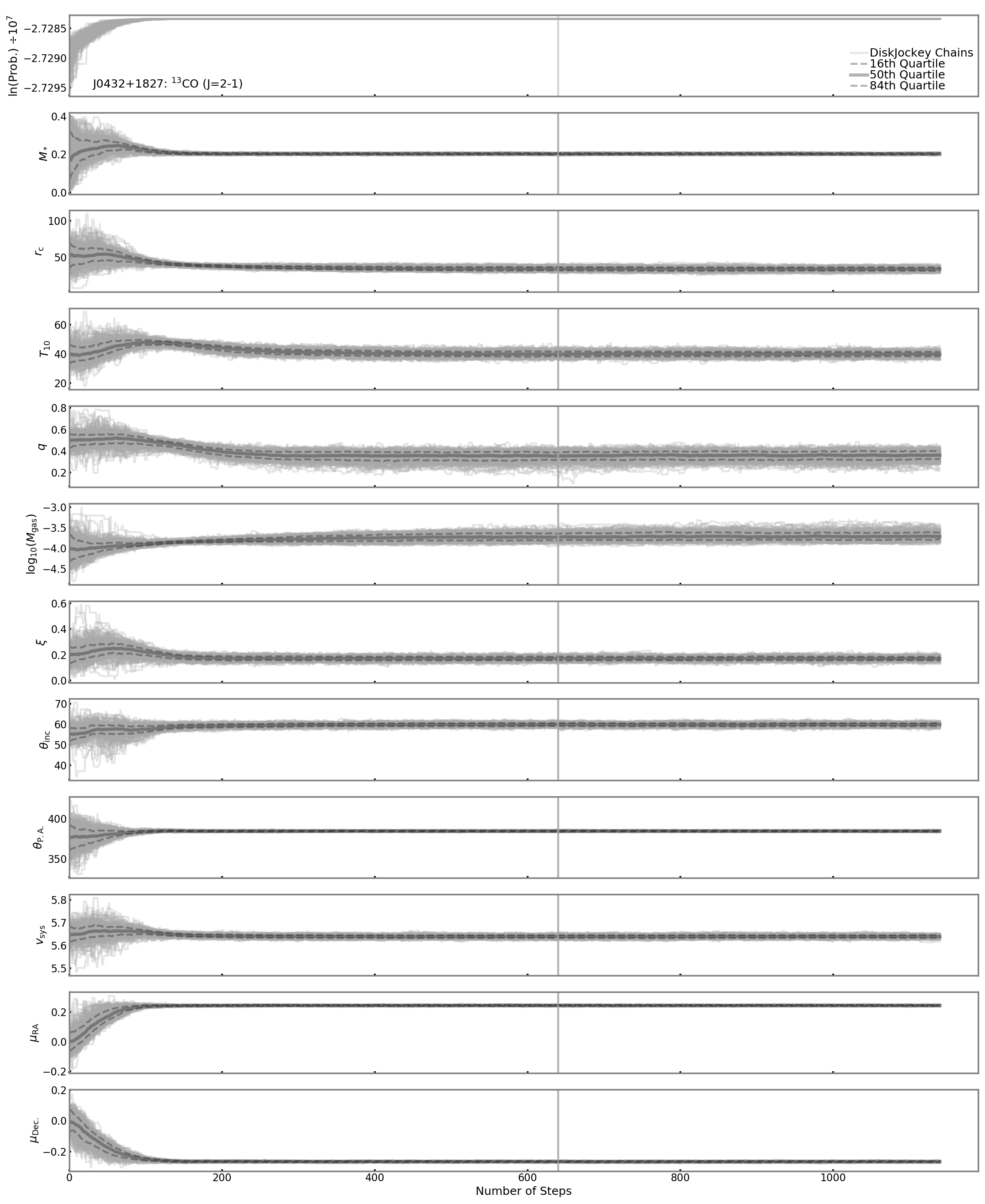}}
\caption{MCMC chain plot for $^{13}$CO toward J0432+1827.
\label{fig_C1_3}}
\end{figure*}

\begin{figure*}
\centering
\resizebox{0.99\hsize}{!}{
    \includegraphics{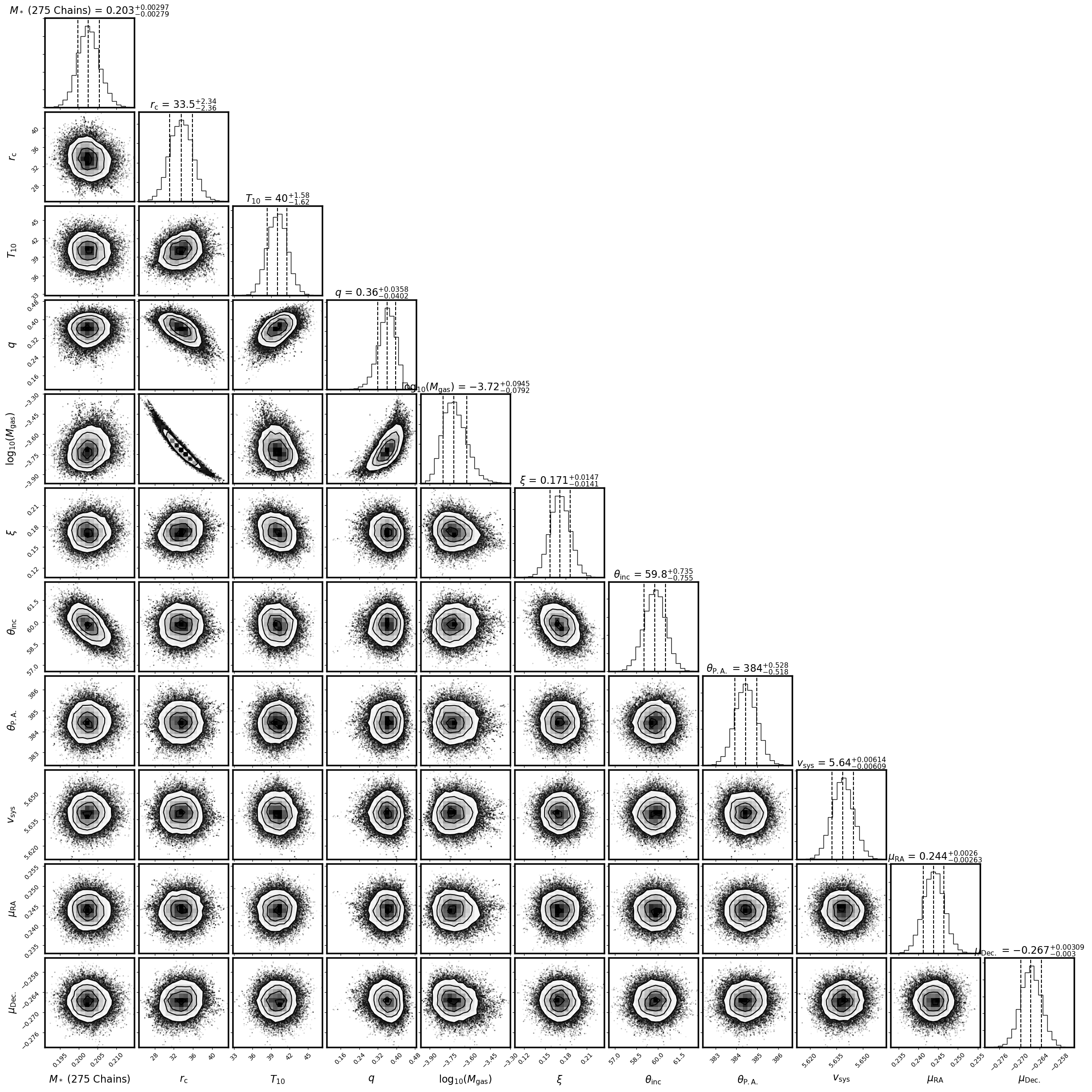}}
\caption{Corner plot for $^{13}$CO toward J0432+1827.
\label{fig_C1_4}}
\end{figure*}

\begin{figure*}
\centering
\resizebox{0.99\hsize}{!}{
    \includegraphics{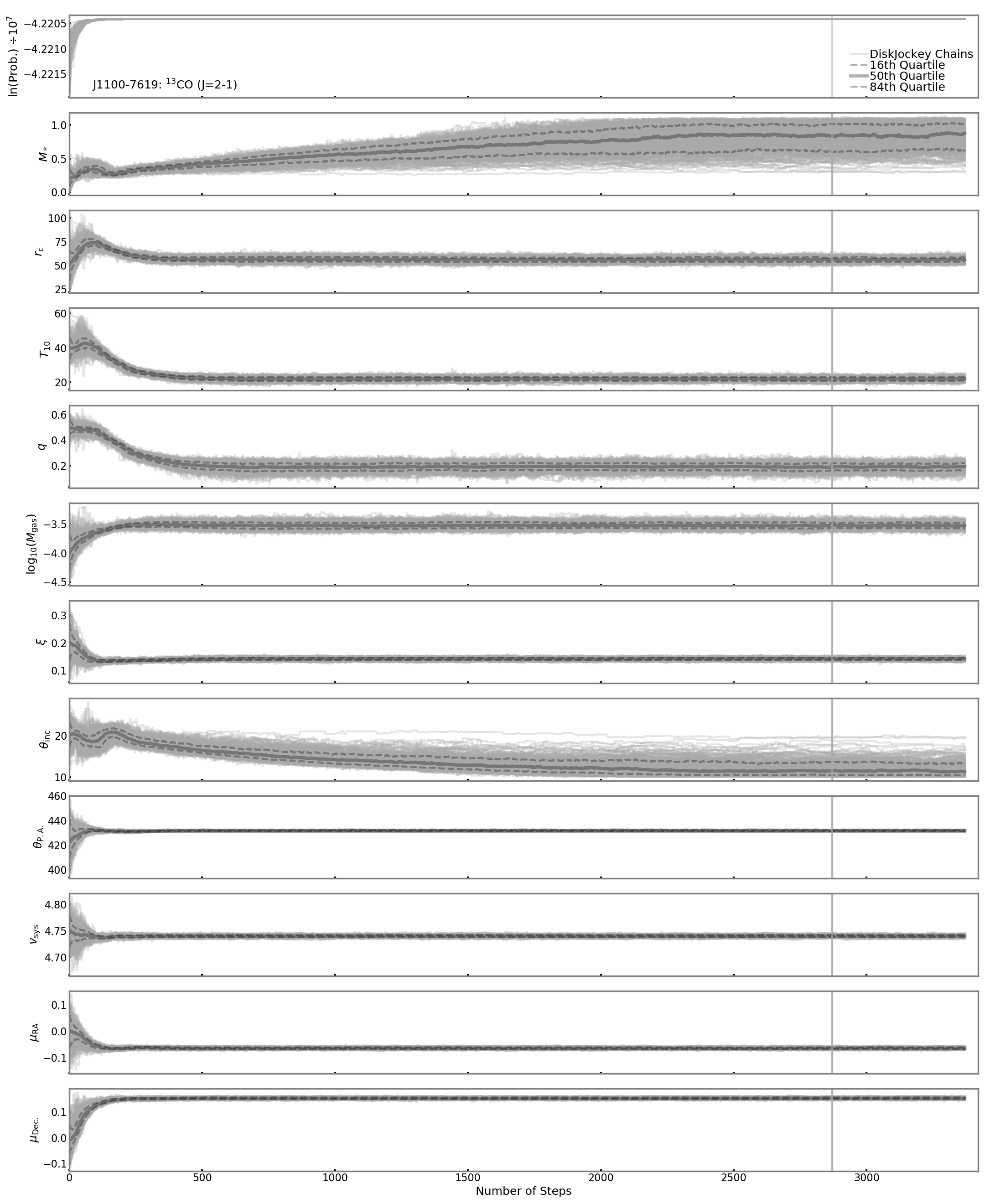}}
\caption{MCMC chain plot for $^{13}$CO toward J1100-7619.
\label{fig_C1_5}}
\end{figure*}

Figures~\ref{fig_C1_1} through~\ref{fig_C1_5} present the MCMC chain plots and corner plots for the $^{13}$CO 2--1 emission.

%


%

\clearpage

\section{Structural Parameter Constraints}
\label{sec_appendix_params}

Table~\ref{table_paramsminor} lists the final constraints on the disk structural parameters.  We stress that the $^{12}$CO and $^{13}$CO emission is optically thick toward our three star+disk systems, and so these structural parameters are likely \textit{not} true measurements of each disk's surface density and midplane temperature profiles.  Therefore, these parameters describe only the optically thick surfaces of the observed emission and should \textit{not} be interpreted as measurements of disk structure.
%

\section{The Correlation between Dynamical Mass and Inclination Angle}
\label{sec_appendix_Msin}

\begin{figure*}
\centering
\resizebox{0.99\hsize}{!}{
    \includegraphics[trim=9pt 9pt 9pt 10pt, clip]{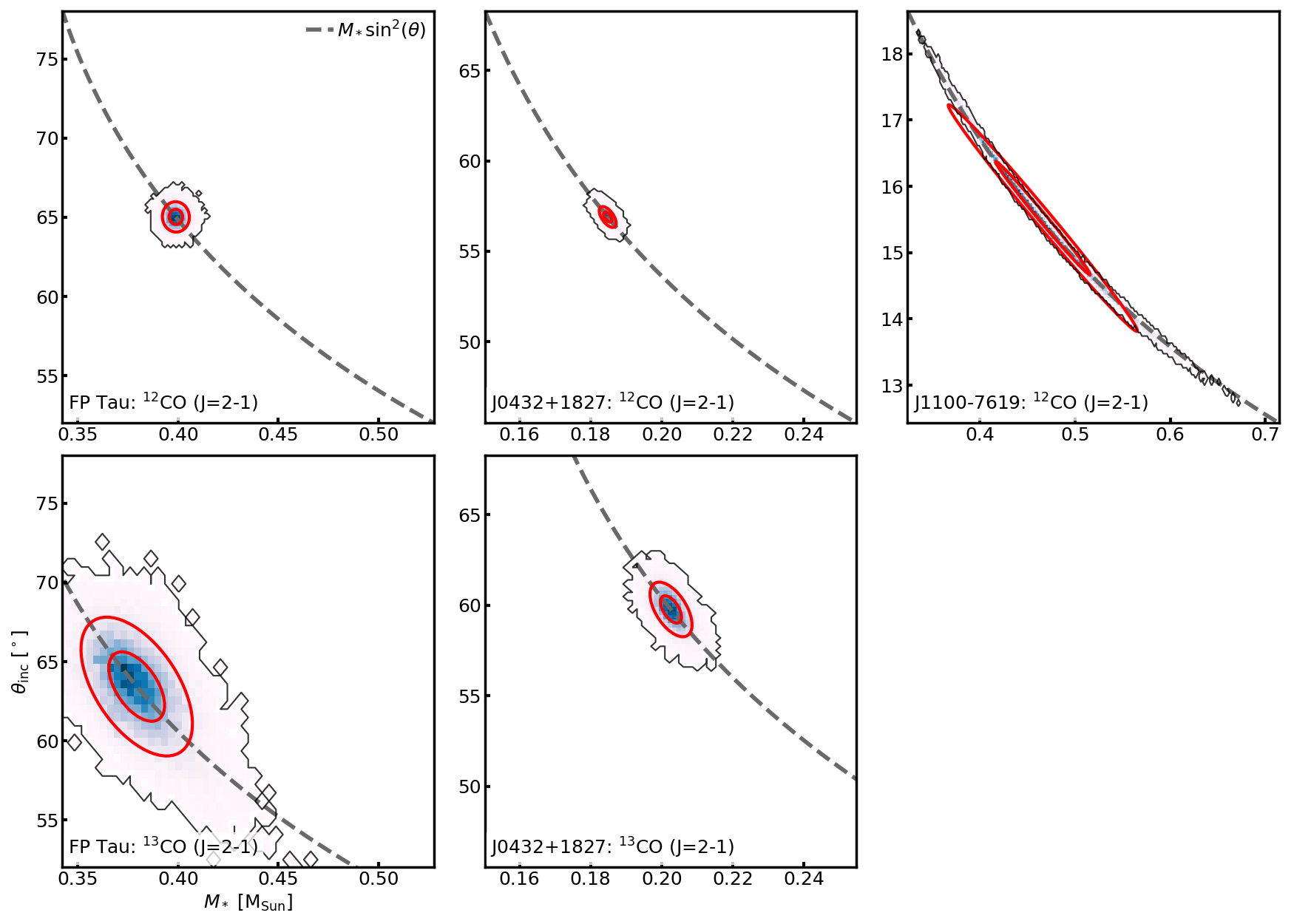}}
\caption{The constrained $M_*$ vs. $\theta_\mathrm{inc}$ sampling distributions for $^{12}$CO (top row) and $^{13}$CO (bottom row) toward FP Tau (left), J0432+1827 (middle), and J1100-7619 (right).  The sampling distributions are shown as 2D histograms, where the darker colors indicate a higher density of points.  The distributions are outlined in black for visual clarity.  The red ellipses show the 1$\sigma$ and 2$\sigma$ contours, which contain 68\% and 95\% of the data, respectively, assuming the data is Normally distributed.  Each dashed gray line draws the ($A$=$M_* \sin^2 (\theta)$) contour, where $A$ is a constant, through the median M$_*$ and $\theta_\mathrm{inc}$ values of the distributions.  For the $^{12}$CO panels, the y-axes span $\pm$20\% of the median $\theta_\mathrm{inc}$ values, and the x-axes encompass the resulting $A$=$M_* \sin^2 (\theta)$ contour.  The axes for the $^{13}$CO panels are fixed to match the $^{12}$CO panels.
\label{fig_Msin}}
\end{figure*}

For Keplerian disk-based dynamical mass measurements, the line-of-sight velocities are proportional to $(\sqrt{M_*} \sin \theta_\mathrm{inc})$~\citep[e.g.,][]{cite_dutreyetal1994}.  Figure~\ref{fig_Msin} illustrates this correlation between $M_*$ and $\theta_\mathrm{inc}$ for the sampling distributions of FP Tau, J0432+1827, and J1100-7619.  Figure~\ref{fig_Msin} also illustrates the effect on the mass measurements if our estimates of the inclination angles were changed by up to 20\%.  For $^{12}$CO toward FP Tau, the stellar mass measurement would change by at most -14\% or +32\%, while for $^{12}$CO toward J0432+1827 the mass measurement would change by at most -19\% or +38\%.  The effect would be most significant for J1100-7619, where a change in inclination angle by up to 20\% would change the stellar mass measurement by at most -30\% or +55\%.

\section{Starspot Stellar Evolutionary Model Mass Predictions}
\label{sec_appendix_fspot}

Table~\ref{table_fspot} lists the mass predictions, based on tracks from the starspot stellar evolutionary model~\citep{cite_spots}, for the combined sample of low-mass M-stars from this work and from~\cite{cite_simonetal2019}.
For the stars from the literature, the spectral types were taken directly from~\cite{cite_herczegetal2014}.  The stellar effective temperatures were taken from the~\cite{cite_herczegetal2014} conversion table, interpolated in~\cite{cite_simonetal2019} and used here assuming 0.02dex uncertainties ($\sim$1 subclass for these spectral types).  The stellar luminosities were taken from~\cite{cite_herczegetal2014}, scaled to \textit{Gaia} distances, and given 0.2dex uncertainties~\citep[the luminosity uncertainty quoted in][]{cite_herczegetal2014}.

\begin{deluxetable*}{|l|cc|cc|c|}
\tablecaption{Structural Parameter Constraints. \label{table_paramsminor}}
\tablehead{
\                   & \multicolumn{2}{c|}{FP Tau}      & \multicolumn{2}{c|}{J0432+1827}      & \multicolumn{1}{c|}{J1100-7619}    \\
\ & $^{12}$CO   & $^{13}$CO & $^{12}$CO   & $^{13}$CO & $^{12}$CO   
}
\startdata
$r_\mathrm{c}$ (AU)                   & 7.5$^{+0.56}_{-0.52}$        & 4.75$^{+1.1}_{-0.5}$             & 28.6$^{+0.69}_{-0.67}$       & 33.5$^{+2.3}_{-2.4}$        & 43.4$^{+0.9}_{-0.88}$      \\
$T_{10}$ (K)                    & 60$^{+1.3}_{-1.2}$           & 30.8$^{+2.3}_{-2.1}$             & 49.1$^{+0.57}_{-0.56}$       & 40$^{+1.6}_{-1.6}$          & 23.9$^{+0.35}_{-0.34}$            \\
$q$                                 & 0.559$^{+0.0093}_{-0.0094}$  & 0.345$^{+0.063}_{-0.077}$        & 0.466$^{+0.0053}_{-0.0054}$  & 0.36$^{+0.036}_{-0.04}$     & 0.178$^{+0.006}_{-0.006}$   \\
$\Sigma_\mathrm{c}^*$ (g   cm$^{-2}$) & 56.4$^{+84}_{-33}$           & 1.38e+03$^{+3.9e+03}_{-1.2e+03}$ & 0.196$^{+0.039}_{-0.032}$    & 0.24$^{+0.1}_{-0.065}$      & 0.115$^{+0.019}_{-0.016}$   \\
$\xi$ (km s$^\mathrm{-1}$)                & 0.1$^{+0.02}_{-0.022}$       & 0.361$^{+0.067}_{-0.055}$        & 0.17$^{+0.0041}_{-0.004}$    & 0.171$^{+0.015}_{-0.014}$   & 0.122$^{+0.0018}_{-0.0018}$  
\enddata
\tablecomments{The final constraints from \textsc{DiskJockey} on the structural parameters of the three star+disk systems.  All parameters are described in Section~\ref{sec_methodology_model}.  These were measured from $^{12}$CO and $^{13}$CO independently.  No values are shown for $^{13}$CO toward J1100-7619 because the corresponding model was unconstrained.  These parameter values should \textit{not} be interpreted as measurements of disk structure.  $*$: $\Sigma_\mathrm{c}$ was calculated from the $\log_{10}(M_\mathrm{gas})$ values determined by \textsc{DiskJockey}.  We display the former here because it is more interpretable with $r_\mathrm{c}$.}
\end{deluxetable*}

\begin{deluxetable*}{|l|cccccccccccc|}
\rotate
\tablecaption{Starspot Stellar Evolutionary Model Mass Predictions. \label{table_fspot}}
\tablehead{
Disk        & Dist   & Spectral   & $L_*$ & $T_{\mathrm{eff}}$ & $M_\mathrm{dyn}$ & $M_\mathrm{fid}$ & $M_\mathrm{spot}$  & $M_\mathrm{spot}$  & $M_\mathrm{spot}$  & $M_\mathrm{spot}$  & $M_\mathrm{spot}$  & $M_\mathrm{spot}$  \\
        &   & Type   &  &  &  &  &  $f_\mathrm{spot}$=0.00 & $f_\mathrm{spot}$=0.17 & $f_\mathrm{spot}$=0.34 & $f_\mathrm{spot}$=0.51 & $f_\mathrm{spot}$=0.68 & $f_\mathrm{spot}$=0.85 \\
         &  {(}pc{)} &   &  ($L_\Sun$)    &  (K) & ($M_\Sun$) & ($M_\Sun$) & ($M_\Sun$) & ($M_\Sun$) & ($M_\Sun$) & ($M_\Sun$) & ($M_\Sun$) & ($M_\Sun$)
}
\startdata
FP Tau     & 128$^{+2}_{-2}$ & M4         & 0.269$^{+0.033}_{-0.029}$ & 3311$^{+156}_{-149}$ & 0.395$^{+0.012}_{-0.012}$ & 0.240$^{+0.091}_{-0.066}$ & 0.240$^{+0.099}_{-0.054}$ & 0.302$^{+0.087}_{-0.088}$ & 0.355$^{+0.092}_{-0.104}$ & 0.407$^{+0.117}_{-0.105}$ & 0.479$^{+0.152}_{-0.116}$ & 0.575$^{+0.183}_{-0.139}$ \\
J0432+1827 & 141$^{+3}_{-3}$ & M4.75      & 0.076$^{+0.024}_{-0.018}$ & 3162$^{+149}_{-142}$ & 0.192$^{+0.005}_{-0.005}$ & 0.174$^{+0.083}_{-0.048}$ & 0.170$^{+0.087}_{-0.050}$ & 0.219$^{+0.105}_{-0.071}$ & 0.269$^{+0.120}_{-0.083}$ & 0.363$^{+0.105}_{-0.123}$ & 0.447$^{+0.184}_{-0.207}$ & 0.513$^{+0.263}_{-0.174}$ \\
J1100-7619 & 190$^{+4}_{-4}$ & M4         & 0.145$^{+0.085}_{-0.053}$ & 3311$^{+77}_{-75}$   & 0.461$^{+0.057}_{-0.057}$ & 0.251$^{+0.051}_{-0.042}$ & 0.251$^{+0.051}_{-0.042}$ & 0.302$^{+0.061}_{-0.051}$ & 0.363$^{+0.064}_{-0.054}$ & 0.437$^{+0.076}_{-0.065}$ & 0.550$^{+0.053}_{-0.103}$ & 0.646$^{+0.046}_{-0.109}$ \\
\hline
CX Tau     & 128$^{+1}_{-1}$ & M2.5       & 0.249$^{+0.146}_{-0.092}$ & 3485$^{+164}_{-157}$ & 0.380$^{+0.020}_{-0.020}$ & 0.355$^{+0.124}_{-0.092}$ & 0.355$^{+0.124}_{-0.092}$ & 0.407$^{+0.142}_{-0.105}$ & 0.468$^{+0.163}_{-0.121}$ & 0.550$^{+0.192}_{-0.142}$ & 0.646$^{+0.225}_{-0.167}$ & 0.741$^{+0.259}_{-0.192}$ \\
GO Tau     & 144$^{+1}_{-1}$ & M2.3       & 0.211$^{+0.123}_{-0.078}$ & 3515$^{+166}_{-158}$ & 0.450$^{+0.010}_{-0.010}$ & 0.372$^{+0.130}_{-0.096}$ & 0.380$^{+0.133}_{-0.098}$ & 0.437$^{+0.152}_{-0.113}$ & 0.501$^{+0.175}_{-0.130}$ & 0.589$^{+0.205}_{-0.152}$ & 0.676$^{+0.236}_{-0.175}$ & 0.759$^{+0.265}_{-0.196}$ \\
HO Tau     & 161$^{+1}_{-1}$ & M3.2       & 0.142$^{+0.083}_{-0.052}$ & 3365$^{+159}_{-151}$ & 0.430$^{+0.030}_{-0.030}$ & 0.282$^{+0.098}_{-0.073}$ & 0.288$^{+0.101}_{-0.075}$ & 0.339$^{+0.118}_{-0.088}$ & 0.407$^{+0.142}_{-0.105}$ & 0.479$^{+0.167}_{-0.124}$ & 0.575$^{+0.201}_{-0.149}$ & 0.646$^{+0.225}_{-0.167}$ \\
CY Tau     & 128$^{+1}_{-1}$ & M2.3       & 0.253$^{+0.148}_{-0.093}$ & 3500$^{+165}_{-158}$ & 0.300$^{+0.020}_{-0.020}$ & 0.355$^{+0.124}_{-0.092}$ & 0.363$^{+0.127}_{-0.094}$ & 0.417$^{+0.145}_{-0.108}$ & 0.479$^{+0.167}_{-0.124}$ & 0.562$^{+0.196}_{-0.145}$ & 0.661$^{+0.231}_{-0.171}$ & 0.759$^{+0.265}_{-0.196}$ \\
DE Tau     & 127$^{+1}_{-1}$ & M2.3       & 0.492$^{+0.288}_{-0.182}$ & 3515$^{+166}_{-158}$ & 0.410$^{+0.030}_{-0.030}$ & 0.355$^{+0.124}_{-0.092}$ & 0.355$^{+0.124}_{-0.092}$ & 0.407$^{+0.142}_{-0.105}$ & 0.457$^{+0.160}_{-0.118}$ & 0.525$^{+0.183}_{-0.136}$ & 0.603$^{+0.210}_{-0.156}$ & 0.692$^{+0.241}_{-0.179}$ \\
FM Tau     & 131$^{+1}_{-1}$ & M4.5       & 0.071$^{+0.042}_{-0.026}$ & 3085$^{+145}_{-139}$ & 0.360$^{+0.020}_{-0.020}$ & 0.151$^{+0.067}_{-0.037}$ & 0.148$^{+0.066}_{-0.036}$ & 0.178$^{+0.079}_{-0.043}$ & 0.224$^{+0.100}_{-0.054}$ & 0.282$^{+0.126}_{-0.068}$ & 0.355$^{+0.158}_{-0.086}$ & 0.457$^{+0.204}_{-0.110}$
\enddata
\tablecomments{Starspot stellar evolutionary model predictions for the low-mass M-stars in this work (top three rows), and for a subset of Taurus stars from~\cite{cite_simonetal2019} (bottom six rows).  All distances (Dist.) are from \textit{Gaia}~\citep[e.g.,][]{cite_gaia2016, cite_gaia2018b}.  The spectral types, stellar luminosities ($L_*$), and stellar effective temperatures ($T_\mathrm{eff}$) were derived from~\cite{cite_herczegetal2014} as described in the text.  The dynamical masses are either from this work (top three rows) or from~\cite{cite_simonetal2019} (bottom six rows).  The fiducial stellar evolutionary model mass predictions ($M_\mathrm{fid}$) were all computed using tracks from the MIST code~\citep{cite_mist1, cite_mist2}.  The starspot stellar evolutionary model mass predictions ($M_\mathrm{spot}$) were computed using tracks from the SPOTS code~\citep{cite_spots} and the fractional starspot coverage values ($f_\mathrm{spot}$) listed in the header.}
\end{deluxetable*}


\bibliography{projectbib}

\end{document}